%% file: main.tex
\documentclass[acmsmall
,nonacm
,screen
]{acmart}

\newcommand{\Scategorysurround}{\mathbb{S}}
\newcommand{\cospan}[1]{\mathbf{Cospan}(#1)}
\newcommand{\Span}[1]{\mathbf{Span}(#1)}
\newcommand{\cobord}[1]{\mathbf{Cob}(#1)}
\newcommand{\games}[1]{\mathbf{Games}(#1)}
\newcommand{\spanbicat}[1]{\mathbf{Span}(#1)}
\newcommand{\source}[1]{\textit{in}}
\newcommand{\target}[1]{\textit{out}}
\newcommand{\inl}{\textit{inl}}
\newcommand{\inr}{\textit{inr}}
\newcommand{\setofmachinestates}{\mathit{State}}
\newcommand{\anchortensor}{\anchor^{\|}}
\newcommand{\anchorstate}{\anchor_{S}}
\newcommand{\anchorstatetensor}{\anchor_{S}^{\|}}
\newcommand{\anchorproof}{\anchor_{Sep}}
\newcommand{\anchorlogic}{\anchor_{L}}
\newcommand{\semin}[1]{\sem{#1}_{\textit{in}}}
\newcommand{\semout}[1]{\sem{#1}_{\textit{out}}}
\newcommand{\semsupport}[1]{\sem{#1}_{\textit{support}}}
\newcommand{\seminS}[1]{\sem{#1}_{S,\textit{in}}}
\newcommand{\semoutS}[1]{\sem{#1}_{S,\textit{out}}}
\newcommand{\semsupportS}[1]{\sem{#1}_{S,\textit{support}}}
\newcommand{\seminL}[1]{\sem{#1}_{L,\textit{in}}}
\newcommand{\semoutL}[1]{\sem{#1}_{L,\textit{out}}}
\newcommand{\semsupportL}[1]{\sem{#1}_{L,\textit{support}}}
\newcommand{\semSepin}[1]{\sem{#1}_{\textit{Sep,in}}}
\newcommand{\semSepout}[1]{\sem{#1}_{\textit{Sep,out}}}
\newcommand{\semSepsupport}[1]{\sem{#1}_{\textit{Sep,support}}}
\newcommand{\Cplayer}{\mathbf{C}}
\newcommand{\Fplayer}{\mathbf{F}}
\newcommand{\machinemodel}{\anchor^{\bullet}}
\newcommand{\statefulmachinemodel}{\anchor^{\bullet}_S}
\newcommand{\statelessmachinemodel}{\anchor^{\bullet}_L}
\newcommand{\machinemodeltwoplayers}{\anchor^{\circ\bullet}}
\newcommand{\statefulmachinemodeltwoplayers}{\anchor^{\circ\bullet}_S}
\newcommand{\statelessmachinemodeltwoplayers}{\anchor^{\circ\bullet}_L}
\newcommand{\statefulmachinemodelthreeplayers}{\anchor^{\circ\circ\bullet}_S}
\newcommand{\statelessmachinemodelthreeplayers}{\anchor^{\circ\circ\bullet}_L}
\newcommand{\machinemodelthreeplayers}{\anchor^{\circ\circ\bullet}}
\newcommand{\Sepmodel}{\anchor^{\bullet}_\textit{Sep}}
\newcommand{\Sepmodeltwoplayers}{\anchor^{\circ\bullet}_\textit{Sep}}
\newcommand{\Sepmodelthreeplayers}{\anchor^{\circ\circ\bullet}_\textit{Sep}}

\usepackage{etoolbox}
\newtoggle{enable comments}

\input{prelude}

\begin{document}

\newtheorem{remark}[theorem]{Remark}


\title[Concurrent Separation Logic Meets Template Games]{Concurrent Separation Logic Meets Template Games}

\author{Paul-Andr\'e Melli\`es}
\email{mellies@irif.fr}          
\author{L\'eo Stefanesco}
\email{leo.stefanesco@irif.fr}

\affiliation{
  \institution{IRIF, CNRS \& Universit\'e de Paris}
  \streetaddress{8 place Aur\'elie Nemours}
  \city{Paris}
  \postcode{75013}
  \country{France}
}


\begin{abstract}
  An old dream of concurrency theory and programming language
  semantics has been to uncover the fundamental synchronization
  mechanisms which regulate situations as different as game semantics
  for higher-order programs, and Hoare logic for concurrent programs
  with shared memory and locks.
  In this paper, we establish a deep and unexpected connection between
  two recent lines of work on concurrent separation logic (CSL) and on
  template game semantics for differential linear logic (DiLL).
  Thanks to this connection, we reformulate in the purely conceptual
  style of template games for DiLL the asynchronous and interactive
  interpretation of CSL designed by Melli{\`e}s and Stefanesco.
  We believe that the analysis reveals something important about the
  secret anatomy of CSL, and more specifically about the subtle
  interplay, of a categorical nature, between sequential composition,
  parallel product, errors and locks.
\end{abstract}

\maketitle

\section{Introduction}
In this paper, we explore an instructive and
fascinating analogy between 
the notion of Hoare triple $\{P\}C\{Q\}$ in programming language
semantics, and the notion of cobordism in mathematical physics,
conveniently formulated in the categorical language of cospans \cite{BaezStay,Grandis,SassoneSobocinski,DayStreet2004,Mellies-dialogue-frobenius}.
Consider the typical situation of a code $C$ written in an imperative
and sequential programming language, and executed on a set
$\setofmachinestates$ of machine states.
By construction, the code $C$ comes equipped with a set $C_{in}$ of
input states and a set $C_{out}$ of output states, together with a
pair of labeling functions
\begin{equation}\label{equation/lambdain-lambdaout}
\lambda_{in} : C_{in}\to \setofmachinestates
\quad\quad\quad
\lambda_{out}: C_{out}\to \setofmachinestates
\end{equation}
which assign to every input state $s\in C_{in}$ and output state
$s'\in C_{out}$ of the code $C$ its underlying machine state
$\lambda_{in}(s)\in \setofmachinestates$ and
$\lambda_{out}(s')\in \setofmachinestates$.
Following the philosophy of Hoare logic, the predicates $P$ and $Q$
describe specific subsets of the set $\setofmachinestates$ of states
of the machine, which induce (by inverse image) predicates $P_{in}$ on
$C_{in}$ and $Q_{out}$ on $C_{out}$.
The Hoare triple $\{P\}C\{Q\}$ then expresses the fact that the code
$C$ transports every input state $s\in C_{in}$ which satisfies the
predicate $P_{in}$ into an output state $s'\in C_{out}$ which
satisfies the predicate $Q_{out}$.
Looking at the situation with the eyes of the physicist, the
set~$C_{in}$ of input states can be depicted as a two-dimensional disk
(in gray) with the predicate~$P_{in}$ represented as a subset or
subdisk (in blue) living inside~$C_{in}$ ; and similarly for the
predicate~$Q_{out}$ on the set~$C_{out}$ of output states:
\begin{equation}\label{equation/input-output-disk}
\raisebox{-2.4em}{\includegraphics[height=5em]{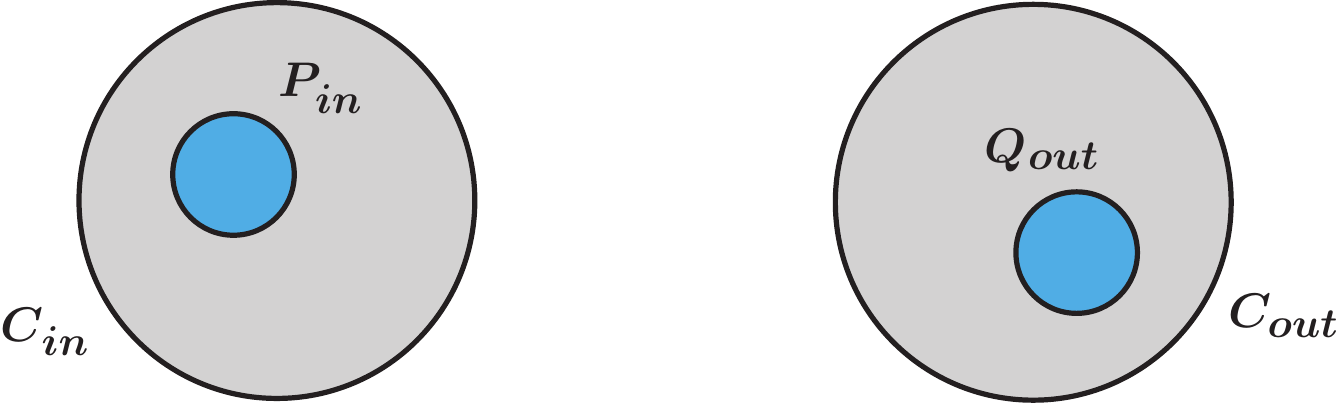}}
\end{equation}
Since the purpose of the code $C$ is to transport the input states
$s\in C_{in}$ to output states $s'\in C_{out}$, it makes sense to
depict $C$ (in white) as a three-dimensional tube or cylinder
connecting the input disk~$C_{in}$ to the output disk~$C_{out}$.
Here, the three-dimensional cylinder $C$ should be understood as a
geometric object living in space and time, and describing the
evolution in time of the internal states of the code $C$ in the course
of execution:
\begin{equation}\label{equation/Sin-Sout-C-tube}
\raisebox{-2.4em}{\includegraphics[height=5.5em]{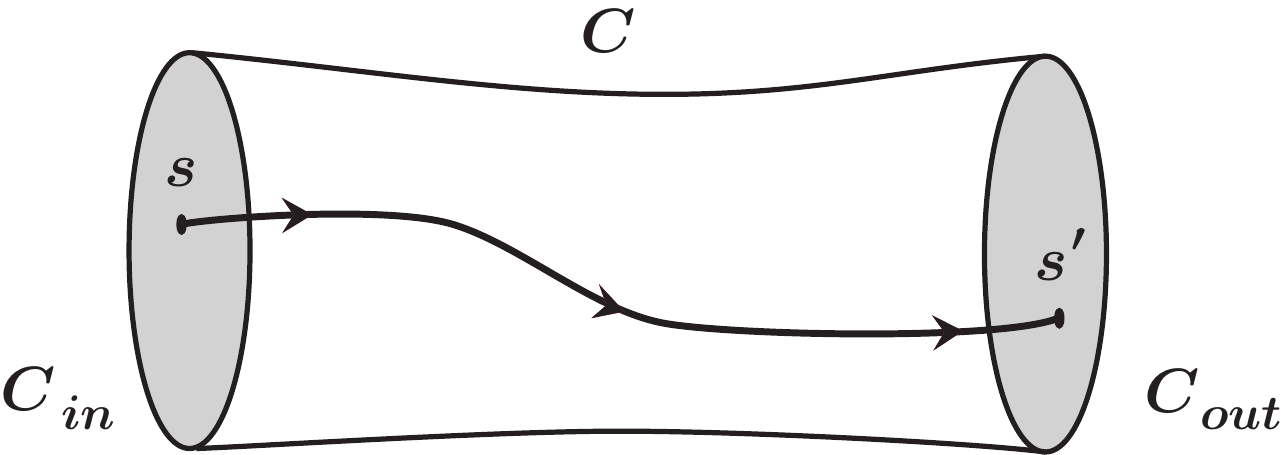}}
\end{equation}
%
The physical intuition that the code $C$ transports states
$s\in C_{in}$ to states $s'\in C_{out}$ may be expressed by equipping
the cylinder with a (three-dimensional) vector field describing how a
state evolves in time through the code $C$.
The description is reminiscent of fluid mechanics, and the way one
describes the trajectory of a particle along a flow in the cylinder
$C$.
In that graphical representation, the Hoare triple $P\{C\}Q$ indicates
that the flow of execution described by the vector field on the
cylinder $C$ transports every state $s\in C_{in}$ starting from
$P_{in}$ to a state $s'\in C_{out}$ exiting in $Q_{out}$:
\begin{equation}\label{equation/cobordism-from-P-to-Q}
\raisebox{-3.5em}{\includegraphics[height=7.5em]{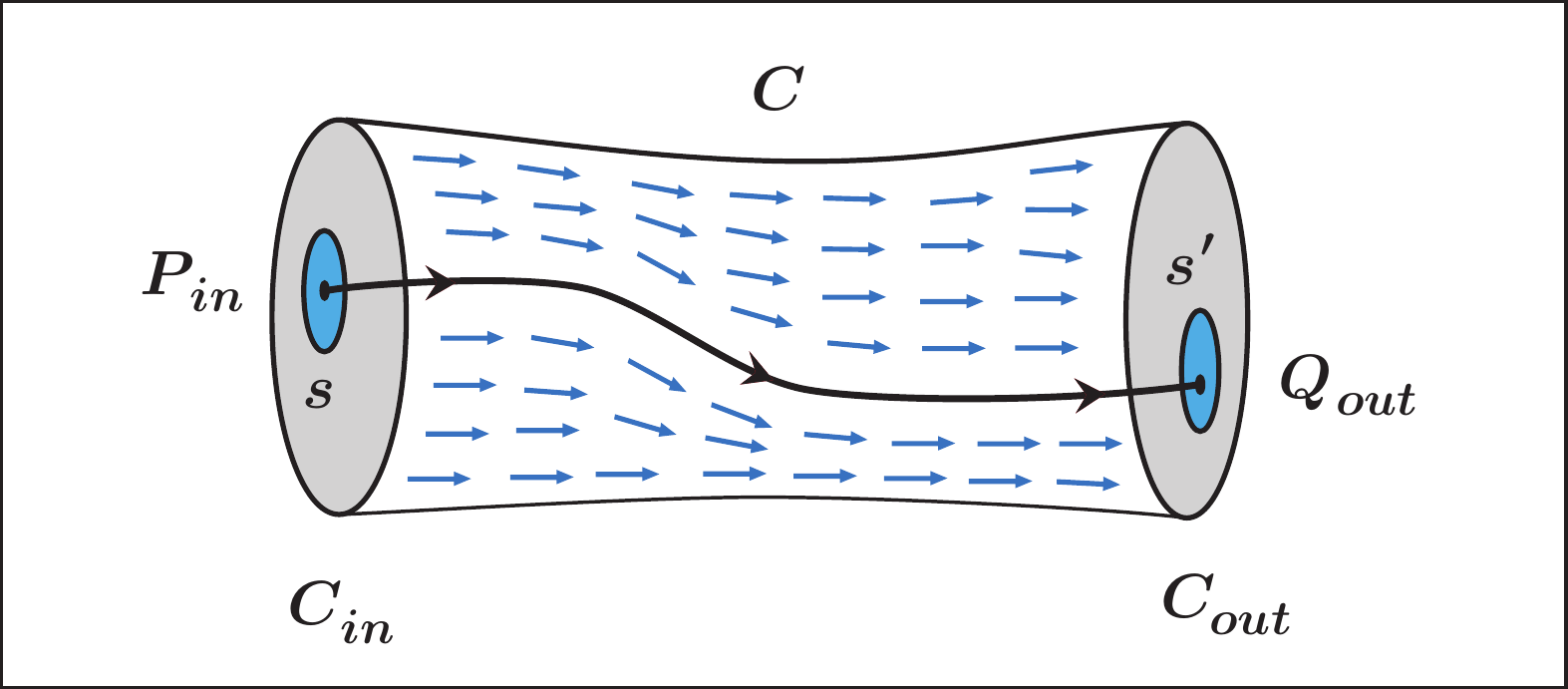}}
\end{equation}
%
This intuition underlies the sequential rule of Hoare logic
\begin{equation}\label{equation/sequential-composition}
\AxiomC{$\{P\}C\{Q\}$}
\AxiomC{$\{Q\}D\{R\}$}
\RightLabel{\quad sequential composition}
\BinaryInfC{$\{P\}C;D\{R\}$}
\DisplayProof
\end{equation}
which states that Hoare triples ``compose well'' in the sense that the
Hoare triple $\{P\}C;D\{R\}$ holds whenever the Hoare triples
$\{P\}C\{Q\}$ and $\{Q\}D\{R\}$ are assumed to hold.
This basic principle of Hoare logic reflects the fact that when the
two codes $C$ and $D$ are executed sequentially as below
\begin{equation}\label{equation/cobordism-from-P-to-Q-to-R}
\raisebox{-3.5em}{\includegraphics[height=7.5em]{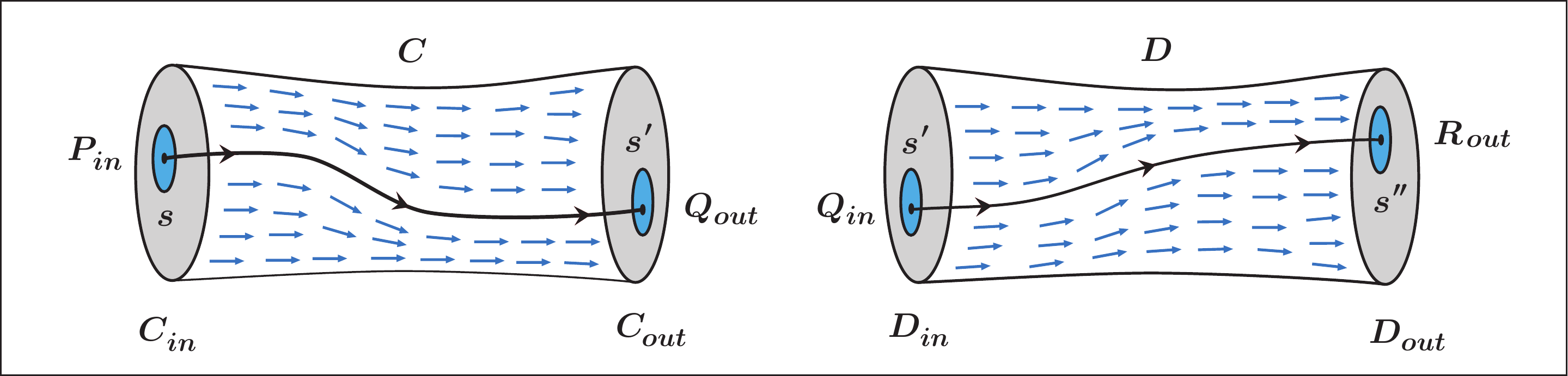}}
\end{equation}
every state $s\in C_{in}$ satisfying the predicate $P_{in}$ is
transported by the code $C$ to a transitory state $s'\in C_{out}$
satisfying the predicate $Q_{out}$; and that the same transitory state
$s'\in D_{in}$ taken now as input is transported by the code $D$ to a
state $s''\in D_{out}$ satisfying the predicate $R_{out}$.
The careful reader will notice that we make here the simplifying and
somewhat unrealistic assumption that the set $C_{out}$ of output
states of the code $C$ coincides with the set $D_{in}$ of input states
of the code $D$; as we will see, this point is interesting and far
from anecdotal, and we will thus come back to it with great attention
at a later stage of the paper, see \S\ref{sec:sequential-composition}.

\paragraph{Cospans of transition systems}
By taking seriously these geometric intuitions and formalizing them in
the language of category theory, we establish in this paper a strong
and unexpected connection between two recent and largely independent
lines of work on concurrent separation logic \cite{mfps,lics18} and on
template games for differential linear logic \cite{popl19,lics19}.
The connection enables us to disclose for the first time a number of
basic and fundamental categorical structures underlying concurrent
separation logic, and more specifically, the proof of the asynchronous
soundness theorem established in \cite{lics18}.
%
Our starting point is 
to define a \textbf{transition system} $(A,\lambda_A)$
on a given labeling graph $\anchor_{label}$ as a morphism
\begin{equation}\label{equation/labeling-map}
\begin{tikzcd}[column sep=1.5em]
{\lambda_A}
\quad : \quad
A
\arrow[rr]
&& 
{\anchor_{label}}
\end{tikzcd}
\end{equation}
in the category $\mathbf{Gph}$ of directed graphs.
The graph $A$ is called the \emph{support} of the transition system
$(A,\lambda_A)$, and $\lambda_A$ its \emph{labeling map}.
It is important here that the transition system $(A,\lambda_A)$ is
labeled with a graph instead of a set, and that it is multi-sorted
with sorts provided by the vertices of the labeling graph
$\anchor_{label}$.
A morphism between two such transition systems
\begin{equation}\label{equation/transition-system-morphism}
\begin{tikzcd}[column sep=1.2em]
(f,\varphi)\quad : \quad (A,\lambda_A)
\arrow[rr]
&& 
(B,\lambda_B)
\end{tikzcd}
\end{equation}
is defined as a pair $(f,\varphi)$ of graph morphisms making the diagram below commute
\begin{equation}\label{equation/morphism-between-transition-systems}
\begin{tikzcd}[column sep=2.5em, row sep=1em]
A
\arrow[dd,"{\lambda_{A}}"{swap}]
\arrow[rr,"{f}"]
&& 
B
\arrow[dd,"{\lambda_{B}}"]
\\
\\
\anchor_{label,1}
\arrow[rr,"{\varphi}"]
&&
\anchor_{label,2}
\end{tikzcd}
\end{equation}
in the category $\mathbf{Gph}$.
The graph morphism $\varphi:\anchor_{label,1}\to\anchor_{label,2}$
between labeling graphs works in the same way as a ``change of base''
and is called the \emph{relabeling map} of the
morphism~(\ref{equation/transition-system-morphism}).
%
%
At this stage, we are ready to formulate the guiding idea of the
paper, which is that the situation of the three-dimensional cylinder
$C$ connecting the sets $C_{in}$ and $C_{out}$ of input and output
states depicted in (\ref{equation/Sin-Sout-C-tube}) can be
conveniently formulated as a \emph{cospan of transition systems}
\begin{equation}\label{equation/first-cospan}
\begin{tikzcd}[column sep=3.5em, row sep=1.2em]
(C_{in},\lambda_{in})
\arrow[rr,"{(\source{},\eta)}"]
&& 
(C,\lambda_{code})
&&
(C_{out},\lambda_{out})
\arrow[ll,"{(\target{},\eta)}"{swap}]
\end{tikzcd}
\end{equation}
defining a commutative diagram in the category $\mathbf{Gph}$:
\begin{equation}\label{equation/second-cospan}
\begin{tikzcd}[column sep=2.5em, row sep=.8em]
C_{in}
\arrow[dd,"{\lambda_{in}}"{swap}]
\arrow[rr,"{\source{}}"]
&& 
C
\arrow[dd,"{\lambda_{code}}"]
&&
C_{out}
\arrow[dd,"{\lambda_{out}}"]
\arrow[ll,"{\target{}}"{swap}]
\\
\\
\anchorstate[0]
\arrow[rr,"{\eta}"]
&&
\anchorstate[1]
&&
\anchorstate[0]
\arrow[ll,"{\eta}"{swap}]
\end{tikzcd}
\end{equation}
%
One important feature of the cospan formulation of
(\ref{equation/Sin-Sout-C-tube}) is the simple and intuitive
definition of the labeling graphs $\anchorstate[0]$ and
$\anchorstate[1]$ and of the relabeling map
$\eta:\anchorstate[0]\to\anchorstate[1]$.
Both labeling graphs $\anchorstate[0]$ and $\anchorstate[1]$ have the
set $\setofmachinestates$ of machine states as set of vertices.
The difference between them is that the graph~$\anchorstate[0]$ is
discrete (that is, it has no edge) and can be thus identified with the
set $\setofmachinestates$ itself, while the graph~$\anchorstate[1]$
has as edges the transitions $inst:s\to s'$ performed by the
instructions $inst\in Inst$ of the machine.
The relabeling map 
\begin{equation}\label{equation/unit-of-the-monad}
\begin{tikzcd}[column sep=1.5em]
\eta
\,\, : \,\,
{\anchorstate[0]}
\arrow[rr]
&& 
{\anchorstate[1]}
\end{tikzcd}
\end{equation}
is the graph morphism which maps every machine state
$s\in \setofmachinestates$ to itself.
The graph $C$ together with the label map
$\lambda_{code}:C\to\anchorstate[1]$ describe the transition system
$(C, \lambda_{code})$ defined by the operational semantics of the
code.
Note in particular that $\lambda_{code}$ labels every vertex (or
internal state) of the graph~$C$ with a machine
state~$\lambda_{code}(s)\in \setofmachinestates$ and every edge (or
execution step) $m:s\to s'$ of the graph~$C$ with a machine
instruction
$\lambda_{code}(m):\lambda_{code}(s)\to \lambda_{code}(s')$.
%
%
The sets $C_{in}$ and $C_{out}$ of input and output states of the code
$C$ are understood in~(\ref{equation/first-cospan}) as discrete
graphs, and similarly for the functions
$\lambda_{in}:C_{in}\to \setofmachinestates$ and
$\lambda_{out}:C_{out}\to \setofmachinestates$
in~(\ref{equation/lambdain-lambdaout}) which are understood as graph
homomorphisms.
Similarly, the source and target maps $\source{}:C_{in}\to C$ and
$\target{}:C_{out}\to C$ are graph morphisms whose purpose is to map
every input $s\in C_{in}$ and output state $s'\in C_{out}$ to the
underlying internal states $\source{}(s)$ and $\target{}(s')$ of the
code~$C$.

\paragraph{The bicategory of cospans}
The discussion leads us to the definition of the bicategory
$\cospan{\Scategorysurround}$ of cospans associated with a category
$\Scategorysurround$ with pushouts.
The bicategory $\cospan{\Scategorysurround}$ has the same objects
$A, B, C$ as the original category $\Scategorysurround$, and its
morphisms are the triples
\begin{equation}\label{equation/cospan}
\begin{tikzcd}[column sep=2em, row sep=1.2em]
(S,\source{},\target{})
\quad
:
\quad
A
\arrow[rr,spanmap]
&&
B
\end{tikzcd}
\end{equation}
consisting of an object $S$
of the category $\Scategorysurround$ together with a pair
$$
\begin{tikzcd}[column sep=2.4em]
A
\arrow[rr,"{\source{}}"]
&& 
S
&& 
B
\arrow[ll,"{\target{}}"{swap}]
\end{tikzcd}
$$
of morphisms of the category $\Scategorysurround$.
Such a triple (\ref{equation/cospan}) 
is called a \demph{cospan} between $A$ and $B$,
with \demph{support} the object $S$ of the category $\Scategorysurround$.
In many situations, the two morphisms $\source{}:A\to S$ and $\target{}:B\to S$ 
can be deduced from the context, and we thus simply write 
$S: A \longrightarrow\hspace{-1.2em}|\hspace{1em}B$ 
for the cospan in that case.
A 2-dimensional cell in the bicategory $\cospan{\Scategorysurround}$
between two such cospans
$$
\begin{tikzcd}[column sep=1em]
\varphi
\quad
:
\quad
(S,\source{},\target{})
\arrow[rr,double,-implies]
&&
(T,\source{},\target{})
\quad
:
\quad
A
\arrow[rr,spanmap]
&&
B
\end{tikzcd}
$$
is defined as a morphism $\varphi:S\to T$
of the original category $\Scategorysurround$, making the diagram commute:
$$
\begin{small}
\begin{tikzcd}[column sep=2.8em, row sep=.2em]
&& S\ar[dddd,"{\varphi}"]
\\
\\
A
\arrow[rruu,"{\source{}}",sloped,pos=.6]
\arrow[rrdd,"{\source{}}"{swap},sloped,pos=.6]
&&&&
B
\arrow[lluu,"{\target{}}"{swap},sloped,pos=.6]
\arrow[lldd,"{\target{}}",sloped,pos=.6]
\\
\\
&& T
\end{tikzcd}
\end{small}
$$
%
%
%
%
Two cospans $S: A \longrightarrow\hspace{-1.2em}|\hspace{1em}B$
and $T:B\longrightarrow\hspace{-1.2em}|\hspace{1em}C$
can be put side by side and composed into the cospan
$T\circ S:A\longrightarrow\hspace{-1.2em}|\hspace{1em}C$
by ``gluing'' their common border $B$, using a \emph{pushout diagram} 
performed in the category $\Scategorysurround$:
$$
\begin{small}
\begin{tikzcd}[column sep=.9em, row sep=.9em]
\color{lightgray}
A
\arrow[rrdd,"\source{}"{swap,pos=.6},sloped,lightgray]
&& 
&& 
B
\arrow[lldd,"\target{}"{swap},sloped,pos=.25]
\arrow[rrdd,"\source{}",sloped,pos=.25]
&&
&&
\color{lightgray}
C
\arrow[lldd,"\target{}",sloped,pos=.7,lightgray]
\\
\\
&& 
S
\arrow[rrdd,"\inl"{swap},pos=.7,sloped]
&&
pushout
&& 
T
\arrow[lldd,"\inr",pos=.7,sloped]
\\
\\
&&&&
T\circ S
&&&&
\end{tikzcd}
\end{small}
$$
\paragraph{Sequential composition as pushout + relabeling}
One good reason for describing codes $C$ and $D$ as cospans
(\ref{equation/first-cospan}) in the category $\mathbf{Trans}$ of
transition systems, is that the very same ``gluing'' recipe based on
pushouts can be applied to compute their \emph{sequential composition}
$C;D$, at least in the specific case where the set $C_{out}$ of output
states of $C$ coincides with the set $D_{in}$ of input states of $D$.
The pushout construction performed in $\mathbf{Trans}$ 
gives rise to the commutative diagram in $\mathbf{Gph}$ below:
\begin{equation}\label{equation/two-cospans-side-by-side}
\begin{small}
\begin{tikzcd}[column sep=1.8em, row sep=1.3em]
\color{lightgray}
C_{in}
\arrow[dd,"{\lambda_{in}}"{swap},lightgray]
\arrow[rrdd,"{\source{}}",sloped,near start,lightgray]
&& 
&&
C_{out}=D_{in}
\arrow[dd,"{\lambda_{out/in}}"]
\arrow[lldd,"{\target{}}"{swap},sloped,pos=.4]
\arrow[rrdd,"{\source{}}",sloped,pos=.4]
&&
&&
\color{lightgray}
D_{out}
\arrow[dd,"{\lambda_{out}}",lightgray]
\arrow[lldd,"{\target{}}"{swap},sloped,near start,lightgray]
\\
\\
\color{lightgray}
\anchorstate[0]
\arrow[rrdd,"{\eta}",lightgray]
&&
|[alias=C]|
C
\arrow[dd,"{\lambda_{C}}"]
&&
\anchorstate[0]
\arrow[lldd,"{\eta}"{swap},near start]
\arrow[rrdd,"{\eta}",near start]
&&
D
\arrow[dd,"{\lambda_{D}}"]
\arrow[lldd,crossing over]
&&
\color{lightgray}
\anchorstate[0]
\arrow[lldd,"{\eta}"{swap},lightgray]
\\
\\
&&
\anchorstate[1]
\arrow[rrdd,"{\inl}"{swap},sloped]
&&
|[alias=CD]|
{C;D}
\arrow[dd,"{\lambda_{po}}",pos=.3]
&&
\anchorstate[1]
\arrow[lldd,"{\inr}",sloped]
&&
\\
\\
&&&&
\anchorstate[2]
&&&&
\arrow[from=C, to=CD, crossing over]
\end{tikzcd}
\end{small}
\end{equation}
It should be noted that the pushout diagram in $\mathbf{Trans}$ is
obtained by computing \emph{independently} in $\mathbf{Gph}$ the
pushout diagram $[a]$ defining the support graph $C;D$ as well as the
pushout diagram $[b]$ defining the labeling graph $\anchorstate[2]$,
as shown below:
\begin{equation}
\begin{small}
\begin{tabular}{ccc}
\begin{tikzcd}[column sep=.8em, row sep=.8em]
&&
C_{in}=D_{out}
\arrow[lldd,"{\target{}}"{swap},sloped,pos=.3]
\arrow[rrdd,"{\source{}}",sloped,pos=.3]
\\
\\
C
\arrow[rrdd,"{\inl}"{swap},sloped,pos=.6]
&&
{pushout [a]}
&&
D
\arrow[lldd,"{\inr}",sloped,pos=.6]
\\
\\
&&
C;D
\end{tikzcd}
&&
\begin{tikzcd}[column sep=.8em, row sep=.8em]
&&
\anchorstate[0]
\arrow[lldd,"{\eta}"{swap}]
\arrow[rrdd,"{\eta}"]
\\
\\
\anchorstate[1]
\arrow[rrdd,"{\inl}"{swap},sloped,pos=.6]
&&
{pushout [b]}
&&
{\anchorstate[1]}
\arrow[lldd,"{\inr}",sloped,pos=.6]
\\
\\
&&
{\anchorstate[2]}
\end{tikzcd}
\end{tabular}
\end{small}
\end{equation}
Pictorially, the purpose of the pushout $[a]$ is to ``glue'' together
the two cylinders $C$ and $D$ represented in
(\ref{equation/cobordism-from-P-to-Q-to-R}) along their common border
$C_{out}=D_{in}$, so as to obtain the cylinder $C;D$ describing the
result of composing the two codes $C$ and $D$ sequentially:
\begin{equation}\label{equation/cobordism-before-glueing}
\raisebox{-3.5em}{\includegraphics[height=7.5em]{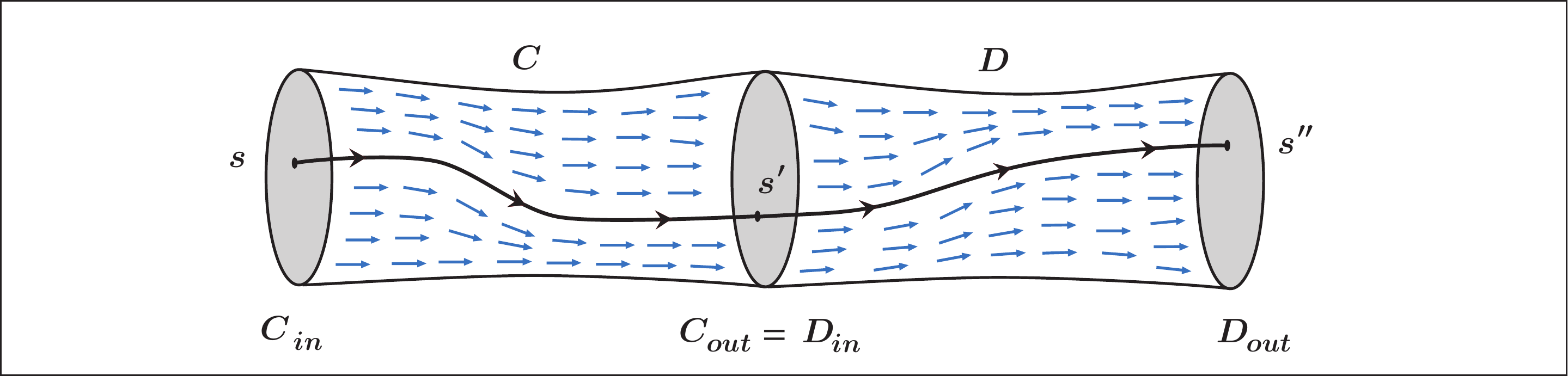}}
\end{equation}
One main contribution and novelty of the paper is the idea
that the pushout diagram $[a]$ performed on the graphs $C$ and $D$
should come together with a pushout diagram $[b]$ performed this time
at the level of their labeling graphs.
The labeling graph $\anchorstate[2]$ produced by the pushout $[b]$ is
easy to compute: its vertices are the machine states
$s\in \setofmachinestates$ and there is a pair of edges noted
$inst_l, inst_r:s\to s'$ for every edge $inst:s\to s'$ of the graph
$\anchorstate[1]$.
The intuition is that the code $C$ performs the instructions of the
form $inst_l$ in $\anchorstate[2]$ while the code $D$ performs the
instructions of the form $inst_r$.
The labeling graph $\anchorstate[2]$ comes together with a relabeling map
\begin{equation}\label{equation/multiplication-of-the-monad}
\begin{tikzcd}[column sep=1.5em]
\mu
\,\, : \,\,
{\anchorstate[2]}
\arrow[rr]
&& 
{\anchorstate[1]}
\end{tikzcd}
\end{equation}
which maps every edge of the form $inst_l, inst_r:s\to s'$ of
$\anchorstate[2]$ to the underlying edge $inst:s\to s'$ of
$\anchorstate[1]$.
The transition system $(C;D,\lambda_{C;D})$ defining the sequential
composition is simply obtained by relabeling along $\mu$ the
transition system $(C;D,\lambda_{po})$ computed by the pushout of
$(C,\lambda_C)$ and $(D,\lambda_D)$ in $\mathbf{Trans}$, in order to
obtain the transition system with labeling map:
$$
\begin{tikzcd}[column sep=1.5em]
C;D
\arrow[rr,"{\lambda_{C;D}}"]
&& 
{\anchorstate[1]}
\end{tikzcd}
\quad = \quad
\begin{tikzcd}[column sep=1.5em]
C;D
\arrow[rr,"{\lambda_{po}}"]
&& 
{\anchorstate[2]}
\arrow[rr,"{\mu}"]
&& 
{\anchorstate[1]}
\end{tikzcd}
$$
The construction establishes that:

\begin{equation}\label{equation/slogan-sequential}
\fbox{
Sequential composition \,\, = \,\, gluing by pushout \,\,+ \,\,relabeling
}
\end{equation}


\paragraph{Template games and internal opcategories}
The fact that the relabeling along
$\mu:\anchorstate[2]\to\anchorstate[1]$ plays such a critical role in
the definition of sequential composition $C,D\mapsto C;D$ is
reminiscent of a similar observation in the work by Melli{\`e}s on
template games for differential linear logic (DiLL).
In that case, the construction of the bicategory
$\mathbf{Games}(\anchor)$ of template games, strategies and
simulations relies on the assumption that the synchronization
template~$\anchor$ defines an \emph{internal category} in the
underlying category~$\Scategorysurround$.
An important fact observed for the first time by
\citet{intro-to-bicategories} is that an internal category~$\anchor$
in a category~$\Scategorysurround$ with pullbacks is the same thing as
a monad in the associated bicategory $\spanbicat{\Scategorysurround}$
of spans, see \cite{eberhart-hirschowitz-fscd-2019} for a discussion.
Since we are working in a category~$\Scategorysurround$ with pushouts,
it makes sense to dualize the situation, and to call \textbf{internal
  opcategory} $\anchor$ a monad in the bicategory
$\cospan{\Scategorysurround}$.
In other words, an internal opcategory $\anchor$ is defined as a pair
of objects $\anchor[0]$ and $\anchor[1]$ equipped with a cospan
and a pair of morphisms:
\begin{equation}\label{equation/cospan-of-an-opcategory}
\begin{tikzcd}[column sep=1.4em]
{\anchor[0]}
\arrow[rr,"\source{}"]
&& 
{\anchor[1]}
&& 
{\anchor[0]}
\arrow[ll,"\target{}"{swap}]
\end{tikzcd}
\quad\quad
\begin{tikzcd}[column sep=1.3em]
{\anchor[0]}
\arrow[rr,"{\eta}"]
&& 
{\anchor[1]}
\end{tikzcd}
\quad
\begin{tikzcd}[column sep=1.3em]
{\anchor[2]}
\arrow[rr,"{\mu}"]
&& 
{\anchor[1]}
\end{tikzcd}
\end{equation}
where the object $\anchor[2]$ is defined as the result 
of the pushout diagram in $\Scategorysurround$:
$$
\begin{small}
\begin{tikzcd}[column sep=.8em, row sep=.8em]
\color{lightgray}
{\anchor[0]}
\arrow[rrdd,"\source{}"{swap,pos=.5},sloped,lightgray]
&& 
&& 
{\anchor[0]}
\arrow[lldd,"\target{}"{swap},sloped,pos=.2]
\arrow[rrdd,"\source{}",sloped,pos=.2]
&&
&&
\color{lightgray}
{\anchor[0]}
\arrow[lldd,"\target{}",sloped,pos=.6,lightgray]
\\
\\
&& 
{\anchor[1]}
\arrow[rrdd,"\inl"{swap},pos=.6,sloped]
&&
pushout
&& 
{\anchor[1]}
\arrow[lldd,"\inr",pos=.6,sloped]
\\
\\
&&&&
{\anchor[2]}
&&&&
\end{tikzcd}
\end{small}
$$
In addition, a number of diagrams are required to commute, in order to
ensure that the two morphisms $\eta$ and $\mu$ define the following
2-cells in the bicategory $\cospan{\Scategorysurround}$
$$
\begin{small}
\begin{tikzcd}[column sep=1.5em, row sep=1.5em]
\anchor[0]
\arrow[rrrr,spanmap,"{id}"{xshift=.5ex,yshift=.5ex},""{swap,name=start},bend left]
\arrow[rrrr,spanmap,"{\anchor[1]}"{swap,yshift=-.5ex},""{name=end},bend right]
&&&&
\anchor[0]
\arrow[from=start, to=end, double,-implies,"{\eta}"]
\end{tikzcd}
\quad\quad\quad
\begin{tikzcd}[column sep=1.5em, row sep=1.5em]
&&
|[alias=start]|
\anchor[0]\arrow[rrdd,spanmap,"{\anchor[1]}"{xshift=.2ex,yshift=.2ex}]
\\
\\
\anchor[0]
\arrow[rruu,spanmap,"{\anchor[1]}"{xshift=-.5ex}]
\arrow[rrrr,spanmap,"{\anchor[1]}"{swap,yshift=-.5ex},""{name=end}]
&&&&
\anchor[0]
\arrow[from=start, to=end, double,-implies,"{\mu}"]
\end{tikzcd}
\end{small}
$$
and that these 2-cells satisfy the unitality and associativity axioms
required of a monad
in the bicategory $\cospan{\Scategorysurround}$.
Note in particular that the requirement that $\eta$ defines the
left hand-side 2-cell is very strong, since it implies that the three
morphisms $\eta, \source{},\target{}:\anchor[0]\to\anchor[1]$ are
equal.
%
\paragraph{A bicategory of cobordisms}
We have seen earlier that the labeling graphs
$\anchorstate[0]=\setofmachinestates$ and $\anchorstate[1]$ of machine
states are equipped with morphisms $\eta$ and $\mu$ explicated
in~(\ref{equation/unit-of-the-monad})
and~(\ref{equation/multiplication-of-the-monad}).
Somewhat surprisingly, this additional structure happens to be tightly
connected to the work by Melli{\`e}s on template games, as established
by the following theorem:
\begin{theorem}
The cospan
$$
\begin{tikzcd}[column sep=1.5em]
{\anchorstate[0]}
\arrow[rrr,spanmap,"{\anchorstate[1]}"{yshift=.8ex}]
&&&
{\anchorstate[0]}
\quad\quad
=
\quad\quad
{\anchorstate[0]}
\arrow[rr,"{\eta}"]
&& 
{\anchorstate[1]}
&& 
{\anchorstate[0]}
\arrow[ll,"{\eta}"{swap}]
\end{tikzcd}
$$
together with the graph morphisms $\eta$ and $\mu$
defines an internal opcategory $\anchor$ in the category $\mathbf{Gph}$.
\end{theorem}
This theorem means that the machine model $\anchorstate$
is regulated by almost the same algebraic principles as the
synchronization templates exhibited by Melli\`es in order to generate
his game semantics of differential linear logic (DiLL).
The only difference (it is a key difference though) is that the
original internal category~$\anchor$ in DiLL is ``dualized'' and
replaced in the present situation by an internal opcategory
$\anchor=\anchorstate$ of machine states.
%
Guided by this auspicious connection, we associate a bicategory
$\cobord{\anchor}$ of \textbf{games}, \textbf{cobordisms} and
\textbf{simulations} with every internal opcategory $\anchor$ living in
a category $\Scategorysurround$ with pushouts.
The terminology of \emph{cobordism} pays tribute to our main source of
inspiration for the idea of cospan, which is the construction of the
\emph{category of cobordisms} in algebraic topology and in topological
quantum field theory, see \cite{morton}.
A \textbf{game} $(A,\lambda_A)$ is a pair 
\begin{equation}\label{equation/template-game}
\begin{tikzcd}[column sep=1.5em]
\lambda_A
\quad : \quad
A
\arrow[rr]
&& 
{\anchor[0]}
\end{tikzcd}
\end{equation}
A \textbf{cobordism}
$\sigma:A\longrightarrow\hspace{-1.2em}|\hspace{1em}B$ between two
such games is a quadruple $(S,\source{},\target{},\lambda_{\sigma})$
consisting of an object $S$ and three morphisms making the diagram
below commute:
\begin{equation}\label{equation/cobordism}
\begin{tikzcd}[column sep=2.5em, row sep=1em]
A
\arrow[dd,"{\lambda_{A}}"{swap}]
\arrow[rr,"{\source{}}"]
&& 
S
\arrow[dd,"{\lambda_{\sigma}}"]
&&
B
\arrow[dd,"{\lambda_{B}}"]
\arrow[ll,"{\target{}}"{swap}]
\\
\\
{\anchor[0]}
\arrow[rr,"{\eta}"]
&&
{\anchor[1]}
&&
{\anchor[0]}
\arrow[ll,"{\eta}"{swap}]
\end{tikzcd}
\end{equation}
Here, the object $S$ is called the \demph{support} of the cobordism
$\sigma$.
A \textbf{simulation} between two cobordisms
$$
\begin{tikzcd}[column sep=1em]
\varphi
\quad
:
\quad
\sigma
\arrow[rr,double,-implies]
&&
\tau
\quad
:
\quad
A
\arrow[rr,spanmap]
&&
B
\end{tikzcd}
$$
is defined as a morphism $\varphi:S\to T$
of the original category $\Scategorysurround$, making the three
diagrams commute:
$$
\begin{small}
\begin{tikzcd}[column sep=1em, row sep=1em]
&& A
\arrow[lldd,"{\source{}}"{swap},pos=.5]
\arrow[rrdd,"{\source{}}",pos=.5]
\\
\\
S\arrow[rrrr,"{\varphi}"{swap}]
&&&&
T
\end{tikzcd}
\quad\quad\quad\quad
\begin{tikzcd}[column sep=.8em, row sep=.8em]
&& B
\arrow[lldd,"{\target{}}"{swap},pos=.5]
\arrow[rrdd,"{\target{}}",pos=.5]
\\
\\
S\arrow[rrrr,"{\varphi}"{swap}]
&&&&
T
\end{tikzcd}
\quad\quad\quad\quad
\begin{tikzcd}[column sep=.8em, row sep=.8em]
S
\ar[rrrr,"{\varphi}"]
\ar[rrdd,"{\lambda_{\sigma}}"{swap}]
&&&&
T
\ar[lldd,"{\lambda_{\tau}}"]
\\
\\
&& 
{\anchor[1]}
\end{tikzcd}
\end{small}
$$
The composition of cobordisms $\sigma: A \longrightarrow\hspace{-1.2em}|\hspace{1em}B$
and $\tau: B \longrightarrow\hspace{-1.2em}|\hspace{1em}C$
is defined by pushout and relabeling:
\begin{equation}\label{equation/two-cobordisms-side-by-side}
\begin{small}
\begin{tikzcd}[column sep=1.6em, row sep=1em]
\color{lightgray}
A
\arrow[dd,"{\lambda_{A}}"{swap},lightgray]
\arrow[rrdd,"{\source{}}",sloped,pos=.3,lightgray]
&& 
&&
B
\arrow[dd,"{\lambda_{B}}"]
\arrow[lldd,"{\target{}}"{swap},sloped,pos=.4]
\arrow[rrdd,"{\source{}}",sloped,pos=.4]
&&
&&
\color{lightgray}
C
\arrow[dd,"{\lambda_{C}}",lightgray]
\arrow[lldd,"{\target{}}"{swap},sloped,pos=.3,lightgray]
\\
\\
\color{lightgray}
{\anchor[0]}
\arrow[rrdd,"{\eta}"{swap},sloped,pos=.6,lightgray]
&&
|[alias=S]|
S
\arrow[dd,"{\lambda_{\sigma}}"]
&&
{\anchor[0]}
\arrow[lldd,"{\eta}"{swap},sloped,pos=.05]
\arrow[rrdd,"{\eta}",sloped,pos=.05]
&&
T
\arrow[dd,"{\lambda_{\tau}}"]
\arrow[lldd,crossing over]
&&
\color{lightgray}
{\anchor[0]}
\arrow[lldd,"{\eta}"{sloped},pos=.6,lightgray]
\\
\\
&&
{\anchor[1]}
\arrow[rrdd,"{\inl}"{swap},pos=.6,sloped]
&&
|[alias=SplusT]|
{S+_{B} T}
\arrow[dd,"{\lambda_{po}}",pos=.3]
&&
{\anchor[1]}
\arrow[lldd,"{\inr}",pos=.6,sloped]
&&
\\
\\
&&&&
{\anchor[2]}
\arrow[dd,"{\mu}"]
&&&&
\\
\\
&&&&
{\anchor[1]}
\arrow[from=S, to=SplusT, crossing over]
\end{tikzcd}
\end{small}
\end{equation}
while the identity cobordism $id_A: A \longrightarrow\hspace{-1.2em}|\hspace{1em}A$
is defined as follows:
\begin{equation}\label{equation/cobordisms-identity}
\begin{small}
\begin{tikzcd}[column sep=1.8em, row sep=.7em]
A
\arrow[dddd,"{\lambda_{A}}"{swap}]
\arrow[rr,"{id_A}"]
&&
A
\arrow[dd,"{\lambda_{A}}"{xshift=.5ex}]
&&
A
\arrow[dddd,"{\lambda_{A}}"]
\arrow[ll,"{id_A}"{swap}]
\\
\\
&&
{\anchor[0]}
\arrow[dd,"{\eta}"{xshift=.5ex}]
&&
\\
\\
{\anchor[0]}
\arrow[rr,"{\source{}}"]
&&
{\anchor[1]}
&&
{\anchor[0]}
\arrow[ll,"{\target{}}"{swap}]
\end{tikzcd}
\end{small}
\end{equation}
Given a category $\Scategorysurround$ with pushouts, 
we establish that
\begin{theorem}
Every internal opcategory $\anchor$ in the category $\Scategorysurround$
defines a bicategory $\cobord{\anchor}$.
\end{theorem}
The construction of the bicategory $\cobord{\anchor}$ of cobordisms is
surprisingly similar to the construction of the bicategory
$\games{\anchor}$ performed by \citet{popl19}.
In particular, the composition of strategies in $\games{\anchor}$ and
of cobordisms in $\cobord{\anchor}$ relies in both cases on a
relabeling along the ``multiplication'' morphism
$\mu:\anchor[2]\to\anchor[1]$, while the definition of identities
relies on the ``unit'' morphism $\eta:\anchor[0]\to\anchor[1]$.
There is a simple conceptual explanation for the similitude however,
which is that we construct in both cases the bicategory
$\games{\anchor}$ and $\cobord{\anchor}$ as a bicategory \emph{sliced
  above} a formal monad $\anchor$ living either in the bicategory
$\Span{\anchor}$ (in the case of games) or in the bicategory
$\cospan{\anchor}$ (in the case of cobordisms), see
\cite{eberhart-hirschowitz-fscd-2019} for details.

\paragraph{Parallel product as pullback + relabeling}
We have just described how two cobordisms
$\sigma: A \longrightarrow\hspace{-1.2em}|\hspace{1em}B$ and
$\tau: B \longrightarrow\hspace{-1.2em}|\hspace{1em}C$
can be ``glued'' together and composed sequentially using a pushout
diagram on their common border $(B,\lambda_B)$.
Similarly, given two cobordisms $\sigma: A \longrightarrow\hspace{-1.2em}|\hspace{1em}B$ 
and $\tau: C \longrightarrow\hspace{-1.2em}|\hspace{1em}D$ interpreting
concurrent imperative programs, we would like to give a nice and conceptual description 
of their parallel product $\sigma\|\tau:A\|C \longrightarrow\hspace{-1.2em}|\hspace{1em} B\|D$.
To that purpose, we shift from a \emph{sequential} to a \emph{concurrent} setting 
by extending our original labeling graphs $\anchorstate[0]$ and $\anchorstate[1]$ 
with transitions performed not just by the Code, but also by the Frame (or the
Environment).
%
When compared to the sequential situation of
(\ref{equation/input-output-disk}), (\ref{equation/Sin-Sout-C-tube})
and (\ref{equation/cobordism-from-P-to-Q}), this extension of the
original labeling graphs $\anchorstate[0]$ and $\anchorstate[1]$
conveys the intuition that a concurrent imperative code~$C$ connects
an input graph~$C_{in}$ to an output graph~$C_{out}$ whose only
transitions (depicted below in red) are performed by the Frame:
\begin{equation}\label{equation/two-borders-with-frame}
\raisebox{-2.8em}{\includegraphics[height=5.6em]{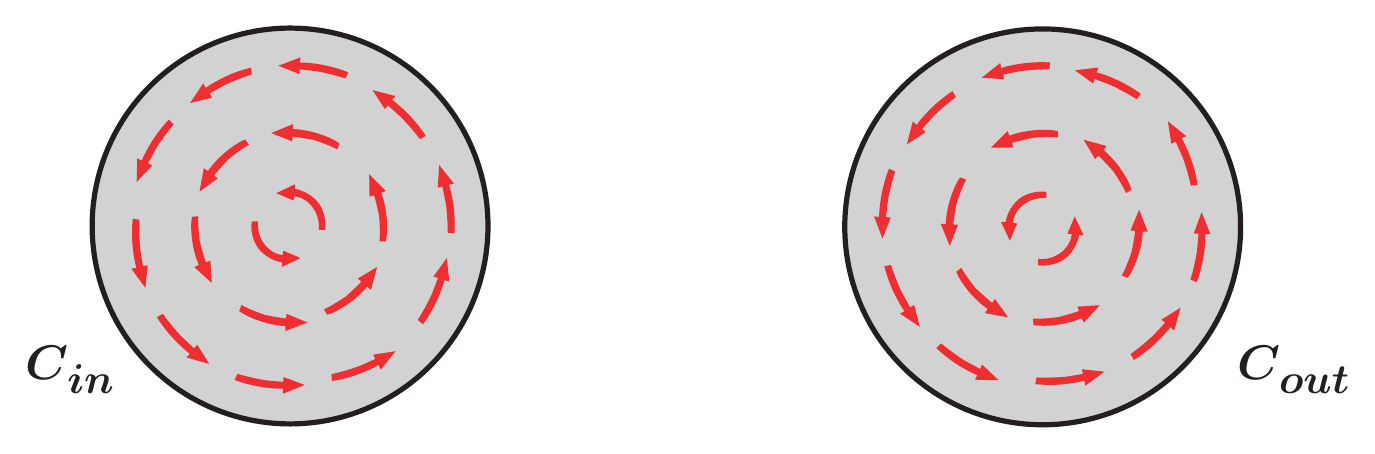}}
\end{equation}
and whose cylinder or cobordism $C:C_{in}\longrightarrow\hspace{-1.2em}|\hspace{1em}C_{out}$
admits transitions performed either by the Code (in blue) or by the Frame (in red),
as shown below:
\begin{equation}\label{equation/input-output-disk-with-frame}
\raisebox{-3.5em}{\includegraphics[height=7.5em]{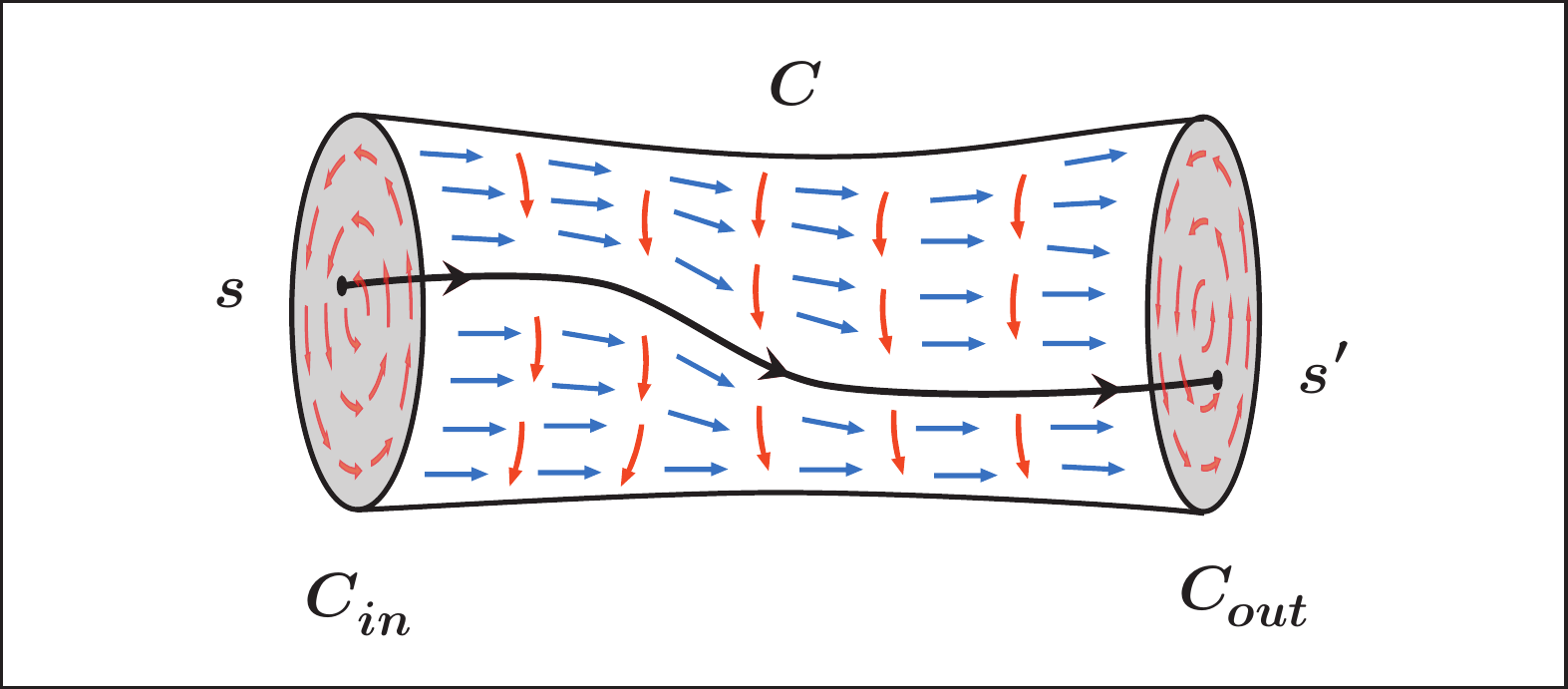}}
\end{equation}
The flexibility of our approach based on cobordism
is illustrated by the great ease in which
%
the shift of paradigm from sequential to concurrent is performed by
moving from the original \emph{one-player} machine model to a
\emph{two-player} machine model where transitions may be performed by
$\Cplayer$ (for the Code) or $\Fplayer$ (for the Frame).
%
In particular, as it stands, the interpretation of the code $C$ as a
cobordism is described by the very same diagram
(\ref{equation/second-cospan}) as in the sequential case, except that
transitions in $\anchor[0]$ are now performed by Frame and those in
$\anchor[1]$ by Frame and by Code.
Finally, in order to reflect the asynchronous structure of
interactions, we take this opportunity to upgrade our model, and shift
from the ambient category $\Scategorysurround=\mathbf{Gph}$ to the
category $\Scategorysurround=\AGraph$ of asynchronous graphs,
described in \S\ref{sec:stateful-stateless}.

At this stage, another striking connection emerges between our
bicategory $\cobord{\anchor}$ of template games and cobordisms and the
original work by Melli{\`e}s on template games for differential linear
logic (DiLL).
Indeed, it appears that the parallel product $A\|B$ and $\sigma\|\tau$
of template games and cobordisms in $\cobord{\anchor}$ can be defined
using the same principles as the tensor product $A\otimes B$ and
$\sigma\otimes\tau$ of template games and strategies in the bicategory
$\games{\anchor}$ for DiLL.
The general recipe is to equip the internal opcategory $\anchor$ with
a pair of internal functors
\begin{equation}\label{equation/span-of-functors}
\begin{tikzcd}[column sep=5em, row sep=1.2em]
    {\anchor \times \anchor} 
    &
    \anchortensor \arrow[l, "{\pick}"{swap}]
    \arrow[r, "{\pince}"]
    &
    \anchor
  \end{tikzcd}
\end{equation}
from which one derives (see \S\ref{sec:parallel-product})
a lax functor $\pull{\pick}$ and a pseudo functor $\push{\pince}$
between bicategories, see \cite{garner-phd-thesis} for definitions:
\begin{equation}\label{equation/pull-push}
\begin{tikzcd}[column sep=1.4em]
\cobord{\anchor \times \anchor} 
\arrow[rrrr, "{\pull{\pick}}"]
&&&&
\cobord{\anchortensor}
\arrow[rrrr, "{\push{\pince}}"]
&&&&
\cobord{\anchor}
\end{tikzcd}
\end{equation}
obtained by \emph{pulling along $\pick$} 
and by \emph{pushing along $\pince$},
respectively.
The parallel product
$$
\begin{tikzcd}[column sep=2em]
\|
\quad
:
\quad
\cobord{\anchor} \times \cobord{\anchor} 
\arrow[rr,"{\cobordmult{\anchor}{\anchor}}"]
&&
\cobord{\anchor \times \anchor} 
\arrow[rr]
&&
\cobord{\anchor}
\end{tikzcd}
$$
is then defined as the lax double functor obtained by precomposing
(\ref{equation/pull-push}) with the canonical pseudo functor
$$
\begin{tikzcd}[column sep=1.4em]
\cobordmult{\anchor_1}{\anchor_2}
\quad : \quad
\cobord{\anchor_1} \times \cobord{\anchor_2}
\arrow[rr]
&&
\cobord{\anchor_1 \times \anchor_2}
\end{tikzcd}
$$
The definition is elegantly conceptual, but not entirely transparent,
and thus probably worth explicating.
The parallel product of two template games $A$ and $B$
is defined by computing the pullback of $\lambda_A\times\lambda_B$
along $\pick[0]$, and then relabeling the resulting labeling map $\lambda_{pb}$
along $\pince[0]$:
\begin{equation}\label{equation/pullback-on-games}
\begin{small}
\begin{tikzcd}[column sep=1.2em,row sep=.6em]
A\times B
\arrow[dd,"{\lambda_{A}\times\lambda_{B}}"{swap}]
&&
A\| B
\arrow[rrrrdd,"{\lambda_{A\|B}}", bend left]
\arrow[ll]
\arrow[dd,"{\lambda_{pb}}"]
\\
&
pullback
&
\\
\anchor[0]\times\anchor[0]
&&
{\anchortensor[0]}
\arrow[ll,"{\pick[0]}"]
\arrow[rrrr,"{\pince[0]}"{swap}]
&&&&
{\anchor[0]}
\end{tikzcd}
\end{small}
\end{equation}
In exactly the same way, the parallel product of
$\sigma\|\tau:A\|C \longrightarrow\hspace{-1.2em}|\hspace{1em} B\|D$
of two cobordisms
$\sigma: A\longrightarrow\hspace{-1.2em}|\hspace{1em}B$ and
$\tau: C\longrightarrow\hspace{-1.2em}|\hspace{1em}D$ with respective
supports $S$ and $T$ has its support $S\|T$ computed by the following
pullback along $\pick[1]$, with resulting labeling map $\lambda_{pb}$
postcomposed with $\pince[1]$:
\begin{equation}\label{equation/pullback-on-cobordisms}
\begin{small}
\begin{tikzcd}[column sep=1.2em,row sep=.6em]
S\times T 
\arrow[dd,"{\lambda_{\sigma}\times\lambda_{\tau}}"{swap}]
&&
S\| T 
\arrow[rrrrdd,"{\lambda_{\sigma\|\tau}}", bend left]
\arrow[ll]
\arrow[dd,"{\lambda_{pb}}"]
\\
&
pullback
&
\\
\anchor[1]\times\anchor[1]
&&
{\anchortensor[1]}
\arrow[ll,"{\pick[1]}"]
\arrow[rrrr,"{\pince[1]}"{swap}]
&&&&
{\anchor[1]}
\end{tikzcd}
\end{small}
\end{equation}
As explained in \S\ref{sec:parallel-product}, the opcategory
$\anchorstatetensor$ describes a \emph{three-player} machine model
where every machine transition may be performed by one of the three
players $\Cplayer_1$, $\Cplayer_2$, $\Fplayer$ (for Frame) involved in
the parallel product $C_1\|C_2$.
The functor $\pick$ describes how the machine model
$\anchorstatetensor$ can be mapped to a pair
$\anchorstate\times\anchorstate$ of two-player machine models, where
$\Cplayer_1$ identifies the transitions of $\Cplayer_2$ as part of its
frame $\Fplayer$ in the left-hand component $\anchorstate$, and
$\Cplayer_2$ identifies the transitions of $\Cplayer_1$ as part of its
frame $\Fplayer$ in the right-hand component $\anchorstate$.
The functor $\pince$ then identifies the transitions of the players
$\Cplayer_1$ and of $\Cplayer_2$ in the three-player machine model
$\anchorstatetensor$ as the transitions of the code $C$ in the
original two-player machine model $\anchorstate$.

To summarize, we see in (\ref{equation/pullback-on-games}) and
(\ref{equation/pullback-on-cobordisms}) that the definition of the
parallel product is performed by a pullback along the functor $\pick$,
whose purpose is to synchronize the transitions performed by
$\Cplayer_1$, $\Cplayer_2$ and $\Fplayer$, followed by a relabeling
along the functor $\pince$.
The construction thus establishes that:
\begin{equation}\label{equation/slogan-parallel}
\fbox{
Parallel product \,\, = \,\, synchronizing by pullback \,\,+ \,\,relabeling
}
\end{equation}
The fact that the sequential composition is computed by a pushout and
thus a colimit (\ref{equation/slogan-sequential}) while the parallel
product is computed by a pullback and thus a limit
(\ref{equation/slogan-parallel}) has the immediate and remarkable
consequence that there exists a natural and coherent family of
morphisms
$$
Hoare_{C_1,C_2,D_1,D_2}
\quad : \quad (C_1 \| C_2);(D_1 \| D_2) \Rightarrow
(C_1 ; D_1)\|(C_2 ; D_2)
$$
which turns $\cobord{\anchor}$ into a lax monoidal bicategory.
This lax monoidal structure of $\cobord{\anchor}$ provides a nice
algebraic explanation for the Hoare inequality principle \cite{Hoare},
since we derive it from the fact that a colimit (sequential
composition) always commutes with a limit (parallel product) up to a
(in that case non reversible) coercion morphism.

\paragraph{A uniform interpretation of the CSL codes and proofs}
One of the main purposes of the present paper is to reunderstand in a more
foundational and axiomatic way the \emph{asynchronous soundness theorem} of
concurrent separation logic (CSL) recently established by \citet{lics18}.
Their proof of soundness is based on the construction of a
game-theoretic and asynchronous interpretation of the codes and of the
proofs of CSL in its original form, see \citet{Brookes:a-semantics}.
One main advantage of our approach based on template games is to
provide a general, uniform and flexible framework to construct models
of CSL, both at the level of codes and of proofs.
We will explain in \S\ref{sec:separated-states} how to construct a
machine model $\mooSep$ based on the notion of \emph{separated states}
introduced by \citet{mfps,lics18} in order to interpret the proofs of
CSL, and not just the codes as before.
We show moreover that there exists a chain of functors between the
three machine models or internal opcategories introduced in the paper:
$$
\begin{tikzcd}[column sep=1.5em]
  \mooSep \arrow[rr,"{u_\semmorphsep}"] 
  &&
  \mooS \arrow[rr,"{u_\semmorph}"] 
  &&
  {\mooL}.
\end{tikzcd}
$$
As it turns out, given a CSL proof $\pi$ of a Hoare triple
$\{P\}C\{Q\}$, this chain of functors induces a chain of translations
$$
\begin{tikzcd}[column sep=1.5em]
  \semSep{\pi} \arrow[rr,"{\semmorphsep}"]
 &&
  \semS{C} \arrow[rr,"{\semmorph}"]
&&
  \semL{C}
\end{tikzcd}
$$
between the interpretations of the CSL proof $\pi$ and of the specified code $C$
in the three template game models associated with $\mooSep$, $\mooS$ and~$\mooL$.
The translation from the cobordism $\semSep{\pi}$ interpreting the CSL
proof $\pi$ above the machine model $\mooSep$ of separated states to
the cobordism $\semS{C}$ providing the operational semantics of the
code $C$ can be depicted as follows:
\begin{center}
\includegraphics[height=14.5em]{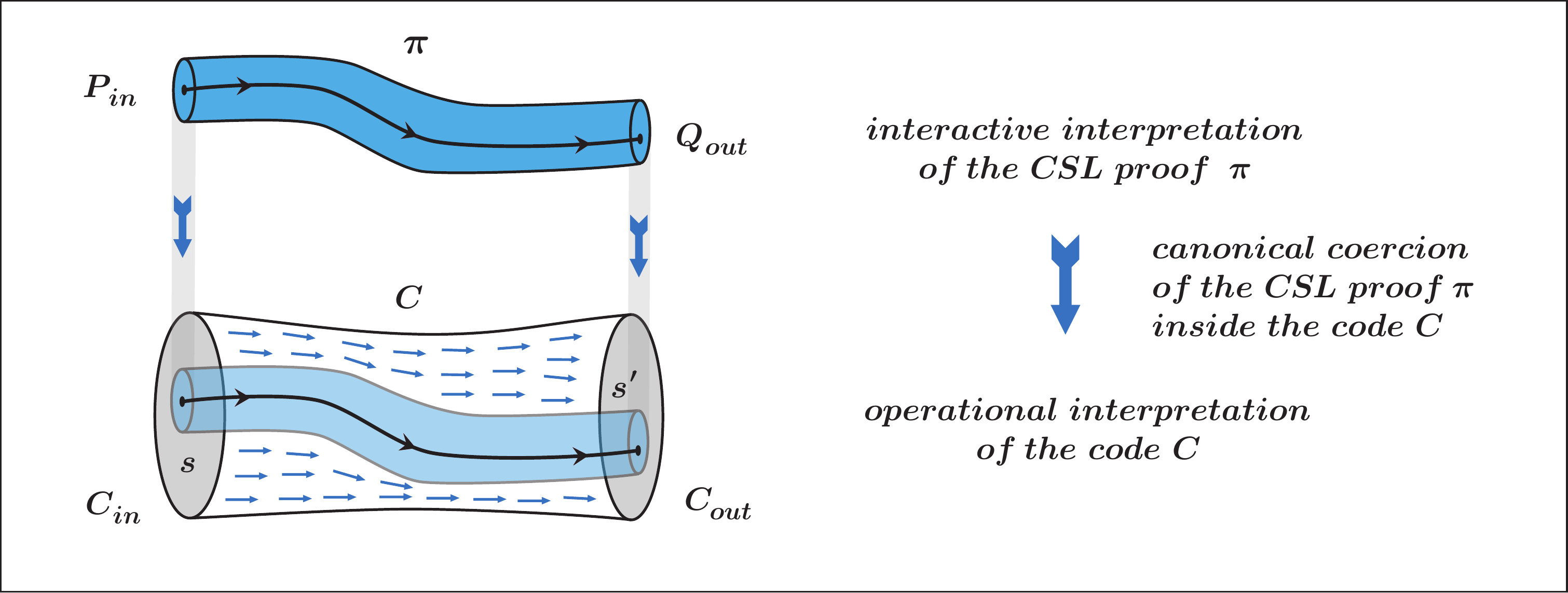}
\end{center}
In the picture, the blue cylinder above describes the asynchronous
graph $\semSep{\pi}$ of separated states where the code $C$ starting
from a state $s\in C_{in}$ of the Code satisfying the predicate $P$
will remain and produce an output $s'\in C_{out}$ satisfying the
predicate $Q$, as long as the Frame (or Environment) does not alter
the part of the separated state owned by the Code, and respects the
invariants required by the CSL judgment.
The interpretation of CSL proofs in the machine model $\mooSep$ of
separated states requires to equip the internal opcategory $\mooSep$
with one asynchronous graph $\mooSep[0,P]$ for each predicate $P$ of
the logic.
For that reason, $\mooSep$ defines what we call a \emph{colored}
$J$-opcategory where $J$ denotes here the set of predicates of the
logic.
Technically speaking, $\mooSep$ defines a \emph{polyad} in the sense
of \citet{intro-to-bicategories} instead of just a monad in the
bicategory $\cospan{\Scategorysurround}$ for
$\Scategorysurround=\AGraph$.
We explain in \S\ref{sec:cobordisms} how to adapt the construction 
of the bicategory $\cobord{\anchor}$ discussed in the introduction,
to the case of a polyad $\anchor$ instead of a monad.
We explain at the end of the paper, see \S\ref{sec:soundness},
how this conceptual toolbox sheds light on
the constructions underlying the recent proof of asynchronous soundness
established by \citet{lics18}.

\paragraph{Synopsis}
We start by introducing in \S\ref{sec:cobordisms} the notion of
internal \emph{colored} opcategory $\anchor$, and explain how to
associate a double category $\cobord{\anchor}$ of games, cobordisms
and simulations with every such synchronization template.
Then, we exhibit in \S\ref{sec:stateful-stateless} two specific
internal opcategories $\anchorstate$ and $\anchorlogic$ living in the
category $\AGraph$ of asynchronous graphs, and providing machine
models based, respectively, on a \emph{stateful} and \emph{stateless}
account of the machine instructions.
We then explain in \S\ref{sec:parallel-product} and
\S\ref{sec:sequential-composition} how to perform the parallel product
and the sequential composition in our template game model.
This includes the description in \S\ref{sec:error-monad} of a
convenient monadic treatment of errors in our asynchronous and
multi-player transition systems.
Once the synchronization template $\anchorproof$ of separated state
has been recalled in \S\ref{sec:separated-states}, we develop in
\S\ref{sec:change-of-locks} an axiomatic and fibrational account of
lock acquisition and release.
This categorical framework enables us to interpret in
\S\ref{sec:uniform-interpretation} codes and CSL proofs in a nice and
uniform way.
We explain in \S\ref{sec:soundness} how to derive from our framework
the asynchronous soundness theorem recently established in
\cite{lics18}.
We discuss the related works in \S\ref{sec:related-works}
and conclude in \S\ref{sec:conclusion}.

\section{The double category $\cobord{\anchor}$ of games and cobordisms}
\label{sec:cobordisms}
We have explained in the introduction how to associate a bicategory 
of template games and cobordisms with an internal opcategory $\anchor$
in a category $\catC$ with pushouts.
We show here that we can in fact associate with $\anchor$ a double
category $\cobord{\anchor}$ instead of just a bicategory, and extend
the construction to the case of \emph{polyads} in $\Cospan(\catC)$
instead of just monads.
We start by recalling the definition of double category, and then of
polyad, introduced by \citet{intro-to-bicategories}.
\subsection{Double categories}

We recall the notion of \demph{(pseudo) double category}, which was
introduced by \citet{categories-structurees}.
Formally, a double category $\dblcat$ is a weakly internal category in
$\Cat$, the 2-category of small categories.
More concretely, it is the data of: a collection of objects, a
collection of vertical morphisms between objects denoted with a simple
arrow $A \to B$, a collection of horizontal morphisms between pairs of
objects denoted with a crossed arrow $A \modto B$, and of 2-cells
filling squares of the form:
\[
  \begin{tikzcd}
    A_1\ar[d, "f"'] \ar[r, module, "F", ""{name=F, swap}] & B_1 \ar[d, "g"]
    \\
    A_2 \ar[r, module, "G"', ""{name=G}] & B_2
    \ar[Rightarrow, "\alpha", from=F, to=G]
  \end{tikzcd}
\]
A 2-cell is called \demph{special} if it is globular, in that $f$
and~$g$ are identities.
Composition of vertical arrows is associative, whereas composition of
horizontal arrows are only associative up to special invertible
2-cells.

The intuition is that, as for most examples, vertical morphisms are
the usual notion of morphisms associated with the objects, for example
ring morphisms for the double category of rings, and the vertical
morphisms are relational structures, such are bimodules in the case of
rings.
The other canonical example, which is relevant for this paper is the
double category of cospans: Given a category~$\catC$ with pushouts, we
consider the double category whose objects and vertical morphisms are
respectively the objects and the morphisms of~$\catC$, and whose
horizontal morphisms between~$A$ and~$B$ are the cospans in~$\catC$ of
the form depicted below left. A two cell is given by the morphism
$S \to S'$ in the right diagram.
\[
  \begin{tikzcd}
    A\ar[r] & S & B\ar[l]
  \end{tikzcd}
  \qquad\qquad
  \begin{tikzcd}
    A\ar[r]  \ar[d]& S  \ar[d]& B\ar[l] \ar[d].
    \\
    A'\ar[r] & S' & B'\ar[l].
  \end{tikzcd}
\]
Composition of horizontal morphisms is given by pushout. Because
pushouts are only unique up to iso, given a choice of pushouts, it is
easy to check that the universality of pushouts implies that
horizontal composition is indeed associative up to an invertible
special two cell.

\subsection{Polyads}
\begin{definition}
  The forgetful functor $|-|:\Cat\to\Set$ which transports every small
  category to its set of objects has a right adjoint
  $\mathbf{chaos}:\Set\to\Cat$ which transports every set~$J$ to its
  \demph{chaotic category} defined as the category $\mathbf{chaos}(J)$
  whose objects are the elements $i, j, k$ of the set~$J$, with a
  unique morphism between each pair of objects.
A \demph{polyad} in a bicategory $\mathcal{D}$ 
is a lax double functor of double category 
  \[
    \mathbf{chaos}(J) \quad\xrightarrow{\qquad\qquad}\quad \mathcal{D}
  \]
from the chaotic category over some set $J$, seen as a double category, to~$\mathcal{D}$.
A polyad is called a \demph{monad} when $J$ is a singleton.
\end{definition}
It is worth mentioning that a polyad $\anchor$ in a bicategory
$\mathcal{W}$ is the same thing as an enriched category $\anchor$ over
the bicategory $\mathcal{W}$ in the terminology of the Australian
school \cite{street-cohomology}.
Given a category $\Scategorysurround$ with pushouts, we call
\demph{internal $J$-opcategory} a polyad $\anchor$ in the double category
$\Cospan(\catC)$.
The definition can be expounded as follows.
An internal $J$-opcategory $\moo$ consists of an object $\anchorint i$
of~$\catC$ for each element $i\in J$, and of a cospan
\[
\anchormap i j
\quad : \quad
\anchorint i \modto \anchorint j
\quad = \quad
  \begin{tikzcd}[column sep=3em]
    \anchorint i \ar[r, "{\anchorinput ij}"] & \anchorcode i j & 
    \anchorint j \ar[l,
    "\anchoroutput ij"{swap}]
  \end{tikzcd}
\]
for each pair $i,j \in J$, together with two coherent families $\eta$
and $\mu$ of 2-cells between cospans:
\[
\begin{small}
\begin{tikzcd}[column sep=1.5em, row sep=1.5em]
\anchor[0,i]
\arrow[rrrr,spanmap,"{\id}"{xshift=.5ex,yshift=.5ex},""{swap,name=start},bend left]
\arrow[rrrr,spanmap,"{\anchor[1,ii]}"{swap,yshift=-.5ex},""{name=end},bend right]
&&&&
\anchor[0,i]
\arrow[from=start, to=end, double,-implies,"{\eta_i}"]
\end{tikzcd}
\quad\quad\quad
\begin{tikzcd}[column sep=1.5em, row sep=1.5em]
&&
|[alias=start]|
\anchor[0,j]\arrow[rrdd,spanmap,"{\anchor[1,jk]}"{xshift=.2ex,yshift=.2ex}]
\\
\\
\anchor[0,i]
\arrow[rruu,spanmap,"{\anchor[1,ij]}"{xshift=-.5ex}]
\arrow[rrrr,spanmap,"{\anchor[1,ik]}"{swap,yshift=-.5ex},""{name=end}]
&&&&
\anchor[0,k]
\arrow[from=start, to=end, double,-implies,"{\mu_{ijk}}"]
\end{tikzcd}
\end{small}
\]
More explicitly, $\anchormu i j k$ is given by the map $g_{ijk}$ in the
following commutative diagram:

\[
 \begin{tikzcd}[row sep=0.4em] 
   \anchorint i 
   \ar[dr, "\anchorinput i j"]
   \ar[dddddrr, "\anchorinput i k"', bend right]
   &&
   \anchorint j
   \ar[dr, "\anchorinput j k"]
   \ar[dl, "\anchoroutput ij"']
   \ar[dd, phantom, "\pushoutdown", very near end] 
   &&
   \anchorint k
   \ar[dl, "\anchoroutput jk"]
   \ar[dddddll, "\anchoroutput i k"', bend left]
   \\
   &
   \anchorcode i j 
   \ar[dr]
   &&
   \anchorcode j k
   \ar[dl]
   \\
   &&
   \anchorpushout ijk
   \ar[ddd, "g_{ijk}", near start]
   \\
   \\
   \\
   &&
   \anchorcode i k
 \end{tikzcd}
\]
Finally, there are some conditions about the associativity of the composition,
and about the identity; they can be found in
\cite[def.~5.5.1]{intro-to-bicategories}.
An \demph{internal opcategory} $\anchor$ is defined as an internal
$J$-opcategory where the set $J$ is a singleton $J=\{j\}$.  We write
in that case $\anchor[0]$ and $\anchor[1]$ for the objects
$\anchor[0,j]$ and $\anchor[1,jj]$ of the ambient category
$\Scategorysurround$, respectively.

\subsection{The double category $\cobord{\anchor}$ of games and cobordisms}
Suppose given an internal $J$-opcategory $\anchor$ in a category
$\Scategorysurround$ with pushouts.  A \demph{$j$-colored game}
$(A,\lambda_A)$ is defined as an object $A$ of $\Scategorysurround$
equipped with a morphism $\lambda_A:A\to \anchor[0,j]$.
An \demph{$ij$-colored cobordism} from a $i$-colored game $(A,\lambda_A)$ to a
$j$-colored game $(B,\lambda_B)$ is defined as a cospan in $\catC$ together with
colors $i,j\in J$ and a map $\lambda_\sigma$ such that the following diagram
commutes:
\[
  \begin{tikzcd}[column sep=4em, row sep=1.5em]
    A \ar[r, "\source{}"] 
    \ar[d, "{\lambda_A}"']
    & 
    S \ar[d, "\lambda_{\sigma}"] 
    &
   B \ar[l, "\target{}"'] \ar[d, "{\lambda_B}"]
    \\
    \anchorint i \ar[r, "\anchorinput i j"] & 
    \anchorcode  i j & 
    \anchorint j \ar[l, "\anchoroutput i j"{swap}]
  \end{tikzcd}
\]
As in the case (\ref{equation/two-cobordisms-side-by-side}) of an
internal opcategory, two such $ij$-colored cobordism
$\sigma: A\longrightarrow\hspace{-1.2em}|\hspace{1em}B$ and
$jk$-colored cobordism
$\tau: B\longrightarrow\hspace{-1.2em}|\hspace{1em}C$ can be composed
into an $ik$-colored cobordism
$\sigma;\tau : A\longrightarrow\hspace{-1.2em}|\hspace{1em}C$ using a
pushout and a relabeling along the 2-cell $\anchormu i j k$ provided
by the internal $J$-opcategory, as shown below:
\[
  \begin{tikzcd}[column sep=5em]
    A
    \ar[r, module, "\acobord", ""{swap, name=LU}]
    \ar[d]
    &
    B
    \ar[r, module, "\acobord'", ""{swap, name=RU}]
    \ar[d]
    &
    C
    \ar[d]
    \\
    \anchorint i
    \ar[r, module, "\anchormap i j"', ""{name=LD}]
    \ar[rr, bend right, module=50, "\anchormap i k"', ""{name=D}]
    &
    \anchorint j
    \ar[r, module, "\anchormap j k"', ""{name=RD}]
    \ar[to=D, Rightarrow, "\anchormu i j k"]
    &
    \anchorint i
    \ar[from=LU, to=LD, Rightarrow]
    \ar[from=RU, to=RD, Rightarrow]
  \end{tikzcd}
\]
In order to give cobordisms over an internal $J$-opcategory~$\ijcat$ a
structure of double category, we need identities, which are given by:
\[
  \begin{tikzcd}[column sep=6em]
    A
    \ar[r, module, "\Id_A", ""{swap, name=U}]
    \ar[d, "\lambda_A"]
    &
    A
    \ar[d, "\lambda_A"]
    \\
    \anchorint i
    \ar[r, module, "\Id_{\anchorint i}"{swap, name=Md}, ""{name=Mu}]
    \ar[r, module, "\anchormap i i"', bend right=55, ""{name=D}]
    &
    \anchorint i
    \ar[from=U, to=Mu, Rightarrow, "\lambda_A"]
    \ar[from=Md, to=D, Rightarrow, "\anchoridentitymap i"]
  \end{tikzcd}
\]
The vertical morphisms are the maps $f: A \to A'$ such that $\lambda_A =
\lambda_B \circ f$, and the two-cells are triples of maps $f_{\mathrm{in}} : A
\to A', f_{\mathrm{out}} : B \to B', f: S \to S'$ such that the following
diagram commutes
\[
  \begin{tikzcd}
    A \ar[r, "\codeinput \acobord"] \ar[d, "f_\mathrm{in}"']& 
    S \ar[d, "f"] &
    B \ar[l, "\codeoutput \acobord"'] \ar[d, "f_\mathrm{out}"]
    \\
    A' \ar[r, "\codeinput \acobord'"] 
                  & S' 
                  & B' \ar[l, "\codeoutput \acobord'"']
  \end{tikzcd}
\]
such that moreover, $f_{\mathrm{in}}$ and $f_{\mathrm{out}}$ are vertical
morphisms, and similarly $\lambda_\sigma = \lambda_{\sigma'} \circ f$.
One obtains:
\begin{theorem}
Every internal $J$-opcategory $\anchor$ induces a double category \/$\Cobord(\ijcat)$
whose objects are the $j$-colored games and whose horizontal maps 
are the cobordisms with composition defined as above.
\end{theorem}

\section{The stateful and stateless synchronization templates}\label{sec:stateful-stateless}
We have seen in \S\ref{sec:cobordisms} how to associate a double
category $\Cobord(\anchor)$ with every internal $J$-opcategory $\anchor$
living in an ambient category~$\Scategorysurround$ with pushouts.
As a matter of fact, all the interpretations performed in the present
paper will live in the same ambient category
$\Scategorysurround=\AGraph$ of \demph{asynchronous graphs}, which are
graphs equipped with two-dimensional tiles.
After describing this category in \S\ref{sec:asynch-cat}, we recall in
\S\ref{sec:asynchronous-machine-models} the stateful and stateless
machine models used in the paper, and simply formulated as
asynchronous graphs $\statefulmachinemodel$ and
$\statelessmachinemodel$, along a recipe initiated in \cite{lics18}.
We explain in
\S\ref{sec:the-stateful-and-stateless-internal-opcategories} how to
derive from $\statefulmachinemodel$ and $\statelessmachinemodel$ the
internal opcategories $\mooS$ and $\mooL$ living in the ambient
category $\Scategorysurround=\AGraph$ of asynchronous graphs, and
associated with the stateful and stateless machine models, respectively.

\subsection{The category of asynchronous graphs}\label{sec:asynch-cat}
A \demph{graph} $G=(V,E,\edgesource, \edgetarget)$ consists of a
set~$V$ of vertices or nodes, a set of $E$ of edges or transitions,
and a source and target functions $\edgesource,\edgetarget:E\to V$.
We call a \demph{square} a pair $(f,g)$ of paths
$P\twoheadrightarrow Q$ of length 2, with the same source and target
vertices.
An \demph{asynchronous graph} $(G,\diamond)$ is a graph~$G$ equipped
with a tuple~$\,\diamond = (T,\partial_\diamond, \sigma)$ of a set $T$
of \demph{permutation tiles}, a function $\partial_\diamond$ from the
set~$T$ to the set of squares of the graph~$G$, and an
endofunction~$\sigma$ on~$T$ such that if a tile $t \in T$ is mapped
by $\partial_\diamond$ to a square $(f,g)$, then
$\partial_\diamond(\sigma(t)) = (g,f)$. We call the function~$\sigma$
the symmetry on~$\diamond$.
A tile which is mapped to a square $(u\cdot v', v \cdot u')$ is
depicted as a 2-dimensional tile between the paths $f=u\cdot v'$ and
$g=v\cdot u'$ as follows:
\begin{equation}\label{equation/permutation-tile}
  \raisebox{-2.8em}{\includegraphics[height=5.5em]{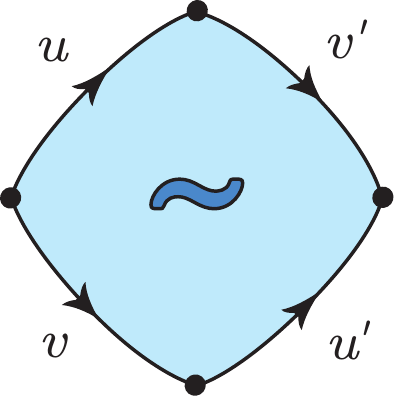}}
\end{equation}
The intuition conveyed by such a permutation tile 
$t:u\cdot v'\diamond v\cdot u'$
is that the two
transitions~$u$ and~$v$ are independent.
For that reason, the two paths~$u\cdot v'$ and~$v\cdot u'$ may be seen
as equivalent up to scheduling.
Two paths $f,g:M\transitionpath N$ of an asynchronous graph are
\demph{equivalent modulo one permutation tile} $h_1\diamond h_2$ when
there exists a tile in $T$ between~$h_1$ and~$h_2$ and when $f$ and
$g$ factor as $f=d\cdot h_1\cdot e$ and $g=d\cdot h_2\cdot e$ for two
paths $d:M\transitionpath P$ and $e:Q\transitionpath N$.
We write $f\sim g$ when the path~$f:M\transitionpath N$ is
``equivalent'' to the path $g:M\transitionpath N$ modulo a number of
such permutation tiles.
Note that the relation $\sim$ is symmetric, reflexive and transitive,
and thus defines an equivalence relation, closed under composition.

\begin{definition}\label{def:homomorphism}
An \demph{asynchronous graph homomorphism}, or \demph{asynchronous morphism},
\begin{equation}\label{equation/asynchronous-graph-homomorphism}
  \mathcal{F} \quad : \quad (G,\diamond_G) \,\, \longrightarrow \,\, (H,\diamond_H)
\end{equation}
is a graph homomorphism $\mathcal{F}: G\to H$ between the underlying
graphs, together with a function $\mathcal{F}_\diamond$ from $T_G$ to
$T_H$ such that
$\mathcal{F} \circ \partial = \partial \circ \mathcal{F}_\diamond$,
where we extend $\mathcal{F}$ in the usual way to paths and squares.
\end{definition}

\noindent
This defines the category $\AGraph$ of asynchronous graphs and
asynchronous morphisms.
As it is needed for the cobordism construction, this category has all
small limits and colimits, since it is a presheaf category, as is it
shown below.
The limits and colimits are computed pointwise for nodes, edges and
tiles.
We detail below the case of pullbacks, and consequently of Cartesian
products.
The pullback of a cospan
\[
  A_1 \xrightarrow{\quad f \quad} B \xleftarrow{\quad g \quad} A_2
\]
of asynchronous graphs is defined as the asynchronous graph~$A_1 \times_B A_2$ below.
Its nodes are the pairs $(x_1,x_2)$ consisting of a node $x_1$ in $A_1$
and of a node $x_2$ in $A_2$, such that $f(x_1)=g(x_2)$.
Its edges $(u_1,u_2):(x_1,x_2)\to(y_1,y_2)$ are the pairs consisting
of an edge $u_1: x_1 \to y_1$ in~$A_1$ and of an edge
$u_2: x_2 \to y_2$ in~$A_2$, such that $f(u_1)=g(u_2)$.
In the same way, a tile
$(\alpha_1,\alpha_2):(u_1,u_2)\cdot(v'_1,v'_2)\diamond(v_1,v_2)\cdot(u'_1,u'_2)$
is a pair consisting of a tile
$\alpha_1:u_1\cdot v'_1\diamond v_1\cdot u'_1$ of the asynchronous
graph $A_1$ and of a tile
$\alpha_2:u_2\cdot v'_2\diamond v_2\cdot u'_2$ of the asynchronous
graph $A_2$, such that the two tiles $f(\alpha_1)$ and $g(\alpha_2)$
are equal in the asynchronous graph $B$.
The Cartesian product $A_1\times A_2$ of two asynchronous graphs is
obtained by considering the special case when $B$ is the terminal
asynchronous graph, with one node, one edge, and one tile.
The definition of $A_1\times A_2$ thus amounts to forgetting the
equality conditions in the definition of the pullback
$A_1\times_B A_2$ above.

\begin{remark}
  The category $\AGraph$ of asynchronous graphs can be seen as the
  category of presheaves over the category presented by the graph:
\[
 \begin{tikzcd}
 {[0]} \ar[r, bend right=10, "t"']\ar[r, bend left=10, "s"] &
 {[1]} \ar[r, bend right=15, "ur" description]
       \ar[r, bend left=15, "dr" description]
       \ar[r, bend right=38, "ul" description]
       \ar[r, bend left=38, "dl" description]
   &
   {[2]} \ar[loop right, "\sigma"]
 \end{tikzcd}
\]
and with the equations:
\begin{align*}
 dl \circ s &= ul \circ s &
 dl \circ t &= dr \circ s &
 dr \circ t &= ur \circ t &
 ul \circ t &= ur \circ s
 \\
 \sigma \circ dl &= ul &
 \sigma \circ ul &= dl &
 \sigma \circ dr &= ur &
 \sigma \circ ur &= dr
\end{align*}
\end{remark}

\subsection{The stateful and stateless machine models}\label{sec:asynchronous-machine-models}
Following \citep{lics18}, we define the stateful and the stateless
\emph{machine models} as asynchronous graphs $\statefulmachinemodel$
and $\statelessmachinemodel$, describing the primitive semantics of
the machine: the states, the transitions between them, and the tiles
which explain how to permute the order of execution of two transitions
performed in parallel.  Note that each model depends on an implicit
set $\Locks$ of free locks ; we will come back to this point and make
it precise in \S\ref{sec:change-of-locks}.

\subsubsection{The stateful model $\statefulmachinemodel$}\label{sec:stateful-model}
A \demph{memory state} is an element $\mu = (s,h)$ of the set
\( (\Var \parfinmap \Val) \times (\Loc \parfinmap \Val) \) with
$\NN \subseteq \Loc \subseteq \Val$, where the finite partial map $s$
stands for the stack and $h$ for the heap.
A \demph{machine state} is a pair $\mstate = (\mu, L)$ of a memory
state and of a subset of~$\Locks$, which represents the available
locks.
A \demph{machine state footprint} 
\[
  \fp \in \powerset(\Var+\Loc) \times
  \powerset(\Var+\Loc) \times
  \powerset(\Locks) \times
  \powerset(\Loc)
\]
is, made of:
(i) $\readarea{\fp}$, the part of the memory that is \emph{read},
(ii) $\writearea{\fp}$, the part of the memory that is \emph{written},
(iii) $\lock(\fp)$, the locks that are \emph{touched},
and (iv) $\allocarea{\fp}$ the addresses that are \emph{allocated} or
\emph{deallocated}.
Two footprints $\fp$ and $\fp'$ are declared \demph{independent} when:
\[
\begin{array}{c}
(\,\readarea{\fp\phantom{'}} \cup \writearea{\fp\phantom{'}}\,) \cap \writearea{\fp'} \,\,=\, \emptyset
\\
(\,\readarea{\fp'} \cup \writearea{\fp'}\,) \cap \writearea{\fp\phantom{'}} \,\,=\, \emptyset
\end{array}
\;
\begin{array}{c}
\lock(\fp) \cap \lock(\fp') = \emptyset
\\
\allocarea{\fp} \cap \allocarea{\fp'} = \emptyset
\end{array}
\]
The \demph{stateful model} $\statefulmachinemodel$ is defined as the
following asynchronous graph: its nodes are the machine states
and~$\Error$, its transitions
\[
  (\mu,L) \yrightarrow{m} (\mu',L')
  \quad \mbox{or} \quad
  (\mu,L) \yrightarrow{m} \Error
\]
are given by the semantics of the instructions, which are of the form:
\begin{align*}
  m ::=\; &x \coloneqq E
      \mid x \coloneqq \deref{E}
      \mid \deref{E} \coloneqq E'
      \mid \ktest(B)
      \mid \knop
      \mid x \coloneqq \kalloc(E,\location)
      \mid \kdispose(E)
      \mid P(r)
      \mid V(r)
\end{align*}
where $x\in\Var$ is a variable, $r\in\RVar$ is a resource name,
$\location$~is a location, and $E,E'$ are arithmetic expressions,
possibly with ``free'' variables in~$\Var$.
For example, the instruction $x\coloneqq E$ executed in a machine
state $\mstate=(\memstate, L)$ assigns to the variable~$x$ the value
$E(\mu)\in\Val$ when the value of the expression $E$ can be evaluated
in the memory state~$\mu$, and produces the runtime error~$\Error$
otherwise.
The instruction $P(r)$ acquires the resource variable $r$ when it is
available, while the instruction $V(r)$ releases it when~$r$ is
locked, as described below:
\[
\begin{small}
\begin{array}{cc}
 \prftree{\evalexpr{E}{\memstate}{v}}
  {\redto{(\memstate, L)}{x \coloneqq E}{(\updmap{\memstate}{x}{v}, L)}}
&
 \prftree{E(\memstate)\,\, \text{not defined}}
  {\redto{(\memstate, L)}{x \coloneqq E}{\Error}}
  \vspace{.4em}
\\
    \prftree{r\notin{}L}
  {\redto{(\memstate, L)}{P(r)}{(\memstate, L\uplus\{r\})}}
&
    \prftree{r\notin{}L}
  {\redto{(\memstate, L\uplus\{r\})}{V(r)}{(\memstate, L)}}
  \end{array}
\end{small}
\]
The inclusion $\Loc \subseteq \Val$ means that an expression $E$ may also
denote a location.
In that case, $[E]$ refers to the value stored at location~$E$ in the heap.
The instruction $x \coloneqq \kalloc(E,\location)$ allocates some
memory space on the heap at address $\location\in\Loc$, initializes it
with the value of the expression $E$, and assigns the address
$\location$ to the variable~$x\in\Var$ if $\location$ was free,
otherwise there is no transition.  $\kdispose(E)$ deallocates the
location denoted by~$E$ when it is allocated, and returns~$\Error$
otherwise.
The instruction $\knop$ (for no-operation) does not alter the state.
The instruction $\ktest(B)$ behaves like $\knop$ when the initial
state satisfies~$B$, and is not defined otherwise.
The asynchronous tiles of $\mooS$ are the squares of the form
\[
  \mstate \yrightarrow{m} \mstate_1 \yrightarrow{m'} \mstate' \quad \sim
  \quad \mstate \yrightarrow{m'} \mstate_2 \yrightarrow{m} \mstate'
\]
where their footprints are independent in the sense above.
 
\subsubsection{The stateless model $\statelessmachinemodel$}\label{sec:stateless-model}
A \demph{lock footprint}
\[
  \fp \in \powerset(\Locks) \times \powerset(\Loc)
\]
is made of a set of locks~$\lock(\fp)$ and a set of locations~$\allocarea{\fp}$.
%
Two such footprints are \demph{independent} when their sets are
component-wise disjoint.
The \demph{stateless model} $\statelessmachinemodel$ is defined in the
following way: its nodes are the subsets of $\Locks$, and its
transitions are all the edges of the form (note the non-determinism)
\[
  L \xrightarrow{\,\,P(r)\,\,} L\!\uplus\!\{r\} 
    \quad\;\; 
  L \xrightarrow{\,\,\kalloc(\location)\,\,} L 
   \quad\;\; 
  L \xrightarrow{\;\;\tau\;\;} L 
  \quad\;\;
  L\!\uplus\!\{r\} \xrightarrow{\,\,V(r)\,\,} L
  \quad\;\;
  L \xrightarrow{\,\,\kdispose(\location)\,\,} L 
 \quad\;\;
   L \xrightarrow{\;\;\iwouldprefernotooverline{m}\;\;} \Error
\]
where $\iwouldprefernotooverline{m}$ is a \demph{lock instruction} of the form:
\[
  P(r) \mid V(r) \mid \kalloc(\location) \mid \kdispose(\location) \mid \tau
\]
for $\location\in\Loc$ and $r\in\Locks$.
The purpose of these transitions is to extract from each instruction
of the machine its synchronization behavior.
An important special case, the transition~$\tau$ represents the
absence of any synchronization mechanism in an instruction like
$x:=E$, $x:=[E]$ or $[E]:=E'$.
The asynchronous tiles of $\mooL$ are the squares of the form
\[
L \yrightarrow{x} L_1 \yrightarrow{y} L'
\quad 
\sim
\quad 
L \yrightarrow{y} L_2 \yrightarrow{x} L'
\]
when the lock footprints of $x$ and $y$ are independent.
It is worth noting that $L'$ may be equal to~$\Error$ in such an
asynchronous tile.
%
Note that the asynchronous graph $\statelessmachinemodel$ is more
liberal than $\statefulmachinemodel$ about which footprints commute,
because it only takes into account the locks as well as the allocated
and deallocated locations.
As explained in the introduction, this mismatch enables us to detect
\emph{data races} in the machine as well as in the code, see also
\cite{lics18}.


\subsection{The stateful and the stateless internal opcategories $\mooS$ and $\mooL$}
\label{sec:the-stateful-and-stateless-internal-opcategories}

We have just described in \S\ref{sec:asynchronous-machine-models} the
asynchronous graphs $\statefulmachinemodel$ and
$\statelessmachinemodel$ which we take as stateful and stateless
machine models in the paper.
We explain now how we turn these machine models
$\statefulmachinemodel$ and $\statelessmachinemodel$ into the internal
opcategories $\mooS$ and $\mooL$ which we will use, in the ambient
category $\Scategorysurround=\AGraph$ of asynchronous graphs.
The constructions of the internal opcategories $\moo=\mooS,\mooL$ from
the asynchronous graphs
$\machinemodel=\statefulmachinemodel,\statelessmachinemodel$ follows
exactly the same recipe: both of them are defined as cospans of
monomorphisms in the ambient category $\Scategorysurround=\AGraph$ of
asynchronous graphs
\[
  \begin{tikzcd}[column sep=5em]
    \moo[0] \ar[hook, r, "\anchorinput{}{}"] & \theanchorcode & \moo[0] \ar[l,
    hook', "\anchoroutput{}{}"']
  \end{tikzcd}
\]
describing
$\machinemodel=\statefulmachinemodel,\statelessmachinemodel$ as a
monad (or polyad with a single color $j\in J$) in the double category
$\Cospan(\catC)$, see \S\ref{sec:cobordisms} for details.
We like to think of the asynchronous graphs $\statefulmachinemodel$
and $\statelessmachinemodel$ defined in
\S\ref{sec:asynchronous-machine-models} as ``solipsistic'' games with
only one player: the underlying machine.
Building on this intuition, we follow the conceptual track discussed
in the introduction, and construct an asynchronous graph
$\moo[1]=\mooS[1],\mooL[1]$ with two players (for Code and
Environment) instead of one, both in the stateful case of
$\moo[1]=\mooS[1]$ and in the stateless case $\moo[1]=\mooL[1]$.
%
%
%
%
The shift from one player to two players is performed in a very simple
way.
We consider the functor $\Omega: \textbf{Set} \to \AGraph$ which
transports a given set $\mathcal{L}$ of labels to the asynchronous
graph $\Omega(\mathcal{L})$ with one single node, the elements
of~$\mathcal{L}$ as edges, and one tile for each square.
We are particularly interested in the case when the set of labels
$\mathcal{L}=\{\Cplayer,\Fplayer\}$ contains the two
polarities~$\Cplayer$ and~$\Fplayer$ associated with the Code and to the
Frame (or the Environment), respectively.
%
The asynchronous graph $\Omega(\{ \Cplayer,\Fplayer \})$ enables us to
define the \emph{two-player} stateful and stateless machine
models~$\statefulmachinemodeltwoplayers$ and
$\statelessmachinemodeltwoplayers$ as the Cartesian product of
asynchronous graphs
\[
  \statefulmachinemodeltwoplayers = \statefulmachinemodel \times \Omega(\{ \Cplayer,\Fplayer \})
\quad\quad
\statelessmachinemodeltwoplayers = \statelessmachinemodel \times \Omega(\{ \Cplayer,\Fplayer \})
\]
Note that the resulting asynchronous graph
$\machinemodeltwoplayers=\statefulmachinemodeltwoplayers,\statelessmachinemodeltwoplayers$
has the same nodes
as~$\machinemodel=\statefulmachinemodel,\statelessmachinemodel$ and
two edges, the first one labeled with a polarity~$\Cplayer$ for Code,
the second one labeled with a polarity~$\Fplayer$ for Frame, for each
edge in the original asynchronous graph
$\machinemodel=\statefulmachinemodel,\statelessmachinemodel$.
The two circles $\circ$ and $\bullet$ in the notation
$\machinemodeltwoplayers$ are mnemonics designed to remind us that
there are two players in the game: the Code playing the white side
($\circ$) and the Environment playing the black side ($\bullet$).
%
We have accumulated enough material at this stage to define the
internal opcategories $\moo=\mooS,\mooL$ in the ambient category
$\Scategorysurround=\AGraph$.
The asynchronous graph $\moo[1]$ is defined as the two-player machine
model $\machinemodeltwoplayers$ while the asynchronous graph $\moo[0]$
is defined as the one-player machine model.
In summary:
$$
\moo[1]=\machinemodeltwoplayers
\quad\quad
\moo[0]=\machinemodel.
$$
The asynchronous graph $\moo[0]$ with one player
is then embedded in the asynchronous graph $\moo[1]$ with two players
by the homomorphism $\anchorinput{}{}=\anchoroutput{}{}:\moo[0]\to\moo[1]$ 
obtained by transporting every node of $\moo[0]$ to the corresponding node 
in $\moo[1]$, and every edge of $\moo[0]$ to the corresponding edge 
in $\moo[1]$ with the polarity $\Fplayer$ of the Frame.

\section{The parallel product}\label{sec:parallel-product}
We describe below in full detail how to construct the parallel product
$C_1\| C_2$ of two codes $C_1$ and $C_2$
by applying a \emph{push} and \emph{pull} functors along a pair
(\ref{equation/span-of-functors}) of internal functors.
%
We start by formulating in \S\ref{sec:internal-functor} the notion of
\demph{internal functor} from an internal $J$-opcategory~$\ijcat$ to
an internal $J'$-opcategory~$\ijcat'$ living in the same ambient
category $\Scategorysurround$ with pushouts.
We then introduce in \S\ref{sec:acute-spans} the notion of \emph{acute
  span}, which leads us to the notion of \emph{span-monoidal} internal
$J$-opcategory formulated in \S\ref{sec:span-monoidal}.
The parallel product of $\|$ of two codes is then described in
\S\ref{sec:illustration-parallel-product}.

\subsection{Internal functors between internal $J$-opcategories}\label{sec:internal-functor}
The notion of internal functor between internal $J$-opcategories is
defined as expected.
\begin{definition}[Internal functor]
\label{def:internal-functor}
  An internal functor $F=(f, F[0,\cdot], F[1,\cdot])$
  from~$\ijcat$ to~$\ijcat'$ is a triple consisting of a function $f: J \to J'$ 
  between the sets of colors, together with:
  \begin{itemize}
    \item for each $i\in J$, a map $F[0,i]$ in the ambient category~$\Scategorysurround$:
      \[
        F[0,i] : \anchorint i \longrightarrow \anchorpint {f(i)},
      \]
    \item for each $i,j \in J$, a map $F[1,ij]$ in the ambient category~$\Scategorysurround$:
      \[
        F[1,ij] : \anchorcode i j \longrightarrow \anchorpcode{f(ij)}{}
      \]
  \end{itemize}
  where we use the lighter notation $f(ij)$ for $f(i)f(j)$.
  One asks moreover that the diagram below commutes
  \[
    \begin{tikzcd}[column sep=5.5em]
      \anchorint i \ar[d, "{F[0,i]}"']\ar[r,"{\anchorinput{i}{j}}"]&
      \anchorcode i j \ar[d, "{F[1,ij]}"]&
      \anchorint j \ar[d, "{F[0,j]}"]\ar[l,"{\anchoroutput{i}{j}}"{swap}]
      \\
      \anchorpint {f(i)} \ar[r,"{\anchorinput{f(i)}{f(j)}}"] &
      \anchorpcode {f(ij)}{} &
      \anchorpint {f(j)} \ar[l,"{\anchoroutput{f(i)}{f(j)}}"{swap}]
    \end{tikzcd}
  \]
  and that the internal functor is compatible with the identities: the
  diagram on the right below commutes; and with composition: the maps $F[2, ijk]
  : \anchorpushout ijk \to \anchorppushout{f(ijk)}{}{}$ induced by 
  universality of the pushout must make the left diagram commute.
  \[
    \begin{tikzcd}
      \anchorpushout ijk \ar[d, "{F[2,ijk]}"']\ar[r]&
      \anchorcode ik \ar[d, "{F[1, ik]}"]
      \\
      \anchorppushout {f(ijk)}{}{} \ar[r]&
      \anchorpcode {f(ik)}{}
    \end{tikzcd}
    \qquad
    \qquad
    \begin{tikzcd}
      \anchorint i \ar[d, "{F[0,i]}"']\ar[r, "\eta_i"]&
      \anchorcode ik \ar[d, "{F[1, ii]}"]
      \\
      \anchorint {f(i)}{}{} \ar[r, "\eta'_{f(i)}"']&
      \anchorpcode {f(ii)}{}
    \end{tikzcd}
  \]
\end{definition}

\noindent
The internal $J$-opcategories and internal functors form a category
$\InternalCat \catC$.
This category admits products: the index set of the product of two internal
categories is the product of their index sets, and $(\moo \times
\moo^{\!'})[0,(i,j)] = \anchorint i \times \anchorpint j$. This fact will play a
central role in the sequel.

\subsection{Plain internal functors}
An important family of internal functors are \demph{plain internal functors},
which are functors whose action on the colors is the identity.
With the notations of Definition~\ref{def:internal-functor}, they are the
internal functors such that $f=\id$.
Plain internal functors are useful because they define a pull
operation when $\catS$ has pullbacks, given by the obvious three
pullbacks in the following diagram:
\[
  \begin{tikzcd}
    && B \ar[dl] \ar[dd]
    \\
    & S \ar[dd]
    \\
    A\ar[dd]\ar[ru] && \moo_1[j] \ar[dl]
    \\
    & \moo_1[ij] && \moo_2[j] \ar[ul] \ar[dl]
    \\
    \moo_1[i] \ar[ur] && \moo_2[ij] \ar[ul]
    \\
    & \moo_2[i] \ar[ul] \ar[ur]
  \end{tikzcd}
\]
This construction induces a lax double functor between
$\Cobord(\moo_2)$ and $\Cobord(\moo_1)$.
The converse operation of pushing exists for any internal functor.
\begin{lemma}\label{lem:push-pull-internal-functors}
  A plain internal functor $F: \moo_1 \to \moo_2$ induces a lax double
  functor of double category defined by taking the pointwise pullbacks
  along the components of~$F$
  \[
    \pull{F} : \Cobord(\moo_2) \longrightarrow \Cobord(\moo_1).
  \]
  Conversely, any internal functor $F: \moo_1 \to \moo_2$ induces a
  pseudo functor by post-composition
  \[
    \push{F} : \Cobord(\moo_1) \longrightarrow \Cobord(\moo_2).
  \]
\end{lemma}

\subsection{Acute spans of internal functors}\label{sec:acute-spans}
We start by introducing the notion of \demph{acute span} of internal functors,
and then describe how transport of structure works along an acute span.

\paragraph{Acute spans} 
An \demph{acute span} between an internal $J_1$-opcategory $\moo_1$
and an internal $J_2$-opcategory $\moo_2$ is defined as a span of internal functors
\[
  \begin{tikzcd}[column sep=5em]
    \moo_1 & \moo_0 \ar[l, "=", "F"'] \ar[r, "G"]& \moo_2
  \end{tikzcd}
\]
where $\moo_0$ is an internal $J_0$-opcategory and the $=$ sign
indicates that the internal functor $F$ is plain.
Here, we suppose that the ambient category $\Scategorysurround$ has
pushouts and pullbacks.
In that case, the definition gives rise to a bicategory $\AcuteSpan$,
whose objects are the internal $J$-opcategories in the ambient
category $\Scategorysurround$, whose $1$-cells are acute spans, and
whose $2$-cells are commuting diagrams of the form:
\[
  \begin{tikzcd}[row sep=0.3em, column sep=3em]
    & \moo_0 \ar[dl, "="{sloped}] \ar[dr]
    \ar[dd, "\phi", "="{sloped} ]
    \\
    \moo_1 && \moo_2
    \\
    & \moo^{\!'}_0 \ar[ul, "="{sloped,swap}] \ar[ur]
  \end{tikzcd}
\]
where the internal functor~$\phi:\anchor_0\to\anchor'_0$ is plain.
Acute spans are composed using pullbacks of plain internal functors
along internal functors, in the category of internal $J$-opcategories.
The pullback is defined as follows.
Assume that we are given two internal functors:
\[
  \begin{tikzcd}[column sep=5em]
    \moo_1\ar[r, "{(f,F)}"] & \moo_2 & \moo_3 \ar[l, "{(\id,G)}"', "="]
  \end{tikzcd}
\]
between an internal $J_1$-opcategory $\moo_1$ and a $J_3$-opcategory $\moo_3$.
Their pullback is defined as the internal $J_1$-opcategory $\moo_1 \times_{\moo_2} \moo_3$ 
described below.
At the level of its components and objects of $\Scategorysurround$, for each $i\in J_1$, 
the two internal functors give the following diagram:
\[
  \begin{tikzcd}[column sep=5em]
    \moo_1[0,i] \ar[r, "{F[0,i]}"]
    &
    \moo_2[0,f(i)]
    &
    \moo_3[0,f(i)] \ar[l, "{G[0,f(i)]}"']
  \end{tikzcd}
\]
and we simply define $(\moo_1 \times_{\moo_2} \moo_3)[0,i]$ as the
pullback of that diagram in the ambient category $\Scategorysurround$.
We proceed similarly to define the
$(\moo_1 \times_{\moo_2} \moo_3)[1,ij]$; and the universality of the
pullbacks in $\Scategorysurround$ gives the structural maps between the
two.

\paragraph{Transport along acute spans}
The ultimate \emph{raison d'{\^e}tre} of acute spans is to 
induce an operation of transport by ``pull-then-push'' along the two legs of a span. 
This operation plays a fundamental role in the paper, in particular because we derive 
our definition of parallel product from it.
This operation can be formulated as a pseudo functor $\Cobord:\AcuteSpan\to\DblCatLax$ 
of bicategories, from the bicategory $\AcuteSpan$ just defined to the bicategory $\DblCatLax$ 
of double categories and lax double functors.
It is defined on objects and $1$-cells by:
\begin{align*}
  \Cobord \quad : \quad \AcuteSpan \;&\longrightarrow\; \DblCatLax
  \\
    \moo \;&\longmapsto\; \Cobord(\moo)
    \\
    (F,G) \; &\longmapsto\; \push G \circ \pull F
\end{align*}
To prove that this induces a pseudo functor, one needs in particular
to check that, given two composable acute spans $F$ and $G$,
$\Cobord(F \circ G) \isomorphic \Cobord(F) \circ \Cobord(G)$.
This is indeed the case because, in the following commutative diagram,
the map from $B$ to $A_2$ is an isomorphism, for the top face of the
cube is a pullback according to the pasting lemma for pullbacks.
\[
  \begin{tikzcd}[column sep=7em, row sep=.5em]
    A \ar[ddd] && A_1 \ar[ddd]
    \\
    {\vphantom{A}}
    \\
    & A_1 \ar[ddd] \ar[luu] \ar[ruu, equal] && A_2 \ar[luu] \ar[ddd]
    \\
    \vphantom{A}\moo_1 &&\moo_3 \ar[ldd, leftarrow] \ar[rdd, leftarrow] && \moo_5
    \\
    && B \ar[ddd] \ar[luu, crossing over] \ar[ruu, crossing over]
    \\
    &A \moo_2 \ar[luu] & & \moo_4 \ar[ruu]
    \\
    {\vphantom{A}}
    \\
    && \moo_8 \ar[luu] \ar[ruu]
  \end{tikzcd}
\]

\subsection{Span-monoidal internal $J$-opcategories}\label{sec:span-monoidal}

We use the span operation above to define the parallel product of two programs.
Categorically, we construct a lax-monoidal structure on $\Cobord(\moo)$, that is
to say a lax double functor
\[
  \parallel \;: \Cobord(\moo) \times \Cobord(\moo) \to \Cobord(\moo).
\]
By lax double functor, we mean that there exists a natural
and coherent family of maps:
\[
  (C_1 \parallel C_2)\: ; (D_1 \parallel D_2) 
  \;\;\longrightarrow\;\; 
  (C_1 ; D_1) \parallel (C_2 ; D_2).
\]
Since the composition of cobordisms corresponds to the sequential composition of
programs, these coercion maps capture the famous Hoare inequality \citep{Hoare}
\[
(C_1 \parallel C_2)\: ; (D_1 \parallel D_2) 
 \quad\!\subseteq\!\quad 
  (C_1 ; D_1) \parallel (C_2 ; D_2).
\]
In our setting, this follows from Proposition~\ref{lem:push-pull-internal-functors},
and therefore from the fact that there is always a map from a colimit of
limits to the corresponding limit of colimits.

This tensor product is built the same way as in the case of template games
\citep{popl19}, using the notion of internal span-monoidal opcategory.
We remarked below Definition~\ref{def:internal-functor},
the category $\InternalCat{\Scategorysurround}$ of internal opcategories
in the ambient category $\Scategorysurround$ is a Cartesian category.
As a consequence, the lax functor $\Cobord$ is lax monoidal, where we equip both
$\AcuteSpan$ and $\DblCatLax$ with the monoidal structure induced by their Cartesian
products.
In particular, there exist coercions living in the Cartesian category $\DblCatLax$,
and thus provided by pseudo functors of double categories
\[
  \begin{array}{ccccc}
    \cobordmult{\moo_1}{\moo_2} &:& \Cobord(\moo_1) \times \Cobord(\moo_2)
                                &\longrightarrow&
    \Cobord(\moo_1 \times \moo_2)
    \\
    \cobordunit &:& \mathbb{1} &\longrightarrow & \Cobord(\mathbf{1})
  \end{array}
\]
The first coercion is obtained in the natural way by taking the
``pointwise'' Cartesian product of the two cobordisms, and the second is the
trivial cobordism, given by the initial opcategory $\mathbf 1$.
\begin{definition}\label{def:span-monoidal}
  A \demph{span-monoidal internal $J$-opcategory} $(\moo, \parallel, \eta)$ is
  a symmetric pseudomonoid object in the symmetric monoidal bicategory
  $\AcuteSpan$.
  In particular, $\parallel$ and~$\eta$ are two acute spans:
  \[
    \begin{tikzcd}[column sep=4em]
      \moo \times \moo &
      \mootensor \ar[l, "\pick"']\ar[r, "\pince"] &
      \moo
    \end{tikzcd}
    \qquad\qquad
    \begin{tikzcd}[column sep=4em]
      \mathbf{1} &
      \moo^\eta \ar[l]\ar[r] &
      \moo.
    \end{tikzcd}
  \]
  together with invertible $2$-cells that witness the associativity
  of~$\parallel$, and the fact that~$\eta$ is is a left and right identity of~$\parallel$.
  See \citep[\S3]{monoidal-bicategories-and-Hopf-algebroids} for the complete
  definition.
\end{definition}

\noindent
Now, combining the external operations $\cobordmult{\moo}{\moo}$ and
$\cobordunit$ on the one hand, and the internal operations ${\parallel},\eta$ on
the other, we get a lax-monoidal structure on~$\Cobord(\moo)$ whose
multiplication and unit are respectively given by:
\[
  \begin{tikzcd}[column sep=5em, row sep=0.5em]
    \Cobord(\moo) \times \Cobord(\moo)
    \ar[r, "\cobordmult\moo\moo"]
    &
    \Cobord(\moo \times \moo) 
    \ar[r, "\Cobord({\parallel})"]
    &
    \Cobord(\moo)
    \\
    \mathbb{1}
    \ar[r, "\cobordunit"]
    &
    \Cobord(\mathbf 1) 
    \ar[r, "\Cobord(\eta)"]
    &
    \Cobord(\moo)
  \end{tikzcd}
\]
\begin{theorem}
  Every span-monoidal internal $J$-opcategory $\moo$ induces a symmetric
  lax-monoidal double category $\Cobord(\moo)$.
\end{theorem}

\subsection{Illustration: parallel product of codes}\label{sec:illustration-parallel-product}

Now that we have a general method for equipping the double category $\Cobord(\moo)$ 
with a lax-monoidal structure, we use it to define the parallel product 
for the stateful and the stateless semantics of the Code.
The case of separated states is a bit more involved and will be
treated later, see \S\ref{sec:separated-states/span-monoidal} for
details.
The basic idea is to think of the code $C_1 \parallel C_2$ as a situation
where (1) there are \emph{three} players involved: the Code of~$C_1$, 
the Code of~$C_2$ and the overall Frame~$F$, but where (2)
we have forgotten the identities of the codes~$C_1$ and~$C_2$ 
by considering both of them to be the Code~$C$.
This idea leads us to define the three-player machine models~$\statefulmachinemodelthreeplayers$
and~$\statelessmachinemodelthreeplayers$, with polarities $\Cplayer_1, \Cplayer_2$ and~$\Fplayer$. 
In the same way the two-player versions of the machine models are
deduced from their one-player versions in
\S\ref{sec:the-stateful-and-stateless-internal-opcategories} using a
product, we use here the pullback:
\[
  \begin{tikzcd}[column sep=3em, row sep=1em]
    \machinemodelthreeplayers 
    \ar[rr, "{\pi^{(12)(\Fplayer)}_{\mathrm{pol}}}"]
    \ar[dd, "{\pi^{(12)(\Fplayer)}_{\mathrm{state}}}"'] 
    &&
    \Omega\{\Cplayer_1, \Cplayer_2,\Fplayer\}
    \ar[dd, "{\Omega (12)(\Fplayer)}"]
    \\
    \\
    \machinemodeltwoplayers 
    \ar[rr, "\pi_{\mathrm{pol}}"] 
    && \Omega\{\Cplayer,\Fplayer\}
  \end{tikzcd}
\]
where $(12)(\Fplayer)$ denotes
the function $\{\Cplayer_1,\Cplayer_2,\Fplayer\} \to \{\Cplayer,\Fplayer\}$
which maps the polarities~$\Cplayer_1$ and $\Cplayer_2$ to~$\Cplayer$, 
and the polarity~$\Fplayer$ to itself.
%
More explicitly, $\machinemodelthreeplayers$ has three copies for each
transition of $\machinemodel$, one copy for each of the three
polarities $\Cplayer_1,\Cplayer_2,\Fplayer$.
The span-monoidal structures on the internal
opcategories~$\moo=\mooS,\mooL$ are defined in the same way:
\[
  \begin{tikzcd}[column sep=3em]
    \machinemodeltwoplayers \times \machinemodeltwoplayers &&
    \machinemodelthreeplayers \ar[ll, "{\left< \pi^{(1)(2\Fplayer)}_{\mathrm{state}}, \pi^{(2)(1\Fplayer)}_{\mathrm{state}}  \right>}"']
    \ar[r, "\pi^{(12)(\Fplayer)}_{\mathrm{state}}"] &
    \machinemodeltwoplayers
  \end{tikzcd}
  \qquad\quad
  \begin{tikzcd}[column sep=1em]
    \mathbf{1} &&
    \machinemodel \ar[ll]\ar[rr, "{\iota_{\Fplayer}}"] &&
    \machinemodeltwoplayers.
  \end{tikzcd}
\]
where, for instance, $\pi^{(1)(2\Fplayer)}_{\mathrm{state}}$ maps
every transition with polarity $\Cplayer_1, \Cplayer_2, \Fplayer$ to
the corresponding transition with polarity given by the map
$\Cplayer_1\mapsto\Cplayer$ and $\Cplayer_2, \Fplayer\mapsto\Fplayer$
; while the homomorphism $\iota_{\Fplayer}$ embeds the asynchronous
graph $\machinemodel$ into $\machinemodeltwoplayers$ with every edge
transported to the corresponding edge of polarity~$\Fplayer$.
The resulting lax-monoidal product $\parallel$ is essentially the same
as the parallel product which was defined by hand in \citet{lics18}.
The product synchronizes transitions of the Code in~$C_1$ with transitions of the
Environment in~$C_2$ which are mapped to the same transition in~$\moo^{\!}$. 
The intuition here is that a transition of $C_1$ is seen by~$C_2$
as a transition of its Environment.
Note that the parallel product preserves tiles from $C_1$ and
from~$C_2$, and adds Code/Code tiles when one transition comes from
$C_1$ and the other transition from $C_2$---meaning that the two
instructions are executed on two different ``threads''---and their
image in~$\machinemodeltwoplayers \times \machinemodeltwoplayers$
forms a tile---meaning that these two instructions are independent.

\section{Sequential composition}\label{sec:sequential-composition}

So far, to compose cobordisms horizontally, the target of the first cobordism
must exactly match the source of the second. This is not a reasonable
assumption, for the initial and the final states of a program are part of its
internal state.
To summarize, in general, the two cobordisms we wish to compose look like:
\[
  \begin{tikzcd}
    A\ar[rr, module]\ar[d] &  & B\ar[d, "\lambdaout"] & & B'\ar[d, "\lambdain"']
    \ar[rr, module]& & C' \ar[d]
    \\
    \anchorint i
    \ar[rr, module, "\anchorcode{i}{j}"]
                   &&
    \anchorint j&
                &
    \anchorint {i'}
    \ar[rr, module, "\anchorcode{i'}{j'}"]
                &&
    \anchorint {j'}
  \end{tikzcd}
\]
In practice, in all the cases we consider, $\anchorint j$ and
$\anchorint{i'}$ will be equal, but $B$ and~$B'$ will be different.
To bridge the gap between $(B,\lambdaout)$ and $(B',\lambdain)$, 
we will use a \demph{filling system} over the internal $J$-opcategory~$\moo$.
With each pair of interfaces~$\lambdaout : B \to \anchorint j$ and
$\lambdain : B' \to \anchorint{i'}$, the filling system associates a
cobordism
\[
(B, \lambdaout) \xrightarrow{\quad \quad} \Fill{(B,\lambdaout)}{(B',\lambdain)} \xleftarrow{\quad \quad} (B', \lambdain)
\]
from $(B, \lambdaout)$ to $(B', \lambdain)$.
Thanks to this mediating cobordism, it becomes possible to compose the two cobordisms 
using the usual composition of cobordism.
%
Given such a filling system, we write $C_1 \fancyseq C_2$ 
for this generalized form of composition.
\begin{proposition}\label{prop:Hoare-still-holds}
  Suppose that there exists a map
  $\Fill{\lambda \parallel \lambda'}{\mu \parallel \mu'} \to \Fill
  \lambda \mu \parallel \Fill{\lambda'}{\mu'}$.  In that case, the
  Hoare inequality holds:
  \(
    (C_1 \parallel C'_1) \fancyseq (C_2 \parallel C'_2) \to (C_1 \fancyseq C_2)
    \parallel (C'_1 \fancyseq C'_2)
  \).
\end{proposition}
\begin{proof}
  The Hoare inequality for the usual composition applied twice gives
  us the map below, from which we can conclude using the hypothesis on
  the filling system.
  \[
    (C_1 \parallel C'_1) ; (\Fill \lambda \mu \parallel
\Fill{\lambda'}{\mu'}) ; (C_2 \parallel C'_2) \to (C_1 \fancyseq C_2)
    \parallel (C'_1 \fancyseq C'_2) \qedhere
  \]
\end{proof}

\noindent
When the base category~$\catC$ has pullbacks as well as pushouts,
which is the case in the examples we are considering, a filling system always
exists. It is defined by the following diagram:
\[
  \begin{tikzcd}[row sep=0.1em]
    &
    B_j \times_{\anchorcode j {j'}} A_{j'}\ar[dl]\ar[dr]
    \\
    B_j \ar[dr]\ar[dd, "\lambda"]
    &&
    A_{j'} \ar[dl]\ar[dd, "\lambda'"]
    \\
    &{B_j \cup_{\anchorcode j {j'}} A_{j'}}\ar[dd, dashed]
    \\
    \anchorint j\ar[dr, "\anchorinput j{j'}", near start]
    &&
    \anchorint {j'}\ar[dl, "\anchorinput j{j'}"', near start]
    \\
    &
    \anchorcode j {j'}
  \end{tikzcd}
\]
where $B_j \times_{\anchorcode j {j'}} A_{j'}$ is a pullback, and where $B_j
\cup_{\anchorcode j {j'}} A_{j'}$ is a pushout above that pullback.
Intuitively, it identifies all nodes of~$B_j$ and of~$A_{j'}$ that
have the same underlying state in $\anchorcode j {j'}$.
This is the filling system which will be used in our interpretation
of sequential composition.
We establish in the Appendix \S\ref{sec:fill-Hoare} that the
hypothesis of Proposition~\ref{prop:Hoare-still-holds} holds in the
case of the stateful and stateless templates~$\mooS$ and~$\mooL$.
Interestingly, this is not necessarily the case for the template
$\mooSep$ of separated states regulating the interpretation of CSL
proofs.
The reason is that there are several ways to decompose a
given separated state between two players $\Cplayer, \Fplayer$ in
$\Sepmodeltwoplayers$ into a separated state between three players
$\Cplayer_1,\Cplayer_2,\Fplayer$ in $\Sepmodelthreeplayers$.

\section{The error monad}\label{sec:error-monad}

We want to keep track of the errors in our transition systems, in
particular we want to carefully control how errors propagate in the
various constructions.
Our method is to use the \demph{error monad}~$\monT$ over the category
of asynchronous graphs, which adds an isolated node~$\bullet$.
Then we swap our base category from the category $\catS$ to its
category $\EM \catS \monT$ of $\monT$-algebras.
In the case of asynchronous graphs, a $\monT$-algebra is simply a
pointed asynchronous graph $(x,G)$ where $x$~is a node of~$G$, and a
morphism of algebras is an asynchronous morphism which maps the
distinguished node of its source to that of its target.
In our case, the distinguished node will correspond to the error.
As we will see, this category $\EM \catS \monT$ of $\monT$-algebras
inherits from~$\catS$ the properties that are needed to define the
operations that we use to interpret programs and proofs.
This principled approach to error handling ensures that the
error is represented as a single node in the transition systems.

\subsection{Liftings}

Both the stateful and the stateless machine models have natural
$\monT$-algebra structures: the distinguished node is the error
node~$\Error$.
Given an opcategory~$\moo$ in the category of algebras
$\EM\catS\monT$, the monad $\monT$ can be lifted to a monad $\monTT$ on
$\Cobord(\moo)$, where we omit the writing of the forgetful functor
from the algebra to the underlying category.
The idea is to define the image under $\monTT$ of a cobordism to be
the cobordism
\[
  \begin{tikzcd}
    \monT A \ar[r, "\monT\source{}"] 
    \ar[d, "{\monT\lambda_A}"']
    & 
    \monT S \ar[d, "\monT\lambda_{\sigma}"] 
    &
   \monT B \ar[l, "\monT\target{}"'] \ar[d, "{\monT\lambda_B}"]
    \\
    \monT\theanchorint \ar[d, "a_0"'] \ar[r, "\monT\anchorinput{}{}"] & 
    \monT\theanchorcode \ar[d, "a_1"] & 
    \monT\theanchorint \ar[d, "a_0"] \ar[l, "\monT\anchoroutput{}{}"{swap}]
    \\
    \theanchorint \ar[r, "\anchorinput{}{}"] & 
    \theanchorcode & 
    \theanchorint \ar[l, "\anchoroutput{}{}"{swap}]
  \end{tikzcd}
\]
where $a_0$ and $a_1$ are the structural morphism of $\moo[0]$ and
$\moo[1]$ respectively.
The multiplication $\dot\mu$ is given for games by the diagram:
\[
  \begin{tikzcd}
    \monT\monT A \ar[d, "\monT\monT \lambda_A"]\ar[r, "\mu"]& \monT A\ar[d, "\monT \lambda_A"]
    \\
    \monT\monT\theanchorint \ar[d, "\monT a_0"] \ar[r, "\mu"] & \monT
    \theanchorint \ar[d, "a_0"]
    \\
    \monT\theanchorint \ar[d, "a_0"] \ar[r, "a_0"] & \theanchorint
    \\
    \theanchorint \ar[ur, "\id"']
  \end{tikzcd}
\]
and similarly for cobordisms.
The case of the unit~$\dot\eta$ follows the same idea.
The fact that $\monTT$ is compatible with horizontal composition and that it
satisfies the monad laws follows from the fact that the error monad~$T$ is a
cocartesian monad: as a functor, it preserves pushouts, and all the naturality
squares of $\mu$ and of $\eta$ are pushout squares.
Further, we have an inclusion from cobordisms in the ambient category
of $\monT$-algebras into the algebras of the lifted monad~$\monTT$ on
cobordisms:
\[
  \Cobord_{\EM\catS\monT}(\moo) \hookrightarrow \EM{\Cobord_\catS(\moo)}{\monTT}
\]
By monad on a double category~$\dblcat$, we mean a monad in the
vertical 2-category $\DblCatVert$ of double categories (see
\cite{garner-phd-thesis}).

In the case of the error monad on asynchronous graphs, the category of
$\monT$-algebras is complete (as the forgetful functor always creates
all limits), cocomplete (as $\monT$ preserves reflexive coequalizers)
and adhesive (as the forgetful functor creates pullbacks and
pushouts).

\subsection{Tensor product}

In order to define the parallel product in this new ambient category,
we need a tensor product in the category of $\monT$-algebras.
We achieve this by lifting the Cartesian product into the \demph{smash
  product} of pointed asynchronous graphs
$(x_1, G_1) \Ttimes (x_2, G_2)$ which identifies any pair of nodes of
the form $(x_1, x_2), (x_1, {-})$ or $({-}, x_2)$ and make this
node the distinguished node of the smash product.
In our case, the smash product is a lifting to the Eilenberg-Moore
category of the the Cartesian product on asynchronous graphs, in that
they commute with both the forgetful functor and the free algebra
functors in the expected way.  and the forgetful functor in the
expected ways.
See Lemma 2.20 in \citep{mulry}, and \citep{jonhstone-adjoint-lifting}
for more details.

\section{Separated states}\label{sec:separated-states}

We now define the last of the three internal $J$-opcategories: $\mooSep$, that
will be used to interpret the proofs of CSL.
Its structure is richer than the other two as it is a proper internal
$J$-opcategory, indexed by the set of predicates of CSL.
First, we define the notion of separated states, then we define the machine
model $\Sepmodel$, and finally we define the $J$-internal category
structure~$\mooSep$.

\subsection{Separated states}
We use the same notion of separated states as \citet{mfps,lics18}.
We suppose given an arbitrary partial cancellative commutative monoid $\Perm$
which we call the \demph{permission monoid}, with a top element~$\top$,
following \citet{Bornat:permissions}.
We require~$\top$ to admit no multiples: $\forall x\in\Perm, \top \cdot x$ is
not defined.
The set $\LStates$ of logical states is defined in much the same way
as the set of memory states, with the addition of permissions to each
variable and heap location:
\[
  (\Var \parfinmap (\Val \times \Perm)) \times
  (\Loc \parfinmap (\Val \times \Perm))
\]
Permissions enable us to define a \emph{separation product} $\state * \state'$
between two logical states $\state$ and $\state'$ which generalizes disjoint
union.
When it is defined, the logical state $\state * \state'$ is defined as a partial
function with domain
\[
\dom(\state * \state) = \dom(\state) \cup \dom(\state')
\]
in the following way: for $a\in\Var\amalg\Loc$,
\[
  \state*\state'(a) =
  \begin{cases*}
   \,\, \state(a) & if $a \in \dom(\state)\setminus\dom(\state')$\\
    \,\, \state'(a) & if $a \in \dom(\state')\setminus\dom(\state)$\\
   \,\, (v,p\cdot{}p') & if $\state(a) = (v, p)$ and $\state'(a) = (v,p')$\\
  \end{cases*}
\]
The separation product $\state * \state'$ of the two logical states $\state$ and
$\state'$ is not defined otherwise.
In particular, the memory states underlying $\state$ and $\state'$ agree on the
values of the shared variables and heap locations when the separation product is
well defined.

The predicates of CSL are predicates on logical states. Their grammar is the
following, it consists mainly in first order logic enriched with the separating
conjunction:
\begin{align*}
  P,Q,R,J \Coloneqq \; &\emp \mid \True \mid \False \mid P \vee Q \mid
                         P \wedge Q \mid \neg P \\
  \mid\; \forall v. P
         \mid \; \exists v. &P
  \mid\; P * Q \mid v \pointsto{p} w \mid \own_p(x) \mid E'_1=E'_2
\end{align*}
where $x\in\Var$, $p\in\Perm$, $v,w\in\Val$, and the $E'_i$ are
arithmetic expressions that can contain metavariables $a,b,c,\ldots$
that are used for quantifiers.
The semantics of these predicates is given by the judgment $\s \vDash P$;
it is defined by induction on the structure of~$P$, some rules are:
\begin{align*}
  (s,h) \vDash v \pointsto p w \;&\Longleftrightarrow\; v\in\Loc \wedge s =\emptyset \wedge h = [v \mapsto (w, p)] \\
  (s,h) \vDash \own_p(x) \;&\Longleftrightarrow\; \exists v\in\Val,  s = [x \mapsto (v, p)] \wedge h = \emptyset\\
  \s \vdash \emp \;&\Longleftrightarrow\; \s = (\emptyset, \emptyset) \\
  \s \vDash P*Q \;&\Longleftrightarrow\; \exists \s_1 \s_2, \, \s = \s_1 {*}
                    \s_2, \s_1 \vDash P \wedge \s_2 \vDash Q\\
  \s \vDash P \wedge Q \;&\Longleftrightarrow\; \s \vDash P \text{ and } \s \vDash Q\\
  \s \vDash E_1 {=} E_2 \;&\Longleftrightarrow\; \sem{E_1} = \sem{E_2} \wedge \fv(E_1=E_2) \subseteq \vdom(s)\\
  \s \vDash \exists a. P \;&\Longleftrightarrow\; \exists x\in\Val, \s \vDash P[v/a]
\end{align*}

We recall the notion of \emph{separated state} whose purpose is to separate the
logical memory state into one region controlled by the Code, one region
controlled by the Frame, and one independent region for each unlocked resource.
In order to define the notion, we suppose given a finite set~$\Locks \subseteq
\RVar$ of locks, and for each lock~$r$, a predicate~$I$. We write $\Gamma =
r_1: I_1, \ldots, r_n:I_n$.
\begin{definition}\label{def:separated-states}
  A \emph{separated state} is a triple
    \[
    (\state_C, \statevector, \state_F) \in \LStates \times (\Locks \to \LStates+\{C,F\}) \times \LStates
  \]
such that the logical state below is defined:
\begin{equation}\label{eqn:sep-state-product}
  \state_C \, * \,\, \Big\{\,\,\bigsprod_{r \in \dom(\statevector)}\statevector(r) \,\, \Big\}\,\, * \, \state_F
  \quad \in \quad \LStates
\end{equation}
where we write $\dom(\statevector)$ for $\statevector^{-1}(\LStates)$, and
similarly for $\domC(\statevector)$ and $\domF(\statevector)$, and such that,
for all $r_k \in \dom(\statevector)$, $\statevector(r) \vDash I_k$.
\end{definition}

The logical state~$\state_C$ describes the part of the (logical and
refined) memory which is owned by the Code; while $\state_F$ describes
the part which is owned by the Frame or the Environment; finally, the
function~$\statevector$ indicates, for each lock~$r$, who is holding
it between Code and Frame, and if nobody is holding it, which part
$\statevector(r)$ of the logical memory the lock is currently
protecting, or owning.
Note that $\statevector(r)$ is the piece of the logical memory which
the Code or the Frame will get when it acquires the lock.
For example, in the traditional example where the monoid of
permissions is defined as $\Perm = (0,1]$ with addition, the situation
where both the Code and the Environment have read access to a
location~$\location$ can be represented by the separated state
$([\location \mapsto (4,\sfrac 1 2)], \emptyset, [\location \mapsto
(4,\sfrac 1 2)])$ where the function~$\statevector$ is empty.

\subsection{The machine model of separated states $\Sepmodel$}
%
In contrast to the stateful and the stateless machine models
$\statefulmachinemodel$ and $\statelessmachinemodel$, which are
one-player games, the machine model of separated states involves two
players Code and Frame.
Similarly to the graphs of states and locks, we define a notion of
footprints, which turn out to be the same as the footprints of
associated with the machine states:
\[
  \fp \in \powerset(\Var+\Loc) \times \powerset(\Var+\Loc) \times
  \powerset(\Locks) \times \powerset(\Loc).
\]
  The \demph{machine model of separated states} $\Sepmodel[\Gamma]$ is the
  asynchronous graph whose nodes are the separated states and whose edges are
  either Code or Frame transitions:
  Code moves are the
    \[
      (\state_C, \statevector, \state_F) \yrightarrow{\,\,\CODE{m}\,\,} (\state'_C, \statevector', \state_F)
      \quad\text{ such that }\quad
      {\mbox{$\bigsprod$}(\state_C, \statevector, \state_F)} \yrightarrow{m}{\mbox{$\bigsprod$}(\state'_C, \statevector', \state_F)}
    \;\;\text{ in $\statefulmachinemodel$}
    \]
    where~$m\in\Instr$ is an instruction, and such that the following
    conditions are satisfied:
    \begin{align*}
     \forall \location\notin&\writearea{m},\;\state_C(\location) = \state'_C(\location) &
      \writearea{m} \cup \readarea{m}&\subseteq \dom(\state_C)\\
      \lock(m) &\subseteq \dom(\statevector) \cup \domC(\statevector) &
      \forall r\notin\lock(m),\; &\statevector(r) {=} \statevector'(r).
    \end{align*}
  Frame moves are of the form
  \(
      (\state_C, \statevector, \state_F) \xrightarrow{\,\,\ENV{m}\,\,}
      (\state_C, \statevector', \state'_F)
      \)
  with symmetric conditions.
  In the same way as the other machine models, the tiles of $\Sepmodel$ are
  the squares such that the footprints of the opposite instructions are
  independent.

\subsection{The internal category $\mooSep$}

Recall that the internal category $\mooSep$ is indexed by the set of
predicates of CSL.
In order to define the asynchronous graph $\moo[0,P]$ for each
predicate $P$, we need to lift the satisfaction relation of a
predicate from logical states to separated states by defining:
\[
  (\state_C, \statevector, \state_F) \vDash P \;\Longleftrightarrow\;
  \state_C \vDash P.
\]
Then, we can define $\mooSep[0,P]$ to be the subgraph of $\Sepmodel$
induced by the separated states that satisfy~$P$, keeping only the
Frame moves.
We are now ready to define the internal $J$-opcategory, the unique support is
the whole two player graph $\mooSep[1] = \Sepmodel$, and the interfaces
is the collection of all the $\mooSep[0,P]$, with the obvious inclusion morphism
into $\mooSep[1]$. The map $\mu_P : \mooSep[2,P] \to \mooSep[1]$ is given by the
universality of the pushout.

\subsection{Parallel product understood as a span-monoidal structure on $\mooSep$}\label{sec:separated-states/span-monoidal}

In order to define the parallel product at the level of proofs, which
we will use to interpret the CSL rule for the parallel product, we
endow $\mooSep$ with a span-monoidal structure.
The main piece of this structure is the asynchronous graph
$\Sepmodelthreeplayers$ of \emph{three-player separated states}, which
will be the basis for the support of the span
\[
  \begin{tikzcd}[column sep=4em]
    \mooSep \times \mooSep &
    \mooSep^{\parallel} \ar[l, "\pick"']\ar[r, "\pince"] &
    \mooSep
  \end{tikzcd}
\]
of Definition~\ref{def:span-monoidal}. It is the straightforward generalization
of separated states to the case where there are three players, $\Cplayer_1, \Cplayer_2$
and~$\Fplayer$; its nodes are the states $(\state_1, \state_2, \statevector, \state_F)$, 
such that the product is well defined.
The colors are pairs of predicates, which correspond to $\state_1$ and
$\sigma_2$.
The internal functor $\pince$ has the following action on the supports of the
internal $J$-categories: it maps a three-player state to the (two-player)
separated state $(\state_1 * \state_2, \statevector', \state_F)$, where
$\statevector'$ is the same as~$\statevector$ where $C_1$ and~$C_2$ have been
remapped to $C$. The morphism $\pick$ maps that three-player state to the pair
$\left< (\state_1, \statevector_1, \state_2 * \state_F), (\state_2,
\statevector_1, \state_1 * \state_F)\right>$.
The unit $\mooSep^{\parallel\eta}$ of the structure is the subgraph of $\mooSep[1]$
with only Frame moves.

\section{Change of locks}\label{sec:change-of-locks}

The three machine models $\moo=\mooSep, \mooS, \mooL$
considered in this paper are parameterized by the set 
of \emph{free locks} which the programs can access.
In the case of the two internal opcategories $\mooS$ and $\mooL$, 
the free locks are simply described by a set $\Locks$ of lock names.
In the case of the internal opcategories $\mooSep$ of separated states,
the free locks are described by a context $\Gamma = r_1: I_1, \ldots, r_n:I_n$
which associates with each free lock~$r_k$ the predicate $I_k$ of a CSL invariant.
As we are considering the general case, we write $\Gamma, r$ to denote
a context in either of the two cases.
The operations of \emph{introducing a new lock} and of \emph{creating a critical section}
transport cobordisms across machine models parameterized with different free locks.
The change-of-basis operations are induced by the two acute spans below:
\[
  \begin{tikzcd}[column sep=5em]
    \moo(\Gamma)
    \ar[r, bend left=19, module, "{\ATSwhen r}"]
    &
    \moo(\Gamma,r)
    \ar[l, bend left=19, module, "{\khide[r]}"]
  \end{tikzcd}
\]
where we write explicitly the dependence of the models $\moo(\Gamma)$ on the
contexts (or lists) $\Gamma$ of free locks.
We formalize the situation by defining the graph $\LockGraph$ whose vertices are
the lock contexts $\Gamma$, with edges defined as transitions \emph{adding} or
\emph{removing} one specific lock in the context:
\[
  \begin{tikzcd}[column sep=5em]
    \Gamma
    \ar[r, bend left=19, "\iota_r^\Gamma"]
    &
    \Gamma,r
    \ar[l, bend left=19, "\pi_r^\Gamma"]
  \end{tikzcd}
\]
Consider $\LockGraph^*$ the locally posetal bicategory freely
generated by this graph, with invertible $2$-cells between paths that
are equal up to the reorderings:
$\pi_r \circ \iota_{r'} \sim \iota_{r'} \circ \pi_{r}, \; \pi_r \circ
\pi_{r'} \sim \pi_{r'} \circ \pi_{r}$ and
$\iota_r \circ \iota_{r'} \sim \iota_{r'} \circ \iota_r$, for
$r \neq r'$, and leaving the $\Gamma$'s implicit.
By locally posetal, we mean that there is at most one tile (or
invertible 2-cell) between any pair of morphisms, witnessing that they
are equal up to the reorderings above.
We call a \demph{$\LockGraph$-template} a pseudo functor from this bicategory
to $\AcuteSpan(\catC)$, the bicategory of internal $J$-opcategories and acute
spans we defined in \S\ref{sec:acute-spans}.
The invertible 2-cells in the domain reflect the fact 
that the order of the operations does not matter, up to isomorphism.
The three families $\moo=\mooSep, \mooS, \mooL$ of internal $J$-opcategories 
defined so far are $\LockGraph$-templates; we explain how in the remainder of this section.
Practically, this means that, by post-composing with the lax functor
$\Cobord(\cdot)$, we are able to transport a cobordism defined on the
internal category $\moo(\Gamma)$ to one defined on the internal
category~$\moo(\Gamma, r:I)$, to interpret critical sections, and
back, for resource introduction.
We abuse the notation and write these lax double functors as follows:
\[
  \begin{tikzcd}[column sep=5em]
    \Cobord(\moo(\Gamma))
    \ar[r, bend left=19, "{\ATSwhen r}"]
    &
    \Cobord(\moo(\Gamma,r))
    \ar[l, bend left=19, "{\khide[r]}"]
  \end{tikzcd}
\]

\subsection{Hiding}

To hide a lock~$r$, we proceed in two steps for all the templates that
we consider $\moo = \mooL, \mooS, \mooSep$.
First, we prevent the Environment from touching that lock, and then we
remove this lock from the states and we transform all transitions
$P(r),V(r)$ into $\knop$s.
Formally, hiding is defined by a pull and push operation along the
acute span $\khide[r]$:
\[
\begin{tikzcd}[column sep=4em]
    {\moo(\Gamma,r)}
    & 
    {\moo^{\left< r \right>} (\Gamma, r)} \ar[r, "\khide_C"]
    \ar[l,"\injC"']
    &
    \moo(\Gamma).
  \end{tikzcd}
\]
The support $\moo^{\left< r \right>} (\Gamma, r)$ of the span is
defined to be the same as the template $\moo(\Gamma, r)$, except that
all Environment transitions $P(r),V(r)$ are deleted, and, only in the
case of the template of separated states, we remove the states of the
form $(\s_C, \statevector, \s_F)$, where the lock~$r$ is held by the
Environment.
By definition of $\moo^{\left< r \right>} (\Gamma, r)$ as a
restriction of $\moo(\Gamma, r)$, there is a canonical injection
$\injC$.
The map $\khide_C$ is defined differently depending on the kind of
template we are considering.
In the case of the templates used for the code, respectively $\mooL$
and $\mooS$, we map the states $L \subseteq \Locks$ to
$L \setminus \{r\}$, and $\mstate = (\memstate, L)$ to
$(\memstate, L \setminus \{r\})$ respectively.
In the case of the template of separated states $\mooSep$ used to
interpret proofs, it is defined as follows on separated states:
\[
  (\state_C, \statevector, \state_F) \mapsto
  \begin{dcases*}
    (\state_C * \statevector(r), \statevector \setminus r, \state_F) & if
    $\statevector(r) \in \State$
    \\
    (\state_C, \statevector \setminus r, \state_F) & if
    $\statevector(r) = C$
  \end{dcases*}
\]
In all cases, $\khide_C$ maps $P(r)$ and $V(r)$ to $\knop$, and otherwise
preserves the edges.
The intuition behind the definition of $\khide_C$ for the separated
states is that when we forget about the lock~$r$, even when nobody is
holding it, we need to do something with the resources that is
associated with the lock: we chose to give this resource to the
Code. This means that outside of the binder for~$r$, the resource
associated with~$r$ is not shared, but belongs to the Code.
Remark that the lock~$r$ does not appear in the result, which means
that the semantics of $\resource{r}{C}$ is invariant under
$\alpha$-conversion.

\subsection{Critical sections}

The case of critical sections is more delicate, as the definition
differs between $\mooL,\mooS$ on the one hand, and $\mooSep$ on the
other.
In the case of the semantics of the code, we wish to let the
Environment be wild and change the states of locks whenever it wants,
even if it happens to be held by the Code.
On the other hand, when it comes to the semantics of the proofs, we
enforce the discipline that a lock can only be unlocked by the player
that locked it.
This difference of requirements is reflected by differences in the
definition of $\ATSwhen{r}$.

\paragraph{Stateful and stateless semantics}
We consider the machine model
$\moo_{\left< r \right>}(\Locks\uplus\{r\})$ which is the restriction
of $\moo(\Locks\uplus\{r\})$ where \emph{only the Environment} is able
to use the lock~$r$. This corresponds to the fact that inside a
critical section the Code loses access to the corresponding lock until
the end of the section.
There exists a map~$\nabla$ from
$\moo_{\left< r \right>}(\Locks\uplus\{r\})$ to
$\moo(\Locks\uplus\{r\})$ given by:
\begin{align*}
  (\memstate, L) & \mapsto (\memstate, L \cap \Locks) \\[1ex]
  P(r), V(r) : F &\mapsto \knop \\
  m &\mapsto m
\end{align*}
Then, the lifting of a cobordism for the critical sections is given
by the following acute span:
\[
  \moo(\Locks) \yleftarrow{\;\nabla\;} \moo_{\left< r \right>}(\Locks\uplus\{r\})
  \yrightarrow{\;\text{incl}\;} \moo(\Locks\uplus\{r\}) 
\]
where the right leg is the obvious inclusion.
Intuitively, the operation of pulling along~$\nabla$, which defines
the $\ATSwhen r$ operation, duplicates the whole transition system,
with one version for each state of the lock~$r$. It is the Environment
which is in control of this, and the Code is oblivious to the state of
the lock~$r$.

\paragraph{Separated state semantics}

The lifting operation, that we will use to deal with critical sections, is simply
defined as the push-forward along the map
\[
  \mooSep(\Gamma) \longrightarrow \mooSep(\Gamma,r:I)
\]
which sends a separated state $(\state_C, \statevector, \state_F)$ over $\Gamma$
to the separated state over $\Gamma, r:I$ where the lock is held by the code
$(\state_C, \extmap\statevector r C, \state_F)$.

\section{A uniform interpretation of codes and proofs}\label{sec:uniform-interpretation}
We have carefully studied in previous sections how to express the main
operations of concurrent separation logic (sequential composition,
parallel product and change of lock) in the conceptual language of
template games and cobordisms inspired from \cite{popl19}.
Now that each of these basic operations has been rigorously defined,
the interpretation of the code and of the proofs of concurrent separation logic (CSL)
is uniform and essentially straightforward.

\subsection{Stateful and stateless interpretations of the code}
We begin by describing how the stateful and stateless interpretations~$\semS{C}$ 
and~$\semL{C}$ of a given code~$C$ are computed in our template game model.
Since the interpretation is uniform in $\mooS$ and~$\mooL$
and does only depend on the combinators of $C$, 
we find convenient to write~$\moo$ alternatively for $\mooS$ or~$\mooL$.
We recall below the grammar of the concurrent programming language
with shared memory, dynamic allocation and locks designed by
\citet{Brookes:a-semantics} in his seminal paper on the semantics of
CSL:
\begin{align*}
  B &::=   \True
      \mid \False
      \mid B \wedge B'
      \mid B \vee B'
      \mid E = E'
      \qquad\qquad
  E ::=   0
      \mid 1
      \mid \ldots
      \mid x
      \mid E + E'
      \mid E * E'\\
  C &::=   x \coloneqq E
      \mid x \coloneqq \deref{E}
      \mid \deref{E} \coloneqq E'
      \mid \kskip
      \mid \kdispose(E)
      \mid x \coloneqq \kmalloc(E)
      \mid C ; C'
      \mid C_1 \parallel C_2\\
      & \mid \while{B}{C}
      \mid \resource{r}{C}
     \mid \when{r}{C}
    \mid \ifte{B}{C_1}{C_2}
\end{align*}
The code $C$ is interpreted by structural induction as a (pointed)
cobordism of the form
\begin{equation}\label{equation/C-sem}
\begin{array}{ccc}
\sem{C}
&
\quad = \quad
&
\begin{tikzcd}[column sep=1.5em, row sep=1.7em]
    \semin{C} \ar[r]\ar[d, "\lambda_{\mathrm{in}}"{swap}]
    &
   \semsupport{C}\ar[d]
    &
    \semout{C} \ar[l]\ar[d, "\lambda_{\mathrm{out}}"]
    \\
    \moo(\Locks)[0] \ar[r]
    &
    \moo(\Locks)[1]
    & 
    {\moo(\Locks)}[0] \ar[l].
  \end{tikzcd}
  \end{array}
\end{equation}
%
where the set $L$ of available locks is taken as implicit parameter.
The interpretation of every non-leaf command of the language
corresponds to an operation on cobordisms already defined, 
and it is thus straightforward.
Then, except for $\kmalloc$ which is treated in full details in the
Appendix, every leaf command
$ x \coloneqq E \mid x \coloneqq \deref{E} \mid \deref{E} \coloneqq E'
\mid \kskip \mid \kdispose(E) $ of the language corresponds to a
specific machine instruction~$m$ which is interpreted as a
cobordism~$\sem{m}$ living in the double category $\Cobord(\moo(\Locks))$,
with the set~$\Locks$ of available locks as implicit parameter.
The cobordism~$\sem{m}$ is constructed in the following way: its input
and output borders $\semin{m}$ and $\semout{m}$ are defined as the
asynchronous graph $\moo(\Locks)[0]$, while its support
$\semsupport{m}$ consists of the disjoint union of $\semin{m}$ and
$T\semout{m}$ augmented with an edge $s_1\to s_2$ from
$s_1\in\semin{m}$ to $s_2\in T\semout{m}$ for every machine transition
$m:s_1\to s_2$ performed by the instruction $m$.
Note that $\semin{m}$ and $\semout{m}$ contain only Frame transitions,
and that all the ``transverse'' edges $s_1\to s_2$ 
from $\semin{m}$ to $T\semout{m}$ are Code transitions,
with the state $s_2$ potentially equal to the error state.

We now detail how to give a semantics to any code $C$ as a cobordism $\sem{C}$, by
induction on its structure. This lets us build $\semS{C}$ and $\semL{C}$ in the
same way.
\paragraph{Instructions}
We explain first how to define the cobordism that interprets a single instruction~$m$
using a well chosen pullback.
Consider the following asynchronous graph
\[
  \begin{tikzcd}
    {A=} & \bullet \ar[loop above, "F_1"] \ar[r, "C"] & \bullet \ar[loop above, "F_2"] &
  \end{tikzcd}
\]
with a tile $F_1 \cdot C \sim C \cdot F_2$.
Then, we can construct the pullback, where $\Instr$ is the set of instruction:
\[
  \begin{tikzcd}
    G(m) \ar[r, hook]\ar[d] & A \times \themachinemodel \ar[d, "f"] \\
    \Omega(\{F,m\}) \ar[r, hook] & \Omega(\{F\} \cup \Instr)
  \end{tikzcd}
\]
where the map $f$ sends edges of the form $(F_1, \cdot)$ and $((F_2,\cdot))$ to $F$ and
edges of the form $(C, m')$ to the edge $m'$ in $\Omega(\{ F \} \cup \Instr)$,
for all instructions $m'$.
It is then easy to deduce the maps from the borders using the universal
property of the pullback.

\paragraph{Leaf codes}
For leaf codes that correspond to instructions (all, except for $\kmalloc$),
their semantics is defined to be the same as that of the corresponding
instruction.
For $\kmalloc(E)$, we take the disjoint sum of all the
$\kalloc(E,\location)$:
\[
  \sem{\kmalloc(E)} \quad := \quad \bigoplus_{\ell\in\Loc}\: \sem{\kalloc(E,\location)}
\]

\paragraph{Conditionals}
Conditional branching is interpreted as
\[
  \sem{\ifte{B}{C_1}{C_2}} 
\]
defined as
\begin{equation}\label{equation/the-sum}
\sem{\ktest(B)};\sem{C_1} \graphsum \sem{\ktest(\neg B)};\sem{C_2}
\end{equation}
recomposed with
$$\Fill{\moo[0]\stackrel{id}{\longrightarrow}\moo[0]}{\moo[0]+\moo[0]\stackrel{\nabla}{\longrightarrow}\moo[0]}$$
Here, the purpose of precomposing with the filling is to identify
the two copies of the input $\moo[0]$ appearing on each sides of the disjoint sum (\ref{equation/the-sum}).

\paragraph{Sequential and parallel compositions}
We use the sequential and the parallel product of cobordisms to interpret their
syntactic counterparts:
\[
  \sem{C_1\|C_2} \hspace{.5em} = \hspace{.5em} \sem{C_1}\|\sem{C_2}
  \quad\quad
   \sem{C_1;C_2} \hspace{.5em} = \hspace{.5em} \sem{C_1} \fancyseq \sem{C_2}.
\]

\paragraph{Resource introduction and critical sections}
We use the change of locks operations.
The interpretation of $\resource{r}{C}$ is defined as
\[
  \sem{\resource{r}{C}} = \hide{r}(\sem{C})
\]

The interpretation of $\sem{\when{r}{C}}$ is defined as:
\[
 \sem{P(r)} \fancyseq \ATSwhen{r}\big(\sem{C}\big) \fancyseq\sem{V(r)}.
\]

\paragraph{Loops}
For loops, the interpretation of $C' = \while{B}{C}$ is defined as the unfolding
of
\[
  X \mapsto \sem{\ktest(B)}\fancyseq\sem{C} \oplus
  \sem{\ktest(\neg B)}
\]
see the Appendix for details.

\subsection{Interpretation of the proofs}

\paragraph{Machine instructions}
The rules that correspond to machine instructions $m\in\Instr$ (such as $\LOAD$)
are interpreted in the obvious way, always preserving the permission associated
with affected locations.

\subsection{Interactive and separated interpretations of the proofs}
We recall in Fig.~\ref{fig:inference-rules} the main inference rules
of concurrent separation logic (CSL) as it appears in its original
form, see \citet{Brookes:a-semantics}.
%
%
The inference rule $\RES$ associated with $\resource{r}{C}$ moves to the
shared context~$\Gamma$ a piece of the logical state owned and
potentially used by the Code, so that the new resource $r:J$ can be
accessed concurrently inside the code~$C$.
However, the access to that resource $r:J$ is typically mediated by
the $\kwith$ constructor, which grants temporary access under the
condition that one gives it back.
Note that the rule $\WHEN{}$ has the side condition
$P \Rightarrow \defpred(B)$ which means that if $P$ is true in some
logical state, then it implies that for every free variable $x$ of
$B$, there exists a permission $p$ such that $\own_p(x)$ holds.
\begin{figure*}[t] 
  \centering
  \tiny
\begin{mathpar}
  \prftree[r]{\SEQ}{\triple{\Gamma}{P}{C_1}{Q}}{\triple{\Gamma}{Q}{C_2}{R}}{\triple{\Gamma}{P}{C_1;C_2}{R}}
  \and
  \prftree[r]{\DISJ}{\triple{\Gamma}{P_1}{C}{Q_1}}{\triple{\Gamma}{P_2}{C}{Q_2}}{\triple{\Gamma}{P_1
      \vee P_2}{C}{Q_1 \vee Q_2}}
  \and
  \prftree[r]{\RES}{\triple{\Gamma,
      r:J}{P}{C}{Q}}{\triple{\Gamma}{P*J}{\resource{r}{C}}{Q*J}}
  \and
  \prftree[r]{\WHEN}{P\Rightarrow \defpred(B)}{\triple{\Gamma}{(P*J)\wedge
      B}{C}{Q*J}}{\triple{\Gamma, r:J}{P}{\when{r}{C}}{Q}}
  \and
  \prftree[r]{\PAR}{\triple{\Gamma}{P_1}{C_1}{Q_1}}{\triple{\Gamma}{P_2}{C_2}{Q_2}}{\triple{\Gamma}{P_1*P_2}{C_1\parallel
      C_2}{Q_1 * Q_2}}
  \and
  \prftree[r]{\FRAME}
  {\triple{\Gamma}{P}{C}{Q}}
  {\triple{\Gamma}{P*R}{C}{Q*R}}
\end{mathpar}
\normalsize
\caption{Inference rules of Concurrent Separation Logic}
\label{fig:inference-rules}
\end{figure*}
At this stage, our purpose is to interpret by structural induction
every CSL proof $\pi:\triplestd$ as a cobordism
\begin{equation}\label{equation/pi-sep}
\begin{array}{ccc}
\semSep{\pi}
&
\quad = \quad 
&
  \begin{tikzcd}
    \semSepin{\pi}
    \ar[r]\ar[d]
    &
	\semSepsupport{\pi}
\ar[d]
    &
	\semSepout{\pi} 
	\ar[l]\ar[d]
    \\
    \mooSep(\Gamma)[0,P] \ar[r]
    &
    \mooSep(\Gamma)[1]
    & 
    \mooSep(\Gamma)[0,Q] \ar[l].
  \end{tikzcd}
  \end{array}
\end{equation}
living in the double category $\Cobord(\mooSep(\Gamma))$ associated with the
template $\mooSep(\Gamma)$ of separated states, parameterized by the
context $\Gamma$.
As it stands, the interpretation is essentially straightforward, since
most of the rules of the logic correspond to an operation on
cobordisms already carefully defined.
We refer the reader to the Appendix for the comprehensive definitions.
There is apparently one exception however: the $\FRAME$ rule does not
seem, at least at first sight, to correspond to a basic operation on
cobordisms.
Given a cobordism $\semSep \pi$ which interprets a proof~$\pi$ of the
Hoare triple $\triplestd$, we need to define a new cobordism
associated with the Hoare triple $\triple{\Gamma}{P*R}{C}{Q*R}$.
The solution is not difficult to find however: we define the new
cobordism as the parallel product $\semSep\pi \parallel \mooSep[0,R]$,
where the asynchronous graph $\mooSep[0,R]$ is lifted to the identity
cobordism defined in the expected way in the double category
$\cobord{\anchorproof}$.

\begin{align*}
  \Bigg\llbracket%
  \raisebox{-0.6em}{%
    \prftree%
    {\prfsummary[\raiselabel{$\pi_1$}]{\triple{\Gamma}{P}{C_1}{Q}}}%
    {\prfsummary[\raiselabel{$\pi_2$}]{\triple{\Gamma}{Q}{C_2}{R}}}%
    {\triple{\Gamma}{P}{C_1 ; C_2}{R}}  }%
  \Bigg\rrbracket_{\Sep}%
  \quad&=\quad%
  \semSep{\pi_1}\, \fancyseq \,\semSep{\pi_2}%
\intertext{%
For the parallel product rule \PAR, we use the parallel product of ATSs using
the above notion of \emph{compatibility}:}
  \Bigg\llbracket%
  \raisebox{-0.6em}{%
    \prftree%
    {\prfsummary[\raiselabel{$\pi_1$}]{\triple{\Gamma}{P_1}{C_1}{Q_1}}}%
    {\prfsummary[\raiselabel{$\pi_2$}]{\triple{\Gamma}{P_2}{C_2}{Q_2}}}%
    {\triple{\Gamma}{P_1*P_2}{C_1 \parallel C_2}{Q_1*Q_2}}   }%
  \Bigg\rrbracket_{\Sep}%
  \quad&=\quad%
  \semSep{\pi_1} \,\parallel\, \semSep{\pi_2}%
\\
  \Bigg\llbracket%
  \raisebox{-0.6em}{%
    \prftree%
    {\prfsummary[\raiselabel{$\pi$}]{\triple{\Gamma}{P}{C}{Q}}}%
    {\triple{\Gamma}{P*R}{C}{Q*R}}    }%
  \Bigg\rrbracket_{\Sep}%
  \quad&=\quad%
  \semSep{\pi} \parallel \id_{\mooSep[0,R]}
%
\\
  \Bigg\llbracket%
  \raisebox{-0.6em}{%
    \prftree%
    {\prfsummary[\raiselabel{$\pi$}]{\triple{\Gamma, r:J}{P}{C}{Q}}}%
    {\triple{\Gamma}{P*J}{\resource{r}{C}}{Q*J}}   }%
  \Bigg\rrbracket_{\Sep}%
  \quad&=\quad%
  \hide{r}\big(\sem{\pi}\big)%
\\
%
  \Bigg\llbracket%
  \raisebox{-0.6em}{%
    \prftree%
    {\prfsummary[\raiselabel{$\pi$}]{\triple{\Gamma, r:J}{(P*J)\wedge B}{C}{Q*J}}}%
    {\triple{\Gamma, r:J}{P}{\when{r}{C}}{Q}} }%
  \Bigg\rrbracket_{\Sep}%
  \quad&=\quad%
  \big(\take{r}\big)\,;\,\ATSwhen{r}\big(\semSep{\pi}\big) \,;\, \release{r}%
\\
  \Bigg\llbracket%
  \raisebox{-0.6em}{%
    \prftree%
    {\prfsummary[\raiselabel{$\pi_1$}]{\triple{\Gamma}{P_1}{C}{Q_1}}}%
    {\prfsummary[\raiselabel{$\pi_2$}]{\triple{\Gamma}{P_2}{C}{Q_2}}}%
    {\triple{\Gamma}{P_1 \vee P_2}{C}{Q_1 \vee Q_2}}    }%
  \Bigg\rrbracket_{\Sep}%
  \quad&=\quad%
  \semSep{\pi_1} \cup \semSep{\pi_2}  %
\end{align*}
%
%

%

\section{The asynchronous soundness theorem revisited}\label{sec:soundness}
We briefly explain how our principled and axiomatic description of CSL
leads us to a radically new way to understand and also to prove the
asynchronous soundness theorem recently established by \citet{lics18}.

\subsection{Comparing the three interpretations}\label{sec:comparison}
We have just shown how to recover by purely conceptual means the three
interpretations $\semS{-}$, $\semL{-}$ and $\semSep{-}$ of the codes
and proofs of CSL designed by Melli{\`e}s and Stefanesco in their
proof of the asynchronous soundness theorem.
Given a code~$C$ and a proof~$\pi$ of $\triplestd$, their proof of
asynchronous soundness relies on the existence of a chain of
\emph{translations}
\begin{equation}\label{eqn:two-maps-of-cobordisms}
\begin{tikzcd}[column sep=1.5em]
  \semSep{\pi} \arrow[rr,"{\semmorphsep}"]
 &&
  \semS{C} \arrow[rr,"{\semmorph}"]
&&
  \semL{C}
\end{tikzcd}
\end{equation}
between cobordisms living respectively in $\mooSep(\Gamma)$,
$\mooS(\Locks)$ and~$\mooL(\Locks)$.
Here, we write $\Locks$ for the domain of~$\Gamma$.
In order to clarify the functorial nature of these translations, one
starts by observing the existence of internal functors between the
colored internal opcategories:
$$
\begin{tikzcd}[column sep=1.5em]
  \mooSep(\Gamma) \arrow[rr,"{u_\semmorphsep}"] 
  &&
  \mooS(\Locks) \arrow[rr,"{u_\semmorph}"] 
  &&
  {\mooL(\Locks)}.
\end{tikzcd}
$$
The first internal functor $u_\semmorphsep$ transports every separated
state $(\state_C, \statevector, \state_F)$ into the machine state
obtained by multiplying all its components as
in~(\ref{eqn:sep-state-product}) and by forgetting all the
permissions.  The second internal functor $u_{\semmorph}$ forgets the
memory from a machine state in order to obtain the corresponding lock
state.
The three systems of tiles which equip the synchronization templates
$\anchorproof$, $\anchorstate$ and $\anchorlogic$ were carefully
designed in order to ensure that these internal functors do indeed
exist.
Since we can compare by a simulation two cobordisms defined over the
same internal $J$-opcategory, we can also compare by
``change-of-basis'' two cobordisms over different internal
$J$-opcategories.
As a matter of fact, the translations
in~(\ref{eqn:two-maps-of-cobordisms}) are simulations:
\[
\begin{tikzcd}[column sep=1.5em]
\push{{u_\semmorphsep}}(\semSep \pi) \arrow[rr,"\semmorphsep"]
 && \semS C 
\end{tikzcd}
\quad\text{and}\quad
\begin{tikzcd}[column sep=1.5em]
\push{{u_\semmorph}}(\semS C)
\arrow[rr,"\semmorph"]
&&  \semL C.
\end{tikzcd}
\]
%
The definition of internal $J$-functor implies that we have a natural
family of isomorphisms
\[
\begin{tikzcd}[column sep=1.5em]
{\push u (C \ssep D)}
\arrow[rr,"iso"] 
&& 
{\push u (C) \ssep \push u (D)}
\end{tikzcd}
\]
in the ambient category $\Scategorysurround=\AGraph$. 
Intuitively, the reason why this is an isomorphism is that, in both cases, 
we perform the same pushout for cospan composition $C,D\mapsto C;D$
at the level of the supports of games and cobordisms.
The corresponding (non invertible) comparison map associated with the
parallel product
\[
\begin{tikzcd}[column sep=1.5em]
    \push u (C\parallel D) \arrow[rr] && {\push u (C) \parallel \push u (D)}
\end{tikzcd}
\]
is derived from the fact that the internal functor~$u$ can be equipped
with the structure of a span-monoidal functor, whose main component is
a map~$u^\parallel$ that makes the following diagram commute:
\[
  \begin{tikzcd}
    {\moo \times \moo} \arrow[d, "{u \times u}"{swap}]
    &
    \mootensor \arrow[l, "{\pick}"{swap}]
    \arrow[r, "{\pince}"]
    \arrow[d, "{u^\parallel}"] &
    \moo\arrow[d, "u"]
    \\
    \moo' \times \moo' &
    \mooptensor \arrow[l, "{\pick}"{swap}]
    \arrow[r, "{\pince}"] &
    {\moo'}
  \end{tikzcd}
\]

%

\subsection{The asynchronous soundness theorem}
Suppose given two cobordisms $\sigma$ and $\tau$ with respective supports $S$ and $T$
on one of the three machine models $\moo=\mooSep, \mooS, \mooL$,
together with an asynchronous morphism $f:S\to T$ 
satisfying the expected equation $\lambda_{\sigma}=\lambda_{\tau}\circ f$.
Following the terminology introduced by \citet{lics18}, we declare
that the morphism $f:A\to B$ is
\begin{itemize}
\item a \demph{Code 1-fibration} when every transition $n:f(s)\to s'$
  performed by the Code in $B$ can be lifted to a transition
  $m:s\to s''$ performed by the Code in $A$, such that $f(m)=n$,
\item a \demph{2-fibration} when every tile of the form
  $f(s_1\to s_2\to s_3)\sim f(s_1)\to t'_2\to f(s_3)$ in $B$ can be
  lifted to a tile $s_1\to s_2\to s_3\sim s_1\to s'_2\to s_3$
  in $A$, satisfying in particular $f(s'_2)= t'_2$.
\end{itemize}
We can now state the asynchronous soundness theorem for CSL
established by \citet{lics18} which relies on these two notions of
fibrations in order to express the \emph{safety} and \emph{data-race
  freedom} of the code in a clean topological way.
In order to prove these two properties, the theorem focuses on the
nature of the asynchronous comparison maps between cobordisms
\[
  \semSep \pi \quad \xrightarrow{\quad \semmorphsep \quad} \quad
  \semS C \quad \xrightarrow{\quad \semmorph \quad} \quad
  \semL C
\]
discussed in \S\ref{sec:comparison}.  The theorem states that for
every CSL proof $\pi$ of $\triplestd$,
\begin{theorem}[Asynchronous soundness theorem]\label{thm:the-theorem}
  The comparison map $\semmorphsep : \semSep \pi \to \semS C$ is a Code 1-fibration,
  and the comparison map $\semmorph \circ \semmorphsep : \semSep \pi \to \semL C$ is a 2-fibration.
\end{theorem}
\noindent
The first part of the theorem ensures that a program
specified in CSL does not crash.
Indeed, the cobordism $\semSep \pi$ lives above $\mooSep[1]$ which
does not include the error state~$\Error$.
Since every step performed by the Code in $\semS C$ can be lifted to
$\semSep \pi$ by the 1-fibrational property, the specified Code cannot
produce any error.
The second part of the theorem ensures that a specified program does
not produce nor encounter any data race.
Indeed, every time two instructions are executed in parallel in the
machine, they define a tile in the cobordism~$\semL C$, which can be
lifted by the 2-fibrational property to a tile in the cobordism
$\semSep \pi$ of separated states.
There, the very existence of the tile implies that these two
instructions are independent and do not produce any data race.

The conceptual and axiomatic framework of template games enables us to
reunderstand in a radically new way the original proof of the
asynchronous soundness theorem \cite{lics18} by reducing it to the
preservation properties of the 1-dimensional and 2-dimensional
fibrations with respect to pullbacks, in the hom-categories of
cobordisms.
The proof relies on the observation that the chain of translations
(\ref{eqn:two-maps-of-cobordisms}) induces a sequence of pullback
diagrams between the interpretation of the code in
(\ref{equation/C-sem}) and of the CSL proof in
(\ref{equation/pi-sep}):
\[
 \begin{tikzcd}[column sep=.8em,row sep=.5em]
    \semSepin{\pi}
    \ar[rr]\ar[dd]
    &&
    \semSepsupport{\pi}
    \ar[dd]
    &&
	\semSepout{\pi} 
	\ar[ll]\ar[dd]
    \\
    & \small{pullback} &&  \small{pullback} 
    \\
 \seminS{C} \ar[rr]\ar[dd]
    &&
   \semsupportS{C}\ar[dd]
    &&
    \monT \semoutS{C} \ar[ll]\ar[dd]
\\
&  \small{pullback}  &&  \small{pullback} 
\\
 \seminL{C} \ar[rr]
    &&
   \semsupportL{C}
    &&
    \monT \semoutL{C} \ar[ll]
  \end{tikzcd}
\]
These pullback diagrams indicate that the input and output of the
various cobordisms interpreting $C$ and $\pi$ behave properly from the
point of view of the translations.
This pullback property is a remarkable consequence of the fact that
the ambient category $\AGraph$ of asynchronous graphs is adhesive, as
a presheaf category, see \ref{sec:adhesive} for a definition of
adhesivity.
Then, building on the existence of these pullback diagrams, one
establishes using careful and diagrammatic arguments around the notion
of van Kampen square (see Lemma \ref{lem:horrible}) a key
compatibility property between the 2-dimensional fibrations (stable by
limits) and the pushout (a colimit) defining the sequential
composition.

\subsection{Structural properties of cobordisms}
\label{sec:structural-properties-cobordisms}

The proof of Theorem~\ref{thm:the-theorem} relies on the cobordisms
that interpret programs and on proofs to have strong structural properties.
Most of these properties are of a fibrational nature, ranging in a wider gamut
than what we have introduced so far. In order to handle uniformly this variety
of notions of fibrations, it is quite useful to express them as
right-lifting properties.
Given an asynchronous morphism $f:G\to H$ between two asynchronous graphs whose
edges have $C$ and $F$ polarities, we define the notions of \demph{Code
1-fibration}, \demph{Environment 1-fibration}, \demph{Environment
1-op-fibration} and of \demph{2-fibration} as right lifting properties, denoted
in the four diagrams below, respectively.
Each means that whenever there is such a square that commute, there exists a map
as the one denoted by a dashed arrow that makes the two triangle commute.
\[
  \begin{tikzcd}
    \walkingNode \ar[r]\ar[d,"s"']
    &
    G\ar[d, "f"]
    \\
    \walkingArrowC \ar[r]\ar[ur, dashed]
    &
    H
  \end{tikzcd}
  \qquad\qquad
  \begin{tikzcd}
    \walkingNode \ar[r]\ar[d,"s"']
    &
    G\ar[d, "f"]
    \\
    \walkingArrowF \ar[r]\ar[ur, dashed]
    &
    H
  \end{tikzcd}
  \qquad\qquad
  \begin{tikzcd}
    \walkingNode \ar[r]\ar[d,"t"']
    &
    G\ar[d, "f"]
    \\
    \walkingArrowF \ar[r]\ar[ur, dashed]
    &
    H
  \end{tikzcd}
  \qquad\qquad
  \begin{tikzcd}
    \walkingPathTwo \ar[r]\ar[d,"u"']
    &
    G\ar[d, "f"]
    \\
    \walkingTile \ar[r]\ar[ur, dashed]
    &
    H
  \end{tikzcd}
\]
The asynchronous graph~$\walkingNode$ denotes the graph consisting of a single
node; $\walkingArrowC$ denotes the graph made up of two nodes and a single Code
arrow between them, $\walkingArrow$ denotes the same, but with a Frame polarity;
$\walkingPathTwo$ denotes the asynchronous graph made up of a path of length~2,
and finally $\walkingTile$ is the graph made up of a single tile.
The two maps called~$s$ map the node into the source of the edge, $t$ maps into
the target, and~$u$ maps the path of length~2 into the upper path of the tile.
A map that is both a fibration and an opfibration is called a
\demph{bifibration}.  We say that a \demph{map of cobordism is a fibration} of a
certain kind when its underlying map between the supports of the cobordisms is a
fibration of that kind.
One advantage of this formulation by right-lifting is that it makes is
obvious that all notions of fibration are \emph{stable under pullback},
which will be extremely useful in the sequel.

We now state the structural properties that the cobordisms that
correspond to interpretations of proofs and of programs satisfy.
With the following notations for the cobordism,
\[
  \begin{tikzcd}[column sep=4em, row sep=1.5em]
    A \ar[r, "\source{}"] 
    \ar[d, "{\lambda_A}"']
    & 
    S \ar[d, "\lambda_{\sigma}"] 
    &
   B \ar[l, "\target{}"'] \ar[d, "{\lambda_B}"]
    \\
    \anchorint i \ar[r, "\anchorinput i j"] & 
    \anchorcode  i j & 
    \anchorint j \ar[l, "\anchoroutput i j"{swap}]
  \end{tikzcd}
\]
\begin{enumerate}
\item\label{cond:source-epi} the map $\source{} : A \to \anchorint i$ is an epi,
\item\label{cond:supp-1-fib} the map $\lambda_\sigma : A \to \anchorcode i j$ is an Environment 1-fibration,
\item\label{cond:target-mono} the map $\target{} : B \to S$ is a monomorphism,
\item\label{cond:empty-pullback} the pullback of $\source{}$ along $\target{}$ is the initial object.
\end{enumerate}
Conditions~(1) and~(2) are related to receptivity in game semantics:
the program must let the environment start from any state, and it must
accept any legal mode that the environment wishes to play.
The last two conditions are more technical and are needed to ensure
that sequential composition behaves nicely.
Moreover, we ask of each of our templates that the maps
$\anchorinput i j$ and $\anchoroutput i j$ be monos, Environment
1-fibrations and 2-fibrations

We prove that all interpretations of proofs and of programs satisfy
these conditions by induction on their structures.
The case of the parallel product is quite simple.
Conditions~(1) and~(4) are preserved because epis and monos are stable
under pullbacks (all epis are regular in $\AGraph$), and
$\pince : \mooSep^\otimes[P] \to \mooSep[P]$ is an epimorphism.
Conditions~(2) and~(3) are preserved because fibrations are preserved
by products and pullbacks.

The case of sequential composition is bit more interesting.
Condition (1) is obvious, condition (4) follows from the fact that
monos are preserved under pushouts in an adhesive
category. Preservation of condition~(2) is less obvious, and relies on
the fact that the walking node and the walking arrow $\walkingArrow$
is \demph{tiny}, (which means that the functor
$\Hom(\walkingArrow, \text{--})$ preserves small colimits) since they are
representable. It follows from the following lemma:
\begin{lemma}\label{lem:cube-1-fibration}
  Given the following diagram:
  \[
    \begin{tikzcd}
      & Y\ar[dr, "f'"] \ar[rrrr, "\lambda_2"] &&&& C\ar[dr, "i'"]\ar[ddl, leftarrow, "i"', very near end]
      \\
      && Z \ar[rrrr, dashed, crossing over, "\mu", near start]&&&& D
      \\
      W\ar[dr, "g"', hook] \ar[uur, "f", hook]\ar[rrrr] \ar[urr, "po", phantom]&&&& 
      A \ar[dr, "j"']\ar[urr, "po", phantom]
      \\
      & X \ar[uur, crossing over, "g'"', near start]\ar[rrrr, "\lambda_1"]&&&& B\ar[uur,
      "j'"']
    \end{tikzcd}
  \]
  where $lambda_1$, $\lambda_2$, $i$ and $j$ are Environment 1-fibrations, and $i,i',j,j'$ are monos.
  Then the map $\mu$ induced by the universality of the pushout is an
  Environment 1-fibration as well.
\end{lemma}
\begin{proof}
  Suppose we have a node $x$ in $Z$ which is mapped to the source node
  of an Environment transition $t$ in $D$.
  Because the walking arrow is tiny, either $B$ of $C$ contains a
  transition $t'$ which is mapped to the transition $t$. Without loss
  of generality, suppose that it is in $B$.
  Similarly, the node $x$ has an antecedent $x'$ in either $X$ or $Y$.
  Suppose first that it is in $X$. Then, because $j'$ is a
  monomorphism, it must be that this node is mapped by $\lambda_1$ to
  the source node of $t'$, and then we conclude using the fact that
  $\lambda_1$ is an Environment 1-fibration.
  Suppose now that this node $x'$ is in $Y$. Then the node
  $\lambda_2(x')$ in $C$ is mapped to the same node in $D$ as the
  source node of $t'$, therefore there is a node $y$ in $A$ which is
  mapped under $j$ to the source node of $t'$, and under $i$ to
  $\lambda_2(x')$.
  We get back to the previous case by using the fact that $j$ is an
  Environment 1-fibration.
\end{proof}

We now establish that composing with a filling system also preserves
these structural properties. This is where
condition~(\ref{cond:empty-pullback}) comes into play.
The reason condition~(\ref{cond:target-mono}), which states that the
map from the output states to the support is a mono, is preserved is
that the inputs states~$I$ and the output states~$O$ are disjoints,
which implies that the possible identifications the input states the
that happens during the pushout do not interfere with the output
states.

\begin{lemma}
  Given the following commutative diagram, where the square is a
  pushout and $o'$ is a mono:
  \[
    \begin{tikzcd}
      & M \ar[dl, "o"'] \ar[dr, "i'"] &&
      O' \ar[dl, "o'"', hook]
      \\
      C \ar[dr, "f"'] && C' \ar[dl, "g"]
      \\
      & D
    \end{tikzcd}
  \]
  if the pullback of~$i'$ and~$o'$ is the initial object, then the
  composite $o' \circ g$ is a mono.
  This clearly implies that composing with a filling system preserves
  condition~(\ref{cond:target-mono}).
\end{lemma}
\begin{proof}
  First, we establish that the~$C'$ contains the disjoint union of the
  image of~$M$ under~$i'$ and the image of~$O'$ under~$o'$. For that
  purpose, consider an epi-mono factorization~$i'_2 \circ i'_1$
  of~$i'$ and the following diagram, where the two squares are
  pullbacks:
  \[
    \begin{tikzcd}
      M \ar[d, "i'_1"'] & \emptyset \ar[l]\ar[d]
      \\
      C'_M \ar[d, hook, "i'_2"'] & X \ar[l]\ar[d]
      \\
      C' & O' \ar[l]
    \end{tikzcd}
  \]
  As epimorphisms are preserved under pullbacks in $\AGraph$, the map
  $\emptyset \to X$ is an epi, which means that~$X$ is the initial
  object~$\emptyset$.
  From that, we deduce that the map
  \[
    I' + O' \yrightarrow{} C'_M
  \]
  is a monomorphism, using the fact that colimits are universal. This
  is an instance of the general fact that the union of two subobjects
  is given by the pushout above the pullback.
  Now, consider the following diagram, where all squares are pushouts:
  \[
    \begin{tikzcd}[column sep=3em, row sep=3em]
      C \ar[d] & M \ar[l, "o"] \ar[d, "i'_1"]
      \\
      A \ar[d] & C'_M \ar[l, "h"'] \ar[d, "\iota_1"]
      \\
      A + O' \ar[d, "k"'] & C'_M + O' \ar[l, "h + \id"'] \ar[d, "{[i'_2 , o']}", hook] & O' \ar[l, "\iota_2"']
      \\
      D & C' \ar[l, "g"']
    \end{tikzcd}
  \]
  Because monos are preserved under pushouts, the composite
  $k \circ (h + \id) \circ \iota_2 = k_2 = o'$ is indeed a mono, where
  we write $k = [k_1, k_2]$.
\end{proof}

\subsection{Structural properties of the comparison maps}

As we explained above, the comparisons maps 
\[
  \semSep \pi \quad \xrightarrow{\quad \semmorphsep \quad} \quad
  \semS C \quad \xrightarrow{\quad \semmorph \quad} \quad
  \semL C
\]
of the Theorem satisfy a strong structural property which we call
\demph{strictness}.

\begin{definition}
Suppose given two cobordisms $C \in \Cobord(\moo)$ and $B \in \Cobord(\moo')$,
we say that a map of cobordisms $(f,f_\moo) : A \to B$ is \demph{strict} 
when the two squares in the following diagram are pullbacks.
\[
  \begin{tikzcd}[column sep=2em]
    I \ar[r,"in"] \ar[d,"f_I"{swap}]& \ar[d,"f_C"]C & O \ar[l,"out"{swap}]\ar[d,"f_O"]   
    \\
    I' \ar[r,"in'"{swap}] & C' & O' \ar[l,"out'"]
  \end{tikzcd}
\]
\end{definition}

Again, we prove this fact by induction on the structure of the proof
and of the programs.
The fact that parallel product preserves strictness follows from the
fact that it is defined using pullbacks, and from the usual pullback
pasting lemma; see Lemma~\ref{lem:parallel-strict} in the Appendix for
a detailed proof.

The case of the sequential composition is quite interesting, as it
follows from the fact that the ambient category is adhesive.
The diagrammatic situation is the following:
\begin{equation}\label{eqn:diag-seq-comp-van-kampen}
\begin{small}
\begin{tikzcd}[column sep=1.8em, row sep=1.3em]
  I
\arrow[dd]
\arrow[rrdd,sloped,near start]
&& 
&&
M
\arrow[dd]
\arrow[lldd,sloped,pos=.4]
\arrow[rrdd,sloped,pos=.4]
&&
&&
O
\arrow[dd]
\arrow[lldd,sloped,near start,hook']
\\
\\
I'
\arrow[rrdd]
&&
|[alias=C]|
C_1
\arrow[dd]
&&
M'
\arrow[lldd,near start,hook']
\arrow[rrdd,near start]
&&
C_2
\arrow[dd]
\arrow[lldd,crossing over]
&&
O'
\arrow[lldd,hook']
\\
\\
&&
C_1'
\arrow[rrdd,sloped]
&&
|[alias=CD]|
{C_1;C_2}
\arrow[dd,pos=.3]
&&
C_2'
\arrow[lldd,sloped]
&&
\\
\\
&&&&
C'_1;C'_2
&&&&
\arrow[from=C, to=CD, crossing over]
\end{tikzcd}
\end{small}
\end{equation}
Focusing on the commutative cube, what we need to show is that the
two front faces are pullbacks. This follows from the fact that the
bottom square is a van Kampen square, since it is a pushout along a
monomorphism in an adhesive category. Given that the top face is a
pushout as well, the definition of van Kampen square gives us directly
that the two front faces are pullbacks if and only if the two back
faces are pullbacks, which they are by hypothesis.
%

\subsection{Proof of the Theorem}

Equipped with the structural properties of the two sections above, we
are ready to prove the theorem itself.
As expected for a semantic proof of soundness of a program logic, we
proceed by induction on the structure of the proof~$\pi$ of some Hoare
triple~$\triplestd$.
The Theorem is naturally decomposed into the two following independent
statements, which we prove in turn.
\begin{theorem}[1-soundness]\label{thm:1-soundness}
  The comparison map $\semmorphsep : \semSep \pi \to \semS C$ is a Code 1-fibration.
\end{theorem}
\begin{theorem}[2-soundness]\label{thm:2-soundness}
  The comparison map $\semmorph \circ \semmorphsep : \semSep \pi \to \semL C$ is a 2-fibration.
\end{theorem}
\noindent
We detail the cases of the rules for the parallel product and the
sequential product, first of 2-soundness, and then for 1-soundness,
for the former is simpler.

\paragraph{Proof of the 2-soundness theorem}

To prove that the parallel product preserves 2-fibrations, we use the following
fact:
\begin{lemma}\label{lem:preservation-fibration-mono}
  If $g \circ f$ is a 2-fibration and $g$ is a monomorphism, then $f$ is a 2-fibration.
\end{lemma}
in the following diagram.
\[
  \begin{tikzcd}[column sep=2.5em, row sep=2em]
    \walkingPathTwo \ar[d] \ar[rrr] &&& \walkingTile \ar[d] \\
    { S_1\parallel S_2} \ar[ddd] \ar[dr, "{(\parprojl,\parprojr)}"] \ar[rrr, dashed, "f_1 \parallel f_2"]
    &&&
    { S'_1\parallel S'_2} \ar[ddd] \ar[dl, "{(\parprojl',\parprojr')}"]
    \\
    & { S_1\times S_2} \ar[r, "f_1 \times f_2"] \ar[d] & { S'_1\times S'_2} \ar[d]
    \\
    &  {\moo\times\moo} \ar[r, "u\times u"] & {\moo'\times\moo'}
    \\
    {\mootensor} \ar[d, "\kparproj"] \ar[rrr, "u^\parallel"] \ar[ur, "{(\parprojl,\parprojr)}"]
    &&&
    {\mooptensor} \ar[d, "\kparproj'"] \ar[ul, "{(\parprojl',\parprojr')}"]
    \\
    {\moo} \ar[rrr, "u"] &&& {\moo'}
  \end{tikzcd}
\]
where the map $f_1 \parallel f_2$ is induced by the universal property
of the left pullback, and where $(u, u^\parallel)$ is the
span-monoidal functor structure.
Indeed, in the case of $\moo = \mooSep$, we have the following fact:
\begin{lemma}
  The map $\pick : \mooSep^\parallel \to \mooSep \times \mooSep$ is a $2$-fibration.
\end{lemma}
which means that, by the preservation of fibrations uder pullbacks, the map
$(\lambda_1, \lambda_2): S_1 \parallel S_2 \to S_1 \times S_2$ is one as well.
Similarly, the map $(\lambda_1', \lambda_2'): S_1' \parallel S_2' \to S_1'
\times S_2'$ is a mono since $\pick$ is a mono. We then conclude using
Lemma~\ref{lem:preservation-fibration-mono}.

For the case of sequential composition, we heavily rely on the fact
that van Kampen cubes preserve 2-fibrations using the lemma below.
The preservation result then follows directly from the fact that the
cube in the center of diagram~(\ref{eqn:diag-seq-comp-van-kampen}) is
van Kampen.
\begin{lemma}\label{lem:horrible}
  Consider the following diagram in $\AGraph$:
  \[
    \begin{tikzcd}
      & Y\ar[dr, "f'"] \ar[rrrr, "\lambda_2"] &&&& C\ar[dr, "i'"]\ar[ddl, leftarrow, "i"', very near end]
      \\
      && Z \ar[rrrr, dashed, crossing over, "\mu", near start]&&&& D
      \\
      W\ar[dr, "g"', hook] \ar[uur, "f", hook]\ar[rrrr] \ar[urr, "po", phantom]&&&& 
      A \ar[dr, "j"']\ar[urr, "po", phantom]
      \\
      & X \ar[uur, crossing over, "g'"', near start]\ar[rrrr, "\lambda_1"]&&&& B\ar[uur,
      "j'"']
    \end{tikzcd}
  \]
  such that:
  \begin{enumerate}
    \item $\lambda_1$ and $\lambda_2$ are $2$-fibrations,
    \item the cube is van Kampen,
    \item the two back faces are pullbacks.
  \end{enumerate}
  Then, the asynchronous map~$\mu$ is a $2$-fibration.
\end{lemma}
\begin{proof}
  Let us assume there is a path of length~$2$ in~$Z$ which is mapped by~$\mu$ to
  the top of a tile in~$D$.
  We will use the fact that representables are tiny, in particular, tiles are
  tiny.
  Assume without loss of generality that the tile in $D$ has a preimage $T_C$ in $C$.
  Then the upper path of the tile in $D$ is the target of a path of length 2 in
  $Z$  and the target of a path of length 2 in $C$ (the upper path of $T_C$).
  Since the cube is van Kampen, $Y$ is the pullback of $i'$ along $\mu$, which
  means that there is a path of length 2 in $Y$ that is mapped to the one in
  $C$. We can conclude from there using the fact that $\lambda_2$ is a
  2-fibration.
\end{proof}

\paragraph{Proof of the 1-soundness theorem}

Preservation of Code 1-fibrations by sequential composition follows
directly from a variant of Lemma~\ref{lem:cube-1-fibration} where all
occurrences of Environment 1-fibrations are replaced with Code
1-fibrations.
The proof that the parallel product preserves Code 1-fibrations is a
bit more intricate than the case of 2-fibrations. The reason is that
the map $\pick : \mooS^\otimes[1] \to \mooS[1] \times \moo[1]$ is not
a Code 1-fibration.
Instead, we need to rely on the fact that the map $S \to \mooS[1]$
from the support of the cobordism to the template is an Environment
1-fibration and on the fact that the morphism $\pick$ above never
pairs two Code transitions.

\begin{lemma}[Parallel product preserves $1$-fibrations]
  Given two maps of cobordisms
  \[
    \alpha_i : C_i \to C'_i
  \]
  for $i=1,2$, such that $C_1$ and $C_2$, respectively $C'_1$ and
  $C'_2$, can be sequentially composed, if $\alpha_1$ and $\alpha_2$ are
  $1$-fibrations on Code transitions, then the induced morphism
  \[
    \alpha_1 \parallel \alpha_2 : C_1\parallel C_2 \to C_1' \parallel C_2'
  \]
  is a $1$-fibration on Code transitions as well.
\end{lemma}
\begin{proof}
  Let us focus on the maps between the supports. We call the supports
  $S_1,S_2,S'_1$ and $S'_2$, and we call the maps between that are contained in
  $\alpha_1$ and $\alpha_2$:
  \[
    f_1 : S_1 \longrightarrow S_1' \qquad\qquad
    f_2 : S_2 \longrightarrow S_2'
  \]
  Finally, we write $(\getstate{1}, \getstate{2})$ both for the map $\pick$ in
  the span-monoidal structure, and for the map it induces by pullback at the
  level of the supports of the cobordisms.

  Recall that the map $f_1 \parallel f_2$ is defined by the following
  diagram, where $(u, u^\parallel)$ is the span-monoidal functor structure:
  \[
  	\begin{tikzcd}[column sep=2.5em, row sep=2em]
      \cdot \ar[d] \ar[rrr] &&& \walkingArrowC \ar[d] \\
      { S_1\parallel S_2} \ar[ddd] \ar[dr, "{(\parprojl,\parprojr)}"] \ar[rrr, dashed, "f_1 \parallel f_2"]
      &&&
      { S'_1\parallel S'_2} \ar[ddd] \ar[dl, "{(\parprojl',\parprojr')}"]
      \\
      & { S_1\times S_2} \ar[r, "f_1 \times f_2"] \ar[d] & { S'_1\times S'_2} \ar[d]
      \\
      &  {\moo\times\moo} \ar[r, "u\times u"] & {\moo'\times\moo'}
      \\
      {\mootensor} \ar[d, "\kparproj"] \ar[rrr, "u^\parallel"] \ar[ur, "{(\parprojl,\parprojr)}"]
      &&&
      {\mooptensor} \ar[d, "\kparproj'"] \ar[ul, "{(\parprojl',\parprojr')}"]
      \\
      {\moo} \ar[rrr, "u"] &&& {\moo'}
  	\end{tikzcd}
  \]
  The image of the unique edge of $\walkingArrowC$ in $\mootensor$ is either
  $C_1$ or $C_2$ (recall that the polarities of $\moopoltensor$ are $C_1, C_2$
  and $F$).
  Because the situation is symmetric, we can suppose without loss of generality
  that this polarity is $C_1$.
  In that case, the edge $\walkingArrowC$ is mapped to a Code transition in
  $S'_1$. By assumption, $f_1$ is a $1$-fibration on Code transitions, so we
  can lift this arrow to $S$.
  Moreover, because $\parprojl: \mootensor \to \moo$ is a $1$-fibration on
  Code transitions as well, we can also lift uniquely $\walkingArrowC$
  to~$\mootensor$.
  Therefore, we have the following situation:
  \[
    \begin{tikzcd}
      \cdot \ar[rrr] \ar[dddd]
      &&& S_1 \times S_2 \ar[dd, "{\left< f_1, f_2 \right>}" description]
      \ar[dl, "\pi_1"'] \ar[dr, "\pi_2"]
      \\
      &&S_1 \ar[dd, "\getstate1"'] & & S_2 \ar[dd, "\getstate2"]
      \\
      & \walkingArrowC \ar[ur, dashed, "(1)", bend left=7]&& \moo[1] \times
      \moo[1] \ar[dl, "\pi_1"'] \ar[dr, "\pi_2"]
      \\
      &&\moo[1] & & \moo[1]
      \\
      \walkingArrowPol{C_1}\ar[uur, bend left=7] \ar[rrr] \ar[rrrd, bend right=11]
      &&& \mootensor[1] \ar[ul, "\parprojl"] \ar[ur, "\parprojr"']
      \\
      &&& \walkingArrowF \ar[uur, bend right] \ar[uuruu, bend right=70,
      dashed, "(2)"']
    \end{tikzcd}
  \]
  The map $(1)$ exists because $f_1$ is a $1$-fibration on Code
  transitions, as mentioned above, and the map $(2)$ exists because $\getstate2$
  is a $1$-fibration on Environment moves.
  Therefore, we can lift the arrow $\walkingArrowPol{C_1}$ to $S_1
  \times S_2$.
  We conclude by using the fact that $1$-fibrations on Code transitions are
  stable by pushout, and that $\left< \parprojl, \parprojr \right> : \mootensor
  \to \moo \times \moo$ is a $1$-fibration.
\end{proof}

\section{Related works}\label{sec:related-works}

The first proof of soundness for CSL was established by
\citet{Brookes:a-semantics} using a trace-based and stateless
semantics.
Another proof of soundness was then given a few years later by
\cite{Vafeiadis} and \citet{Views} using a more direct operational
approach.
The first proof of soundness based on a truly concurrent semantics of
CSL was designed by \citet{Hayman-Winskel} using an encoding of the
Code into Petri nets.
The approach Iris \cite{iris-journal} takes is to prove the soundness
of a modal logic, and then defining Hoare triples in this logic.
%
Our present work follows the true concurrency tradition of
interpreting the parallel product as more than a simple interleaving;
more specifically, we use a notion of homotopy to talk about
independence of instructions, in a way closely related to directed
homotopy~\citep{dihomotopy-book,HDA2}.
There is a well-established line of research on cospans by
\citet{cospans-over-reactive-systems, Bonchi} for example, who
generate LTSs using graph rewriting techniques based on cospans.
One main difference with our work is that we use cospans to manipulate
and compose our transition systems, instead of deriving them
explicitly from rewriting systems.



\section{Conclusion}\label{sec:conclusion}
%
One foundational and guiding principle of template game semantics is
that one cannot have a direct access to the internal states of a
program, because this access is necessarily mediated and regulated by
the labels of a specific template $\anchor$ of interest.
This idea inspired from dependent type theory implies that the basic
operations on programs (composition, synchronization, errors, locks,
etc.) should be defined by applying cleverly designed
\emph{change-of-basis} functors on the labeling templates.
We establish in the present paper that, somewhat unexpectedly, the
very same categorical yoga based on \emph{pull} and \emph{push}
functors, works for concurrent separation logic (CSL) and for
differential linear logic (DiLL).
This is achieved by designing a notion of \emph{cobordism}
$\sigma:A\longrightarrow\hspace{-1.2em}|\hspace{1em}B$ based on
cospans for CSL which conveniently replaces the notion of
\emph{strategy} $\sigma:A\longrightarrow\hspace{-1.2em}|\hspace{1em}B$
based on spans for DiLL.
One nice outcome is a categorical explanation for the Hoare inequality
of concurrency, which is derived here from a lax commutation property
between sequential composition (understood as a colimit) and parallel
product (understood as a limit).
We see this healthy convergence between CSL and DiLL as a strong
evidence for the relevance and surprising expressivity of template
games.
One main challenge for future work is to combine these two lines of
research on DiLL and CSL in order to obtain an asynchronous soundness
theorem for a higher-order version of CSL, based on a better
understanding of the relationship with Iris \cite{iris-journal} and
FCSL \cite{FCSL}.

\begin{acks}
  We thank Nicolas Behr and Tom Hirschowitz for the nice and instructive discussions we had with them, and the anonymous reviewers for their helpful remarks.
\end{acks}

\bibliography{citations}





\newpage
\appendix
\input{Appendix}

\end{document}


%% file: prelude.tex
\usepackage{etoolbox}

\usepackage[T1]{fontenc}
\usepackage[utf8]{inputenc}
\usepackage[american]{babel} 

\usepackage{etoolbox}

\usepackage{xfrac}
\usepackage{enumitem}
\usepackage{stmaryrd}
\usepackage{xspace,xcolor}
\usepackage{mathtools}

\usepackage{cleveref}

\usepackage{scalerel}
\usepackage{hyperref}
\usepackage{mathpartir}
\usepackage{prftree}
\usepackage{bussproofs}
\usepackage{lineno}
\usepackage{marvosym}

\usepackage{ellipsis}
\usepackage{placeins} 

\usepackage{tikz}
\usepackage{tikz-cd}

\tikzset{spanmap/.style={
            decoration={markings,
            mark= at position 0.45 with {
                  \node[transform shape] (tempnode) {$|$};
                  }
              },
              postaction={decorate}
}
}

\usepackage{pgfplots}
\usepackage{graphicx}

\usepgfplotslibrary{fillbetween}
\usetikzlibrary{intersections}
\pgfplotsset{compat=1.14}

\usetikzlibrary{arrows}
\usetikzlibrary{arrows.meta}
\usetikzlibrary{decorations.pathmorphing,shapes}

\pgfdeclarelayer{bg}
\pgfsetlayers{bg,main}
\usetikzlibrary{decorations.pathreplacing,decorations.markings}

\definecolor{pamblue}{rgb}{.78, .90, .98}

\newrobustcmd{\demph}[1]{\textbf{{#1}}}

\def\paragraph{\subsubsection*}

\newrobustcmd{\kTileHeight}{.5em}
\newrobustcmd{\kTileColSep}{1.3em}

\newrobustcmd{\up}[1]{\textsuperscript{#1}}

\newrobustcmd{\iwouldprefernotooverline}[1]{#1}

\newrobustcmd{\rname}[1]{[{\sc #1}]}

\newrobustcmd{\NN}{\mathbb{N}}
\newrobustcmd{\ZZ}{\mathbb{Z}}
\newrobustcmd{\RR}{\mathbb{R}}
\newrobustcmd{\BB}{\mathbb{B}}
\newrobustcmd{\FF}{\mathbb{F}}

\newrobustcmd{\s}{\sigma}
\newrobustcmd{\g}{\gamma}
\newrobustcmd{\D}{\Delta}
\renewrobustcmd{\state}{\sigma}
\newrobustcmd{\mstate}{\mathfrak{s}}
\newrobustcmd{\memstate}{\mu}
\newrobustcmd{\statevector}{\boldsymbol{\sigma}}
\newrobustcmd{\statevectorstar}{\boldsymbol{\sigma}^{*}}

\DeclarePairedDelimiter\sem{\llbracket}{\rrbracket}
\DeclarePairedDelimiter\abs{\lvert{}}{\rvert}

\newrobustcmd{\finalnodes}[1]{\abs{#1}}

\newrobustcmd{\restrict}{\!\!\upharpoonright}
\newrobustcmd{\alter}[3]{#1\circledless\{#2\mapsto #3\}}
\newrobustcmd{\register}[1]{\$\hspace{.05em}#1}
\newrobustcmd{\MIPSload}[2]{\texttt{LOAD}\,\,\,#1\,,\,#2}
\newrobustcmd{\MIPSstore}[2]{\texttt{STORE}\,\,\,#1\,,\,#2}
\newrobustcmd{\MIPSadd}[3]{\texttt{ADD}\,\,\,#1\,,\,#2\,,\,#3}
\newrobustcmd{\readarea}[1]{\textrm{rd}(#1)}
\newrobustcmd{\writearea}[1]{\textrm{wr}(#1)}
\newrobustcmd{\allocarea}[1]{\textrm{mem}(#1)}
\newrobustcmd{\readwritearea}[1]{\textsf{ReadWrite}\,(#1)}
\newrobustcmd{\step}[2]{#1[#2]}
\newrobustcmd{\edgesource}[0]{\partial^{-}}
\newrobustcmd{\edgetarget}[0]{\partial^{+}}
\newrobustcmd{\sourcefn}{\partial_0}
\newrobustcmd{\targetfn}{\partial_1}
\newrobustcmd{\sourcecode}[1]{\overline{\partial}_0(#1)}
\newrobustcmd{\targetcode}[1]{\overline{\partial}_1(#1)}
\newrobustcmd{\length}[1]{\mathsf{len}(#1)}
\newrobustcmd{\lengthl}[1]{\mathsf{len}_1(#1)}
\newrobustcmd{\lengthr}[1]{\mathsf{len}_2(#1)}

\newrobustcmd{\aborthandler}{\mathsf{Handler}}
\newrobustcmd{\semhandler}{\mathsf{SemHandler}}
\newrobustcmd{\errorcode}[1]{\mathsf{Error}[#1]}

\newrobustcmd{\State}{\mathbf{State}}
\newrobustcmd{\MStates}{\mathbf{MState}}
\newrobustcmd{\SStates}{\mathbf{SState}}
\newrobustcmd{\SepStates}{\SState}
\newrobustcmd{\LogStates}{{\mathbf{LState}}}
\newrobustcmd{\LStates}{\LogStates}
\newrobustcmd{\Code}{\mathbf{Code}}
\newrobustcmd{\Instr}{\mathbf{Instr}}
\newrobustcmd{\RVar}{\mathbf{Locks}}
\newrobustcmd{\Pred}{\mathbf{Pred}}
\newrobustcmd{\Var}{\mathbf{Var}}
\newrobustcmd{\Reg}{\mathbf{Reg}}
\newrobustcmd{\Loc}{\mathbf{Loc}}
\newrobustcmd{\Val}{\mathbf{Val}}
\newrobustcmd{\Rd}{\mathbf{Rd}}
\newrobustcmd{\rdvar}{\rho\lambda}
\newrobustcmd{\rdvarone}{\rho_1\lambda_1}
\newrobustcmd{\rdvartwo}{\rho_2\lambda_2}
\newrobustcmd{\Traces}{\mathbf{Traces}}
\newrobustcmd{\ATraces}{\mathbf{ATraces}}
\newrobustcmd{\Trans}{\mathbf{Trans}}
\newrobustcmd{\pullbackII}[2]{\mathsf{pull}[#1]\left( #2 \right)}
\newrobustcmd{\pushforwardII}[2]{\mathsf{push}[#1]\left[ #2 \right]}

\newrobustcmd{\returntraces}[1]{\big|\hspace{.08em}#1\hspace{.08em}\big|}
\newrobustcmd{\sequential}{;}

\newrobustcmd{\operationalsem}[1]{[#1]_{\text{oper}}}
\newrobustcmd{\OperationalTraces}{\mathbf{OperationalTraces}}

\newrobustcmd{\Nat}{\mathbb{N}}
\newrobustcmd{\Perm}{\mathbf{Perm}}
\newrobustcmd{\valid}{\mathcal{V}}
\newrobustcmd{\bool}{\mathbf{bool}}

\newrobustcmd{\Shuffles}[2]{\mathbf{Shuffles}(#1,#2)}

\newrobustcmd{\strat}[1]{\mathtt{strat}({#1})}
\newrobustcmd{\totstrat}[1]{\mathtt{Strat}(#1)}
\newrobustcmd{\idstrat}{\mathtt{idstrat}}
\newrobustcmd{\idgame}{G_{\mathrm{id}}}

\DeclareMathOperator{\dom}{dom}
\DeclareMathOperator{\domC}{dom_C}
\DeclareMathOperator{\domF}{dom_F}

\DeclareMathOperator{\vdom}{vdom}

\DeclareMathOperator{\lock}{lock}
\DeclareMathOperator{\True}{\mathbf{true}}
\DeclareMathOperator{\False}{\mathbf{false}}

\newrobustcmd{\reads}{\mathbf{reads}}
\newrobustcmd{\writes}{\mathbf{writes}}

\newrobustcmd{\domain}{\mathbf{d}} 
\newrobustcmd{\domainC}{\mathsf{d}_C}
\newrobustcmd{\domainF}{\mathsf{d}_F}
\newrobustcmd{\domainprime}{\domain'}
\newrobustcmd{\domainCprime}{\domain'_C}
\newrobustcmd{\domainFprime}{\domain'_F}
\newrobustcmd{\minstr}{\mathbf{m}} 

\newrobustcmd{\abstrloc}{\alpha}
\newrobustcmd{\action}[1]{\underline{#1}}

\newrobustcmd{\kw}[1]{\mathbb{\mathtt{#1}}}
\newrobustcmd{\kwhile}{\kw{while}}
\newrobustcmd{\kif}{\kw{if}}
\newrobustcmd{\kthen}{\kw{then}}
\newrobustcmd{\kelse}{\kw{else}}
\newrobustcmd{\kalloc}{\kw{alloc}}
\newrobustcmd{\kmalloc}{\kw{malloc}}
\newrobustcmd{\kdispose}{\kw{dispose}}
\newrobustcmd{\kwith}{\kw{with}}
\newrobustcmd{\kwhen}{\kw{when}}
\newrobustcmd{\kresource}{\kw{resource}}
\newrobustcmd{\knop}{\kw{nop}}
\newrobustcmd{\ktest}{\kw{test}}
\newrobustcmd{\kskip}{\kw{skip}}
\newrobustcmd{\kdo}{\kw{do}}
\newrobustcmd{\ifte}[3]{\kif\: #1\: \kthen\: #2\: \kelse\: #3}
\newrobustcmd{\while}[2]{\kwhile\: #1 \: \kdo \: #2}
\newrobustcmd{\resource}[2]{\kresource\: #1 \: \kdo \: #2}
\newrobustcmd{\when}[2]{\kwith\: #1 \: \kdo \: #2}
\newrobustcmd{\whenII}[2]{\kwith\: #1 \: \kdo \: #2}
\newrobustcmd{\alloc}[2]{#1 \coloneqq \kalloc(#2)}
\newrobustcmd{\deref}[1]{[#1]}

\newrobustcmd{\kwithin}{\kw{within}}
\newrobustcmd{\within}[2]{\kwithin\: #1 \: \kdo \: #2}

\newrobustcmd{\powerset}{\raisebox{.15\baselineskip}{\Large\ensuremath{\wp}}}

\newrobustcmd{\blocked}{\mathtt{blocked}}
\newrobustcmd{\abort}{\mathtt{abort}}
\newrobustcmd{\booleanabort}{\mathbf{abort}}

\DeclareMathOperator*{\bigsprod}{\scalerel*{\circledast}{\sum}}

\newrobustcmd{\GraphofStates}{\mathbf{G(MState)}}
\newrobustcmd{\GraphofSeparatedStates}{\mathbf{G(SState)}}
\newrobustcmd{\SeparationGame}{\mathbf{SGame}}

\newrobustcmd{\pargraphname}{graph of parallel compostition}

\newrobustcmd{\AsynchGraphofStates}{\mathbf{G^{\sim}(State)}}
\newrobustcmd{\AsynchGraphofSeparatedStates}{\mathbf{G^{\sim}(SState)}}
\newrobustcmd{\AsynchGraphofMachineInstructions}{\mathbf{G^{\sim}(Instr)}}
\newrobustcmd{\AsynchGraphofMachineInstructionsfilled}{\mathbf{G^{\approx}(Instr)}}

\newrobustcmd{\PAR}{\text{\sc Par}}
\newrobustcmd{\IF}{\text{\sc If}}
\newrobustcmd{\WHILE}{\text{\sc While}}
\newrobustcmd{\CONJ}{\text{\sc Conj}}
\newrobustcmd{\DISJ}{\text{\sc Disj}}
\newrobustcmd{\WHEN}{\text{\sc With}}
\newrobustcmd{\RES}{\text{\sc Res}}
\newrobustcmd{\AFF}{\text{\sc Aff}}
\newrobustcmd{\SEQ}{\text{\sc Seq}}
\newrobustcmd{\FRAME}{\text{\sc Frame}}
\newrobustcmd{\CONSEQ}{\text{\sc Conseq}}
\newrobustcmd{\STORE}{\text{\sc Store}}
\newrobustcmd{\LOAD}{\text{\sc Load}}
\newrobustcmd{\ALLOC}{\text{\sc Alloc}}
\newrobustcmd{\DISP}{\text{\sc Disp}}

\newrobustcmd{\ds}[0]{\preccurlyeq_S}
\newrobustcmd{\fv}{\mathrm{fv}}

\newrobustcmd{\defpred}{\mathrm{def}}

\DeclareMathOperator{\own}{own}

\makeatletter

\newbox\prf@@fancysummarybox
\newdimen\prf@@fancysymmarylen
\def\prffancysummarybox{%
  \sbox{\prf@@fancysummarybox}{$\vdots$}%
  \prf@@fancysymmarylen\ht\prf@@fancysummarybox%
  \advance\prf@@fancysymmarylen\dp\prf@@fancysummarybox%
  \sbox{\prf@@fancysummarybox}{%
    \raisebox{.25\prf@@fancysymmarylen}[.8\prf@@fancysymmarylen]%
    [0pt]{\usebox{\prf@@fancysummarybox}}}%
  \prf@@fancysymmarylen\wd\prf@summary@label%
  \ifdim\prf@@fancysymmarylen>\z@\relax%
    \prf@@fancysymmarylen\wd\prf@@fancysummarybox%
    \wd\prf@summary@label.4em%
    \hbox to\prf@@fancysymmarylen{%
      \usebox\prf@@fancysummarybox}\kern-.4em%
      \box\prf@summary@label%
  \else\usebox\prf@@fancysummarybox\fi}

\newrobustcmd{\oset}[3][0ex]{%
  \mathrel{\mathop{#3}\limits^{
    \vbox to#1{\kern-2\ex@
    \hbox{$\scriptstyle#2$}\vss}}}}
\makeatother

\newrobustcmd{\raiselabel}[1]{\raisebox{0.4em}{#1}}

\newrobustcmd{\pointsto}[1]{\oset{\text{\tiny $#1$}}{\mapsto}}

\newrobustcmd{\emp}{\mathbf{emp}}

\newrobustcmd\eqdef{\stackrel{\mathclap{\normalfont\mbox{\tiny{def}}}}{=}}

\renewrobustcmd{\O}{\mathcal{O}}

\newrobustcmd{\triple}[4]{#1 \!\vdash\!\{#2\} #3 \{#4\}}
\newrobustcmd{\triplestd}{\triple{\Gamma}{P}{C}{Q}}

\newrobustcmd{\id}{\textit{id}}
\newrobustcmd{\Hom}{\mathbf{Hom}}

\newrobustcmd{\parfinmap}{\rightharpoonup_{\mathit{fin}}}
\newrobustcmd{\parto}{\rightharpoonup}
\newrobustcmd{\extmap}[3]{#1 \uplus{} [{#2} \mapsto{} {#3}]}
\newrobustcmd{\updmap}[3]{#1[#2 \mapsto #3]}
\newrobustcmd{\remmap}[2]{#1 \setminus #2}

\newrobustcmd\evalexpr[3]{#1(#2) = #3}

\newcommand{\kArrowLength}{1cm}

 \newrobustcmd{\redto}[3]{%
     \begin{tikzcd}[ampersand replacement=\&, column sep=\kArrowLength]%
       {#1} \arrow[r, rightsquigarrow, "#2"] \& {#3}%
     \end{tikzcd}%
   }
\newrobustcmd{\yrightarrow}[1]{\,\xrightarrow{\;\;\;#1\;\;\;}\,}
\newrobustcmd{\yleftarrow}[1]{\,\xleftarrow{\;\;\;#1\;\;\;}\,}

\newrobustcmd{\khide}{\mathsf{hide}}
\newrobustcmd{\hide}[1]{\khide[#1]}
\newrobustcmd{\push}[1]{\mathsf{push}[#1]}
\newrobustcmd{\pull}[1]{\mathsf{pull}[#1]}
\newrobustcmd{\inside}[1]{\mathsf{inside}[#1]}
\newrobustcmd{\whentrue}[1]{\mathsf{whentrue}[#1]}
\newrobustcmd{\whenfalse}[1]{\mathsf{whenfalse}[#1]}
\newrobustcmd{\whenabort}[1]{\mathsf{whenabort}[#1]}
\newrobustcmd{\whensimple}[1]{\mathsf{when}[#1]}

\newrobustcmd{\kparproj}{\mathbf{proj}}
\newrobustcmd{\parproj}[1]{\kparproj(#1)}
\newrobustcmd{\parprojl}{\lambda_1}
\newrobustcmd{\parprojr}{\lambda_2}
\newrobustcmd{\splice}[1] {\mathbf{splice}(#1)}
\newrobustcmd{\splicel}[1] {\mathbf{splice}_1(#1)}
\newrobustcmd{\splicer}[1] {\mathbf{splice}_2(#1)}

\newrobustcmd{\verticalproof}[1]{\begin{array}{c}#1\vspace{-.5em}\\ \vdots\end{array}}

\newrobustcmd{\Board}[1]{\mathsf{Board}_{#1}}
\newrobustcmd{\Polarity}[1]{\mathsf{Pol}_{#1}}
\newrobustcmd{\Plays}[1]{\mathsf{Plays}_{#1}}
\newrobustcmd{\Initial}[1]{\mathsf{Init}_{#1}}

\newrobustcmd{\play}{\mathsf{p}}

\newrobustcmd{\transitionpath}{\twoheadrightarrow}

\newrobustcmd{\Wset}[2]{W_{#1}(#2)}
\newrobustcmd{\Rset}[2]{R_{#1}(#2)}
\newrobustcmd{\Tset}[2]{RW_{#1}(#2)}

\newrobustcmd{\histA}{\alpha}
\newrobustcmd{\histB}{\beta}
\newrobustcmd{\semS}[1]{\sem{#1}_S}
\newrobustcmd{\semSclosed}[2]{\sem{#1}_{S}^{#2}}
\newrobustcmd{\semL}[1]{\sem{#1}_L}
\newrobustcmd{\semSep}[1]{\sem{#1}_{\mathit{Sep}}}
\newrobustcmd{\Sep}{Sep}
\newrobustcmd{\Hist}{\mathbf{Hist}}

\newrobustcmd{\getstate}[1]{\lambda_{#1}}
\newrobustcmd{\parprod}{\parallel}
\newrobustcmd{\high}[1]{#1}
\newrobustcmd{\low}[1]{{#1}_{\text{\tiny FP}}}
\newrobustcmd{\mm}{\gamma}
\newrobustcmd{\fp}{\rho}
\newrobustcmd{\fpb}{{\rho'}}
\newrobustcmd{\footprint}{\mathrm{footprint}}
\newrobustcmd{\interstate}[2]{#1 \&\: #2}

\newrobustcmd{\Genvfull}{\mathrm{G}_{env}}

\newrobustcmd{\genstrat}{\mathrm{strat}}
\newrobustcmd{\Error}{\lightning}
\newrobustcmd{\pitilde}{\tilde{\pi}}
\newrobustcmd{\envfn}{\mathbf{env}}

\newrobustcmd{\framing}[1]{\mathrm{frame}[#1]}
\newrobustcmd{\ATSwhen}[1]{\mathrm{when}[#1]}

\newrobustcmd{\take}[1]{\mathrm{acquire}[#1]}
\newrobustcmd{\release}[1]{\mathrm{release}[#1]}

\newrobustcmd{\mooSep}{\genFPgraph_{\mathrm{Sep}}}
\newrobustcmd{\mooS}{\genFPgraph_S}
\newrobustcmd{\mooL}{\genFPgraph_L}
\newrobustcmd{\mooT}{\moo_{\monT}}

\newrobustcmd{\semmorphsep}{\mathcal{S}}      
\newrobustcmd{\semmorph}{\mathcal{L}}     
\newrobustcmd{\semmorphfull}{\semmorph \circ \semmorphsep}

\newrobustcmd{\Locks}{\mathbb{L}}

\newrobustcmd{\ATS}{\mathfrak{A}}
\newrobustcmd{\IOT}{\mathfrak{T}}

\newrobustcmd{\ATSmorph}{\mathcal{F}}

\newrobustcmd{\Image}{\mathrm{Im}}

\newrobustcmd{\genFPgraph}{\text{\Large $\moo$}}

\newrobustcmd{\valuev}{v}
\newrobustcmd{\location}{\ell}

\newrobustcmd{\CODE}[1]{{#1}\, : \,C}
\newrobustcmd{\ENV}[1]{#1 \, : \,F}
\newrobustcmd{\ENVbig}[1]{#1:F}
\newrobustcmd{\CODEbig}[1]{#1:C}

\newrobustcmd{\area}{\mathrm{area}}
\newrobustcmd{\offset}{\mathrm{offset}}
\newrobustcmd{\stack}{\mathfrak{p}}
\newrobustcmd{\stackpointer}{\mathbf{sp}}
\newrobustcmd{\stackarea}{\mathbf{area}}

\def\bendamnt{13}

  \tikzset{
    vertex/.style={text centered, fill, color=black, circle, inner sep=.7pt},
    state/.style={},
    every label/.style={label distance=-2pt},
    every label/.append style={font=\footnotesize},
    labelle/.style={%
      postaction={ decorate,
        decoration={ markings, mark=at position .5 with \node #1;}}}}

\newrobustcmd{\squarediag}[8]{
\begin{tikzpicture}[scale=0.85,]

  \node[vertex, draw] (a) at (-1,0) [label=left:$#1$] {};
  \node[vertex, draw] (d) at (1,0) [label=right:$#2$] {};
  \node[vertex, draw] (b) at (0,1) [label=above:$#3$] {};
  \node[vertex, draw] (c) at (0,-1) [label=below:$#4$] {};

  \begin{pgfonlayer}{bg}
    \fill[pamblue] (a.center) to [bend left=\bendamnt] (b.center)
                   to [bend left=\bendamnt] (d.center)
                   to [bend left=\bendamnt] (c.center)
                   to [bend left=\bendamnt] (a.center);
  \end{pgfonlayer}
  \begin{scope}[line width=.45pt,decoration={
       markings,
       mark=at position 0.56 with {\arrow{Stealth}}}
     ] 
     \draw[postaction={decorate}, labelle={[above left]{\footnotesize $#5$}}] (a.center) to [bend left=\bendamnt] (b.center);
     \draw[postaction={decorate}, labelle={[above right]{\footnotesize $#6$}}] (b.center) to [bend left=\bendamnt] (d.center);
     \draw[postaction={decorate}, labelle={[below left]{\footnotesize $#7$}}] (a.center) to [bend right=\bendamnt] (c.center);
     \draw[postaction={decorate}, labelle={[below right]{\footnotesize $#8$}}] (c.center) to [bend right=\bendamnt] (d.center);
   \end{scope}
\end{tikzpicture}
}

\iftoggle{enable comments}{
  \newrobustcmd{\ls}[1]{\todo[inline, caption={2do},color=lime]{\hspace{.9em}\begin{minipage}{\textwidth}LS: #1\end{minipage}}}
  \newrobustcmd{\pam}[1]{\todo[inline]{LS: #1}}
}{
  \newrobustcmd{\ls}[1]{}
  \newrobustcmd{\pam}[1]{}
}

\tikzset{module/.style={
    decoration={markings,
      mark= at position 0.45 with {
        \node[] (tempnode) {};
        \draw[-,shorten <=.1em,shorten >=.1em,inner
        sep=1pt] (tempnode.north) -- (tempnode.south);
      }
    },
    postaction={decorate}
  }
}

\tikzset{vmodule/.style={
    decoration={markings,
      mark= at position 0.45 with {
        \node[] (tempnode) {};
        \draw[-,shorten <=.1em,shorten >=.1em,inner
        sep=1pt] (tempnode.east) -- (tempnode.west);
      }
    },
    postaction={decorate}
  }
}

\tikzcdset{two fibration/.style={
    two heads
  }}

\newrobustcmd{\op}{\text{op}}

\newrobustcmd{\AGraph}{\mathbf{AsynGph}}
\newrobustcmd{\PAGraph}{\mathbf{AsynGph}_\bullet}
\renewrobustcmd{\Game}{\mathbf{Game}}
\newrobustcmd{\GameMoo}[1]{\mathbf{Game}(#1)}
\newrobustcmd{\AnchoredGraph}{\mathbf{AnchoredGraph}}
\newrobustcmd{\PolAnchoredGraph}{\mathbf{PolAnchoredGraph}}
\newrobustcmd{\PolAGraph}{\mathbf{PolAGraph}}
\newrobustcmd{\PolAGraphFib}{\mathbf{PolAGraph}^{\textrm{$1$F,$2$-fib}}}
\newrobustcmd{\PreAGraph}{\mathbf{PreAGraph}}
\newrobustcmd{\TemplateTAlg}{\mathbf{TemplateTAlg}}
\newrobustcmd{\ATST}{\mathbf{ATS}(\monT)}

\newrobustcmd{\LocksCat}{\mathbf{Locks}}
\newrobustcmd{\Contexts}{\mathbf{Ctxs}}
\newrobustcmd{\CtxPol}{\mathbf{CtxPol}}
\newcommand{\ssep}{\,;\,}
\newrobustcmd\Ttimes{\boxtimes}

\newrobustcmd{\mootensor}{\moo^{\!\!\parallel}}
\newrobustcmd{\mooptensor}{\moo\!'^{\parallel}}
\newrobustcmd{\moontensor}[1]{\moo^{\!\!\parallel}[#1]}
\newrobustcmd{\moon}[1]{\moo[#1]}
\newrobustcmd{\mooplus}{\moo^{\!\!\oplus}}
\newrobustcmd{\moominus}{\moo^{\!\!\ominus}}
\newrobustcmd{\moodot}{\moo_{\!\!\odot}}
\newrobustcmd{\mootensorplus}{\mootensor_{\!\!\oplus}}
\newrobustcmd{\mootensorminus}{\mootensor_{\!\!\ominus}}
\newrobustcmd{\safemoo}{\moo^{\!\!\circ}}
\newrobustcmd{\moopol}{\moo^{\!\mathrm{pol}}_2}
\newrobustcmd{\moopolany}{\moo^{\!\mathrm{pol}}}
\newrobustcmd{\moopoltensor}{\moo^{\!\mathrm{pol}}_{\parallel}}
\newrobustcmd{\pol}{\mathbf{pol}}

\newrobustcmd{\alone}{\mathrm{alone}}

\newrobustcmd{\graphsum}{\oplus}

\newrobustcmd{\inpt}{\iota}
\newrobustcmd{\outpt}{o}

\renewrobustcmd{\mm}{machine model}
\newrobustcmd{\Mm}{Machine model}

\newrobustcmd{\fontforanchoredgraphs}[1]{\mathbb{#1}}

\newrobustcmd{\AG}{G}  
\newrobustcmd{\AAG}{\fontforanchoredgraphs{G}} 
\newrobustcmd{\PAAG}{\mathfrak{G}} 

\newrobustcmd{\HZ}{\mathbb{C}}

\newrobustcmd{\Ainput}{\fontforanchoredgraphs{I}}
\newrobustcmd{\Aoutput}{\fontforanchoredgraphs{O}}

\newrobustcmd{\fibabove}{\sqsubset}

\newrobustcmd{\parallelProduct}{\cdot\!\parallel\!\cdot}

\newrobustcmd\lambdamachin[1]{\lambda_{\mathrm{#1}}}
\newrobustcmd\lambdain{\lambdamachin{in}}
\newrobustcmd\lambdaout{\lambdamachin{out}}

\newrobustcmd\Cmachin[1]{C_{\mathrm{#1}}}
\newrobustcmd\Cin{\Cmachin{in}}
\newrobustcmd\Cout{\Cmachin{out}}

\newrobustcmd{\walkingArrow}{\cdot\!\to\!\cdot}
\newrobustcmd{\walkingArrowPol}[1]{\cdot\!\to_{#1}\!\!\cdot}
\newrobustcmd{\walkingArrowC}{\walkingArrowPol{C}}
\newrobustcmd{\walkingArrowF}{\walkingArrowPol{F}}
\newrobustcmd{\walkingNode}{\text{\boldmath$\cdot$}}

\newrobustcmd{\walkingPathTwo}{\wedge}
\newrobustcmd{\walkingTile}{\blacklozenge}

\newrobustcmd{\isomorphic}{\cong}

\newrobustcmd{\suchthat}[2]{\left\{#2\right\}_{#1}}

\newrobustcmd{\monT}{\mathbf{T}}
\newrobustcmd{\Cat}{\mathbf{Cat}}
\newrobustcmd{\Set}{\mathbf{Set}}
\newrobustcmd\str{\mathsf{t}}
\newrobustcmd{\monTT}{\dot{\monT}}
\newrobustcmd{\EMCat}[2]{{#1}^{#2}}
\newrobustcmd\Kleisli[2]{{#2}_{#1}}

\newrobustcmd{\catC}{\mathbb{S}}
\newrobustcmd{\catS}{\mathbb{S}}
\newrobustcmd{\dblcat}{\mathcal{D}}
\newrobustcmd{\Cospan}{\mathbf{Cospan}}
\newrobustcmd{\CospanH}{\mathbf{Cospan^h}}
\newrobustcmd{\Cobord}{\mathbf{Cob}}
\newrobustcmd{\KlCobord}[2]{\mathbf{Cob(#1)_{#2}}}
\newrobustcmd{\DblCat}{\mathbf{DblCat}}
\newrobustcmd{\DblCatVert}{\mathbf{DblCat}_v}
\newrobustcmd{\DblCatLax}{\mathbf{DblCat}_{\mathrm{lax}}}
\newrobustcmd{\AcuteSpan}{\mathbf{AcuteSpan}}
\newrobustcmd{\IOTemplate}{\mathbf{IOTemplate}}
\newrobustcmd{\ATSCat}{\mathbf{ATS}}
\newrobustcmd\InternalCat[1]{\mathbf{opCat(#1)}}
\newrobustcmd\catOne{\mathbb{1}}
\newrobustcmd\ijcat{\moo}
\newrobustcmd\anchorint[1]{\moo[0,#1]}
\newrobustcmd\anchorcode[2]{\moo[1,#1 #2]}
\newrobustcmd\theanchorcode{\moo[1]}
\newrobustcmd\theanchorint{\moo[0]}
\newrobustcmd\themachinemodel{\underline\moo^{\!1}}
\newrobustcmd\thetwoplayermachinemodel{\underline\moo^{\!2}}
\newrobustcmd\anchoridentitymap[1]{\eta_{#1}}
\newrobustcmd\anchormap[2]{\moo[1,#1 #2]}
\newrobustcmd\anchorpushout[3]{\moo[2,#1 #2 #3]}
\newrobustcmd\anchormu[3]{\mu_{#1#2#3}}
\newrobustcmd\anchorinput[2]{\mathrm{in}_{#1#2}}
\newrobustcmd\anchoroutput[2]{\mathrm{out}_{#1#2}}
\newrobustcmd\theanchorinput[1]{\mathrm{in}_{#1}}
\newrobustcmd\theanchoroutput[1]{\mathrm{out}_{#1}}
\newrobustcmd\Id{\mathbb{1}}
\newrobustcmd\Seed{\mathrm{seed}}

\newrobustcmd\cobordout{\mathit{out}}
\newrobustcmd\cobordin{\mathit{in}}

\newrobustcmd\cobordmult[2]{\mathrm{m}_{#1,#2}}
\newrobustcmd\cobordunit{\mathrm{m}_{\mathbf{1}}}

\newrobustcmd\pince{\text{pince}}
\newrobustcmd\pick{\text{pick}}

\newrobustcmd\anchorpint[1]{\moo{\!'}[0,#1]}
\newrobustcmd\anchorpcode[2]{\moo{\!'}[1,#1 #2]}
\newrobustcmd\theanchorpcode{\moo{\!'}[1]}
\newrobustcmd\anchorppushout[3]{\moo{\!'}[2,#1 #2 #3]}

\newrobustcmd\codeinint[1]{\lambda_{I}}
\newrobustcmd\codeoutint[1]{\lambda_{O}}
\newrobustcmd\codeinput[1]{\mathrm{in}_{#1}}
\newrobustcmd\codeoutput[1]{\mathrm{out}_{#1}}
\newrobustcmd\codecode[1]{\lambda_{#1}}
\newrobustcmd\acobord{\mathbf{C}}

\newrobustcmd\LockGraph{\mathbf{LockGraph}}

\DeclareSymbolFont{bbsymbol}{U}{bbold}{m}{n}
\DeclareMathSymbol{\bbsemicolon}{\mathbin}{bbsymbol}{"3B}

\newrobustcmd\Fill[2]{\mathsf{fill}(#1,#2)}
\newrobustcmd\fancyseq{\bbsemicolon}
 
\newcommand{\pushoutdown}{\mbox{\Huge $\rotatebox[origin=c]{-45}{$\ulcorner$}$}}

\newrobustcmd\modto{%
  \!\!%
  \begin{tikzcd}[column sep=1.4em, ampersand replacement=\&]
    {} \ar[r, module] \& {}
  \end{tikzcd}
  \!\!%
}

\newrobustcmd{\pammodto}{\modto}

\newrobustcmd\xmodto[1]{%
  \!\!%
  \begin{tikzcd}[ampersand replacement=\&]
    {} \ar[r, module, "#1"] \& {}
  \end{tikzcd}
  \!\!%
}

\newrobustcmd{\iotemplate}{\mathbf{io{-}template}}
\newrobustcmd{\template}{\mathbf{template}}

\newrobustcmd{\injprime}{\mathrm{inj}'}
\newrobustcmd{\injC}{\mathrm{inj}_C}
\newrobustcmd{\injF}{\mathrm{inj}_F}

\newrobustcmd{\anchor}{\moo}

\newrobustcmd{\EM}[2]{{#1}^{#2}}


%% file: Appendix.tex

\tableofcontents

\section{Other constructions on cobordisms}

We present three additional constructions, having to do with union,
intersection, and loops.
We need the internal $J$-category to have a $\vee$-structure: a
map $\vee: \moo + \moo \to \moo$. Write $\vee$ for the function on the
indices, and $(l[0,ij], r[0,ij]) : \anchorint i + \anchorint j \to \anchorint {i
\vee j}$ for the maps of the internal functor.

\subsection{Conjunction}

Conjunction is defined as the pointwise pullback along the maps into $\anchorint
{i\vee j}$ that the $\vee$-structure gives us.
The action on the interfaces is as follows; the action on the cobordisms is
defined similarly by pullback.
Given $\lambda : A \to \anchorint i$ and $\lambda' : B \to \anchorint j$, the
$\vee$-structure gives two squares, one from $\lambda$ to $\id_{\anchorint {i
\vee j}}$ and another from $\lambda'$ to $\id_{\anchorint {i
\vee j}}$.
The conjunction of~$\lambda$ and~$\lambda'$ is then the pullback of these two
maps.

\subsection{Disjunction}

Using the same notations as for the conjunction, we define the disjunction
of~$\lambda$ and~$\lambda'$ to be the obvious map $A + B \to \anchorint {i \vee
j}$.

\subsection{Loops}

To interpret a loop $\while B C$, we build its infinite unfolding as the least
fixpoint of the map:
\[
  F(X) = \sem{\ktest(B)} \fancyseq \sem{C} \fancyseq X \;\graphsum\;
  \sem{\ktest(\neg B)}
\]
seen as an endofunctor on the category of arrows of the double
category~$\Cobord(\moo)$ seen as an internal category in $\Cat$.
It exists because that category has all colimits of $\omega$-chains, and $F$
preserves such colimits because it is itself defined using colimits.

\section{Filling systems and the Hoare inequality}\label{sec:fill-Hoare}
In this section, we suppose that our template and internal opcategory $\moo$ 
is either the stateful template $\mooS$ or the stateless template $\mooL$.
Given two games $\lambda: A \to \moo[0]$ and $\mu: B \to \moo[0]$ formulated as asynchronous morphisms, 
let us describe $\Fill\lambda\mu$.
Every node $x$ of $A$ comes with a state $\theanchorinput{\lambda(x)} \in \moo[1]$
which we call the underlying state of $x$.
Similarly, every node $y$ of $B$ comes with an underlying state 
$\theanchorinput{\lambda(y)} \in \moo[1]$.
The pullback $A \times_{\moo[1]} B$ contains all the pairs
$(a,b) \in A \times B$ which share the same underlying state.
Hence, when we perform the pushout, we identify all such nodes~$a$
and~$b$; in particular, in the case where there is another node~$a'$
of~$A$ with the same underlying states, the pullback will also contain
$(a',b)$, and the nodes~$a$ and~$a'$ will be identified in the pushout.
In summary, the support~$S$ of the filling $\Fill\lambda\mu$ is made
of three kinds of nodes:
\begin{enumerate}
  \item the nodes $x$ of $A$ such that no node of $B$ has the same underlying state;
  \item the nodes $y$ of $B$ such that no node of $A$ has the same underlying state;
  \item the states $s\in\moo[1]$ such that there exists nodes in $A$ and nodes in $B$
    whose underlying states is~$s$ ; we use the notation $[s]$ in order to denote these specific nodes.
\end{enumerate}

\noindent
Let us prove now that the filling system
defined at the end of Section~\ref{sec:sequential-composition}
satisfies the property that there exists a map:
\[
  \Fill{\lambda \parallel \lambda'}{\mu \parallel \mu'} \to
  \Fill \lambda \mu \parallel \Fill{\lambda'}{\mu'}
\]
The map is constructed in the following way.
Consider a node of the support of \( \Fill{\lambda \parallel \lambda'}{\mu\parallel \mu'} \). 
As we have just mentioned, there are three possibilities:
\begin{enumerate}
  \item In the first case, the element is a node of $A \parallel A'$, and thus a pair $(x, x') \in A \times A'$ 
consisting of two elements $x\in A$ and $x'\in A'$ with the same underlying state~$s$.
Recall indeed that the asynchronous morphism $\pince[1]: \mootensor[1] \to \moo[1]$ is \emph{injective on states}.
Since we are in the first case, there exists no node of $B \parallel B'$ with underlying state~$s$. 
This means that either:
    \begin{enumerate}
      \item neither $B$ nor $B'$ have nodes whose underlying state is $s$; 
      in that case the node~$x$ is in $\Fill\lambda\mu$, and the node~$x'$ 
      is in $\Fill{\lambda'}{\mu'}$, and we map $(x,x')$ to $(x,x')$.
      \item $B$ has a node whose underlying state is $s$, but not~$B'$; in
        the same way essentially as in the preceding case, 
        we can map $(x,x')$ to $([s], x')$,
        \item $B'$ has a node whose underlying state is $s$, but not~$B$; 
        this case is symmetric to the previous one.
    \end{enumerate}
  \item the second case is symmetric to the previous one.
  \item last case: the node of the support of $\Fill{\lambda \parallel \lambda'}{\mu\parallel \mu'}$ is of the form $[s]$. 
  In that case, there are nodes in each of the four graphs $A,A',B,B'$ whose underlying states is~$s$.
  We are thus allowed to map $[s]$ to the pair $([s],[s])$.
\end{enumerate}

\section{Semantics of the instructions}

\subsection{Stateful instructions}

The \emph{machine instructions} $m\in\Instr$ which label the machine steps are
of the following form:
\begin{align*}
  m ::=\; &x \coloneqq E
      \mid x \coloneqq \deref{E}
      \mid \deref{E} \coloneqq E'
      \mid \knop\\
      \mid \;&x \coloneqq \kalloc(E,\location)
      \mid \kdispose(E)
      \mid P(r)
      \mid V(r)
\end{align*}
where $x\in\Var$ is a variable, $r\in\RVar$ is a resource name, $\location$~is a
location, and $E,E'$ are arithmetic expressions, possibly with ``\emph{free}''
variables in~$\Var$.
For example, the instruction $x\coloneqq E$ executed in a "machine state"
$\mstate=(\memstate, L)$ assigns to the variable~$x$ the value $E(\mu)\in\Val$
when the value of the "expression" $E$ can be evaluated in the memory state~$\mu$,
and produces the runtime error~$\Error$ otherwise.
The instruction $P(r)$
acquires the resource variable $r$ when it is available, while the instruction
$V(r)$ releases it when~$r$ is locked, as described below:
\[
\begin{small}
\begin{array}{cc}
 \prftree{\evalexpr{E}{\memstate}{v}}
  {\redto{(\memstate, L)}{x \coloneqq E}{(\updmap{\memstate}{x}{v}, L)}}
&
 \prftree{E(\memstate)\,\, \text{not defined}}
  {\redto{(\memstate, L)}{x \coloneqq E}{\Error}}
  \vspace{.4em}
\\
    \prftree{r\notin{}L}
  {\redto{(\memstate, L)}{P(r)}{(\memstate, L\uplus\{r\})}}
&
    \prftree{r\notin{}L}
  {\redto{(\memstate, L\uplus\{r\})}{V(r)}{(\memstate, L)}}
  \end{array}
\end{small}
\]
The inclusion $\Loc \subseteq \Val$ means that an expression $E$ may also
denote a location.
In that case, $[E]$ refers to the value stored at location~$E$ in the heap.
The instruction $x \coloneqq \kalloc(E,\location)$ allocates some memory space
on the heap at address $\location\in\Loc$, initializes it with
the value of the expression $E$, and assigns the address $\location$ to the
variable~$x\in\Var$ if $\location$ was free, otherwise there is no transition.
$\kdispose(E)$ deallocates the location denoted by~$E$ when it is allocated, and
returns~$\Error$ otherwise.
Finally, the instruction $\knop$ (for no-operation) does not alter the state.

\subsection{Stateless instructions}

The machine instructions of the stateless semantics as given bellow, with their
action on the lock state.
\[
\begin{array}{ccccc}
  L \xrightarrow{\,\,P(r)\,\,} L\!\uplus\!\{r\} 
    & \quad &
  L \xrightarrow{\,\,\kalloc(\location)\,\,} L 
  & \quad &
  L \xrightarrow{\;\;\tau\;\;} L 
\vspace{.2em}
\\
  L\!\uplus\!\{r\} \xrightarrow{\,\,V(r)\,\,} L
  &&
  L \xrightarrow{\,\,\kdispose(\location)\,\,} L 
  &&
   L \xrightarrow{\;\;\iwouldprefernotooverline{m}\;\;} \Error
\end{array}\]
where $\iwouldprefernotooverline{m}$ is a \emph{lock instruction} of the form:
\[
  P(r) \mid V(r) \mid \kalloc(\location) \mid \kdispose(\location) \mid \tau
\]
for $\location\in\Loc$ and $r\in\Locks$.
The purpose of these transitions is to extract from each instruction of the
machine its synchronization behavior.
An important special case, the transition~$\tau$ represents the absence of any
synchronization mechanism in an instruction like $x:=E$, $x:=[E]$ or $[E]:=E'$.

\section{The full proof system}

The syntax and the semantics of the formulas of Concurrent Separation Logic is
the same as in Separation Logic.
%
%
The grammar of formulas is:
\begin{align*}
  P,Q,R,J \Coloneqq \; &\emp \mid \True \mid \False \mid P \vee Q \mid
                         P \wedge Q \mid \neg P \\
  \mid\; \forall v. P
         \mid \; \exists v. &P
  \mid\; P * Q \mid v \pointsto{p} w \mid \own_p(x) \mid E_1=E_2
\end{align*}
where $x\in\Var$, $p\in\Perm$, $v,w\in\Val$.
Given a logical state $\sigma=(s,h)$ consisting of a logical stack~$s$ and of a
logical heap~$h$, the semantics of the formulas, expressed as the predicate
$\state \vDash P$, is standard:
\begin{align*}
  \s \vDash v \pointsto p w \;&\Longleftrightarrow\; v\in\Loc \wedge s =\emptyset \wedge h = [v \mapsto (w, p)] \\
  \s \vDash \own_p(x) \;&\Longleftrightarrow\; \exists v\in\Val,  s = [x \mapsto (v, p)] \wedge h = \emptyset\\
  \s \vDash E_1 {=} E_2 \;&\Longleftrightarrow\; \sem{E_1} = \sem{E_2} \wedge \fv(E_1=E_2) \subseteq \vdom(s)\\
  \s \vDash P \wedge Q \;&\Longleftrightarrow\; \s \vDash P \text{ and } \s \vDash Q\\
  \s \vDash P{*}Q \;&\Longleftrightarrow\; \exists \s_1 \s_2, \, \s = \s_1 {*} \s_2
  \;\wedge\; \s_1 \vDash P \;\wedge\; \s_2 \vDash Q.
\end{align*}
%
The proof system underlying concurrent separation logic is a sequent calculus,
whose sequents are Hoare triples of the form
\[
\,\,\triplestd\,\,
\]
where $C\in\Code$, $P$, $Q$ are predicates, and $\Gamma$ is a context, defined
as a partial function with finite domain from the set~$\RVar$ of resource
variables to predicates.
Intuitively, the context $\Gamma=r_1:J_1,\dots,r_k:J_k$ describes the invariant
$J_i$ satisfied by the resource variable $r_i$.
The purpose of these resources is to describe the fragments of memory shared
between the various threads during the execution.

\begin{mathpar}
  \prfbyaxiom{\AFF}{\triple{\Gamma}{(\own_\top(x) * P) \wedge E = v}{x \coloneqq
      E}{(\own_\top(x) * P) \wedge x = v}}
  \and
\prfbyaxiom{\STORE}
{\triple
  {\Gamma}
  {E \mapsto -}
  {\deref{E} \coloneqq E'}
  {E \mapsto E'}}
  \and
\prftree[r]{\IF}
     {P \Rightarrow \defpred(B)}
     {\triple{\Gamma}{P\wedge B}{C_1}{Q}} 
     {\triple{\Gamma}{P\wedge \neg B}{C_2}{Q}} 
     {\triple{\Gamma}{P}{\ifte{B}{C_1}{C_2}}{Q}}
     \and
\prftree[r]{\LOAD}
{x \notin \fv(E)}
{\triple
  {\Gamma}
  {E \mapsto^{p} v * \own_{\top}(x)} 
  {x \coloneqq \deref{E}}
  {E \mapsto^{p} v * \own_{\top}(x) * x = v}}
  \and
  \prftree[r]{\SEQ}{\triple{\Gamma}{P}{C_1}{Q}}{\triple{\Gamma}{Q}{C_2}{R}}{\triple{\Gamma}{P}{C_1;C_2}{R}}
  \and
  \prftree[r]{\DISJ}{\triple{\Gamma}{P_1}{C}{Q_1}}{\triple{\Gamma}{P_2}{C}{Q_2}}{\triple{\Gamma}{P_1
      \vee P_2}{C}{Q_1 \vee Q_2}}
  \and
  \prftree[r]{\RES}{\triple{\Gamma,
      r:J}{P}{C}{Q}}{\triple{\Gamma}{P*J}{\resource{r}{C}}{Q*J}}
  \and
  \prftree[r]{\WHEN}{P\Rightarrow \defpred(B)}{\triple{\Gamma}{(P*J)\wedge
      B}{C}{Q*J}}{\triple{\Gamma, r:J}{P}{\when{r}{C}}{Q}}
  \and
  \prftree[r]{\PAR}{\triple{\Gamma}{P_1}{C_1}{Q_1}}{\triple{\Gamma}{P_2}{C_2}{Q_2}}{\triple{\Gamma}{P_1*P_2}{C_1\parallel
      C_2}{Q_1 * Q_2}}
  \and
  \prftree[r]{\FRAME}
  {\triple{\Gamma}{P}{C}{Q}}
  {\triple{\Gamma}{P*R}{C}{Q*R}}
\end{mathpar}

\section{Proof of the Soundness Theorem, continued}

\subsection{Properties of the interpretation}

\subsubsection{Adhesiveness and van Kampen cubes}\label{sec:adhesive}
In order to reason about the sequential composition, we use the fact that
our ambient category $\AGraph$ is adhesive (as is every topos, 
hence every presheaf category).
We recall the notion from \citep{adhesive}.
\begin{definition}[Adhesive category]
  A category $\catC$ is adhesive if (1) it has pushouts along monos, (2) it has
  all pullbacks, and (3) all pushouts along monomorphisms are van Kampen
  squares.
\end{definition}
A \demph{van Kampen square} is a commutative square
\[
  \begin{tikzcd}
    A \ar[r]\ar[d] & B\ar[d]
    \\
    C \ar[r] & D
  \end{tikzcd}
\]
such that for every commutative cube
\[
  \begin{tikzcd}
    Y \ar[rr]\ar[dd]
    &&
    Z \ar[dd]
    \\
    &
    W \ar[rr,crossing over]\ar[ul]
    &&
    X  \ar[dd]\ar[ul]
    \\
    C \ar[rr]
    &&
    D
    \\
    &
    A \ar[rr]\ar[ul]\ar[uu, crossing over, leftarrow]
    &&
    B\ar[ul]
  \end{tikzcd}
\]
whose bottom face is the square above, and whose two back faces are pullbacks,
the top face is a pushout if and only if the two front faces are pullbacks.
We also call such a commutative cube a \demph{van Kampen cube}.

\subsubsection{Proof of the preservation of strictness}

\begin{lemma}\label{lem:parallel-strict}
  The parallel product preserves strictness.
\end{lemma}
\begin{proof}
  First, we remark that, in the definition of the parallel product, the square 
  \[
    \begin{tikzcd}
      I_1 \parallel I_2 \ar[r] \ar[d] & C_1 \parallel C_2 \ar[d]
      \\
      I_1 \times I_2 \ar[r] & C_1 \times C_2
    \end{tikzcd}
  \]
  is a pullback square.
  The reason is that in the following cube, the bottom, right and left faces are
  pullback squares:
  \[
    \begin{tikzcd}
      I_1 \times I_2 \ar[rr]\ar[dd]
      &&
      C_1 \times C_2 \ar[dd]
      \\
      &
      I_1 \parallel I_2 \ar[rr]\ar[dd]\ar[ul]
      &&
      C_1 \parallel C_2  \ar[dd]\ar[ul]
      \\
      \moo[0] \times \moo[0] \ar[rr]
      &&
      \moo[1] \times \moo[1]
      \\
      &
      \mootensor[0] \ar[rr]\ar[ul]
      &&
      \mootensor[1]\ar[ul]
    \end{tikzcd}
  \]
The latteral faces are defined to be pullbacks, and the bottom one is a pullback
for the three templates that we use.
Now, we consider the following cube:
  \[
    \begin{tikzcd}
      I_1 \times I_2 \ar[rr]\ar[dd]
      &&
      C_1 \times C_2 \ar[dd]
      \\
      &
      I_1 \parallel I_2 \ar[rr]\ar[dd]\ar[ul]
      &&
      C_1 \parallel C_2  \ar[dd]\ar[ul]
      \\
      I'_1 \times I'_2 \ar[rr]
      &&
      C'_1 \times C'_2 
      \\
      &
      I'_1 \parallel I'_2 \ar[rr]\ar[ul]
      &&
      C'_1 \parallel C'_2\ar[ul]
    \end{tikzcd}
  \]
  According to the remark above, the top and bottom faces are pullbacks, and by
  hypothesis the back face is a pullback.
  This concludes the proof.
\end{proof}

\subsection{Fibrational properties of change of lock operations}

Recall that change of locks are a push-then-pull operation, where
pulling is achieved by a pullback and pushing by postcomposition.
As fibrations, epis and monos are preserved under pullbacks in our
ambient category, pulling preserves all the structural properties of
our cobordisms.
Hence, the preservation of the properties of the cobordisms under
change of locks will be determined by the properties of the
asynchronous we push along.
\begin{lemma}
  Given an acute span of $J$-opcategories
  \[
    \begin{tikzcd}
      \moo_1 & \moo_2 \ar[l, "F"] \ar[r, "G"] & \moo_3
    \end{tikzcd}
  \]
  and a cobordism $C$ in $\Cobord(\moo_1)$ which satisfies the
  structural properties of \cref{sec:structural-properties-cobordisms}:
  \begin{itemize}
  \item if every map $g[i] : \moo_2[i] \to \moo_3[f(i)]$ is an epi,
    then condition~(\ref{cond:source-epi}) is preserved,
  \item if every map $g[ij] : \moo_2[ij] \to \moo_3[f(ij)]$ is an
    Environment 1-fibration, then condition~(\ref{cond:supp-1-fib}) is
    preserved.
  \end{itemize}
\end{lemma}

In the case of critical sections, we only use the third condition
above, as the other condition is recovered afterward through
the sequential compositions with $\sem{P(r)}$ and $\sem{V(r)}$ which
determine the input and the output states of $\sem{\when{r}{C}}$.
In the case of the semantics of proofs, this corresponds to the fact
that initial states of a critical section must be states where the
lock is held by the code.
The case for hiding is more straightforward as the components of
$\khide_C$ are epimorphisms and Environment 1-fibrations.
Thus, we have established that change of locks preserve the structural
properties of cobordisms.
We now prove that they also preserve the structural properties of maps
between cobordisms, using the following lemma:

\begin{lemma}
  Given a pair of plain internal functors of $J$-opcategories
  \[
    F : \moo_2 \to \moo_1 \qquad\text{and}\qquad
    F' : \moo'_2 \to \moo'_1
  \]
  and two internal functors of $J$-opcategories
  \[
    G_1 : \moo_1 \to \moo'_1 \qquad\text{and}\qquad
    G_2 : \moo_2 \to \moo'_2
  \]
  such that the obvious square that they compose commutes, and given
  two cobordisms $C$ and $C'$ in $\Cobord(\moo_1)$ and
  $\Cobord(\moo_2)$ respectively, together with a comparison map
  $\mathcal L$ between them that sits over~$G_1$, then operation of
  pulling along~$F$ and~$F'$ gives rise to two cobordisms in
  $\Cobord(\moo'_1)$ and $\Cobord(\moo'_2)$ respectively, together
  with a comparison map $\mathcal{L}'$ which sits over the internal
  functor~$G_2$.
  Assuming $\mathcal L$ is strict, the induced map $\mathcal{L}'$ is
  strict as well providing that the following squares are pullbacks:
  \[
    \begin{tikzcd}
      \moo_2[i] \ar[d, "F"'] \ar[r] & \moo_2[ij] \ar[d, "F"]
      \\
      \moo_1[i] \ar[r] & \moo_1[ij]
    \end{tikzcd}
    \qquad\qquad
    \begin{tikzcd}
      \moo_2'[i] \ar[d, "F'"'] \ar[r] & \moo_2'[ij] \ar[d, "F'"]
      \\
      \moo_1'[i] \ar[r] & \moo_1'[ij].
    \end{tikzcd}
  \]
\end{lemma}
\begin{proof}
  We use the pasting lemma for pullbacks in the cube on the left and
  its counterpart with primes, and then again in the right cube.
  \[
    \begin{tikzcd}
      I \ar[rr]\ar[dd]
      &&
      C \ar[dd]
      \\
      &
      I_1\ar[rr]\ar[dd]\ar[ul]
      &&
      C_1 \ar[dd]\ar[ul]
      \\
      \moo_1[i] \ar[rr]
      &&
      \moo_1[ij] 
      \\
      &
      \moo_2[i] \ar[rr]\ar[ul]
      &&
      \moo_2[ij]\ar[ul]
    \end{tikzcd}
    \quad
    \begin{tikzcd}
      I \ar[rr]\ar[dd]
      &&
      C \ar[dd]
      \\
      &
      I_1 \ar[rr]\ar[dd]\ar[ul]
      &&
      C_1  \ar[dd]\ar[ul]
      \\
      I' \ar[rr]
      &&
      C' 
      \\
      &
      I'_1 \ar[rr]\ar[ul]
      &&
      C'_1\ar[ul]
    \end{tikzcd}
  \]
\end{proof}

The last fibrational property we would like the change of lock
construction to preserve is the fact that the comparison morphism
itself being a fibration of some kind.
The following lemma will handle each of our uses of the construction.
\begin{lemma}
  Using the same assumptions and notations as the the preceding lemma,
  if the components
  \[
    F[ij] : \moo_2[ij] \to \moo_1[ij]
  \]
  are fibrations defined by a right lifting property wrt $S_1 \to S_2$
  and if, for every pairs of maps~$l$ and~$r$, if the following
  diagram commutes:
  \[
    \begin{tikzcd}
      S_2 \ar[r, "r"', shift right] \ar[r, "l", shift left] & \moo'_1[1] \ar[r] & \moo'_2[1]
    \end{tikzcd}
  \]
  then $l = r$, then the change of locks construction preserves that
  kind of fibration.
\end{lemma}
\begin{proof}
  The crux of the proof is the rely on the fact that given that the
  following square is a pullback
  \[
    \begin{tikzcd}
      C\ar[d] & C_1 \ar[d] \ar[l]
      \\
      \moo_1[1] & \moo_2[1] \ar[l]
    \end{tikzcd}
  \]
  the map $C_1 \to C$ is a fibration as well using the hypothesis
  on~$F$.
  Then, one concludes by a diagram chasing and the second
  hypothesis.
\end{proof}

\subsection{Other constructions}

It is easy to check that the \emph{coproduct of cobordisms} preserves
all the properties of the cobordisms and of comparison maps between
that we consider.
From this and from the properties established about sequential
composition, we deduce that it is also the case for
\emph{conditionals}, as they are defined as a combination of these
constructions.

For the case of loops, this follows from the fact that loops are defined as am
unfolding of sequential compositions and of sums.